\documentclass[twocolumn]{aastex63}

\usepackage{amsmath}	
\usepackage{enumitem}
\usepackage[space]{grffile}  
\usepackage{bm} 
\usepackage{makecell, booktabs} 
\usepackage{wasysym} 
\usepackage{hyperref}
\usepackage{lineno} 

\def\Kepler{\textit{Kepler}} 
\def\Gaia{\textit{Gaia}} 

\def\SysSim{\textit{SysSim}} 

\def\HMU{HM-U}
\def\HMC{HM-C}

\defcitealias{2019MNRAS.490.4575H}{Paper I}
\defcitealias{2021AJ....161...16H}{Paper II}
\defcitealias{2020AJ....160..276H}{Paper III}
\defcitealias{2020ApJ...891...12N}{NR20}

\def\PaperI{\citetalias{2019MNRAS.490.4575H}} 
\def\PaperII{\citetalias{2021AJ....161...16H}} 
\def\PaperIII{\citetalias{2020AJ....160..276H}} 
\def\NR20{\citetalias{2020ApJ...891...12N}} 

\received{}
\revised{}
\accepted{}
\shorttitle{Planet Radius Valley and Similarity}
\shortauthors{He et al.}
\graphicspath{{./}{Figures/}}

\begin{document}

\title{Architectures of Exoplanetary Systems. IV: A Multi-planet Model for Reproducing the Radius Valley and Intra-system Size Similarity of Planets around Kepler's FGK Dwarfs}

\correspondingauthor{Matthias Y. He}
\email{matthias.y.he@nasa.gov}

\author[0000-0002-5223-7945]{Matthias Y. He}
\altaffiliation{NASA Postdoctoral Program Fellow}
\affiliation{NASA Ames Research Center, Moffett Field, CA 94035, USA}

\author[0000-0001-6545-639X]{Eric B.\ Ford}
\affil{Department of Astronomy \& Astrophysics, Penn State University, 525 Davey Laboratory, 251 Pollock Road, University Park, PA 16802, USA}
\affil{Center for Exoplanets \& Habitable Worlds, Penn State University, 525 Davey Laboratory, 251 Pollock Road, University Park, PA 16802, USA}
\affil{Center for Astrostatistics \& Astroinformatics, Penn State University, 525 Davey Laboratory, 251 Pollock Road, University Park, PA 16802, USA}
\affil{Institute for Computational \& Data Sciences, Penn State University, 224 Computer Building, University Park, PA 16802, USA}
\email{eford@psu.edu}



\begin{abstract}
The \Kepler{}-observed distribution of planet sizes have revealed two distinct patterns: (1) a radius valley separating super-Earths and sub-Neptunes and (2) a preference for intra-system size similarity. We present a new model for the exoplanet population observed by \Kepler{}, which is a ``hybrid" of a clustered multi-planet model in which the orbital architectures are set by the angular momentum deficit (AMD) stability (He et al. 2020) and a joint mass-radius-period model involving envelope mass-loss driven by photoevaporation (Neil \& Rogers 2020). We find that the models that produce the deepest radius valleys have a primordial population of planets with initial radii peaking at $\sim 2.1 R_\oplus$, which is subsequently sculpted by photoevaporation into a bimodal distribution of final planet radii. The hybrid model requires strongly clustered initial planet masses in order to match the distributions of the size similarity metrics. Thus, the preference for intra-system radius similarity is well explained by a clustering in the primordial mass distribution. The hybrid model also naturally reproduces the observed radius cliff (steep drop-off beyond $\sim 2.5 R_\oplus$). Our hybrid model is the latest installment of the \textit{SysSim} forward models, and is the first multi-planet model capable of simultaneously reproducing the observed radius valley and the intra-system size similarity patterns. We compute occurrence rates and fractions of stars with planets for a variety of planet types, and find that the occurrence of Venus and Earth-like planets drops by a factor of $\sim 2$-4 for the hybrid models compared to previous clustered models in which there is no envelope mass-loss.
\end{abstract}



\section{Introduction} \label{sec:intro}


The discovery of thousands of exoplanets by NASA's \Kepler{} mission has transformed our understanding of the architectures of exoplanetary systems and the physical processes that shape them. The \Kepler{} planet candidate catalog continues to drive statistical analyses of the underlying occurrence and distribution of planets, over fifteen years since the initial launch of the \Kepler{} primary mission \citep{2010Sci...327..977B, 2011ApJ...728..117B, 2013ApJS..204...24B}. Detailed studies of this exoplanet ensemble have revealed that planets with sizes between Earth and Neptune ($R_p = 1$-$4 R_\oplus$) are the most common type in the Galaxy and are ubiquitous in the inner environments ($P \lesssim 1$ yr) of main sequence FGK dwarfs \citep[e.g.,][]{2012ApJS..201...15H, 2013ApJ...766...81F, 2013PNAS..11019273P, 2019AJ....158..109H, 2020AJ....159..248K}. Even greater insights were gained from \Kepler{}'s multi-transiting systems, analyses of which have shown that a substantial fraction of systems have multiple planets within 1 au \citep{2011ApJS..197....8L, 2014ApJ...790..146F}. Further analyses of these multi-planet systems have uncovered numerous patterns in their underlying configurations, including a preference for low mutual inclinations ($\sim 1$-$5^\circ$), small period ratios, and uniform spacings \citep[e.g.,][]{2011ApJS..197....8L, 2012ApJ...751...23F, 2012ApJ...758...39J, 2012AJ....143...94T, 2012arXiv1203.6072W, 2014ApJ...790..146F, 2015MNRAS.448.1956S, 2019MNRAS.490.4575H, 2020AJ....160..276H, 2020AJ....159..281G}.

Perhaps the most readily apparent feature\footnote{The radius valley was not always identifiable from the ensemble of transiting exoplanet candidates observed by \Kepler{}. The initially large uncertainties in the stellar radii led to imprecise planet radii, which smeared the gap in the observed radius distribution. It was not until the more precisely measured stellar and planet radii from the California-\Kepler{} Survey (CKS; \citealt{2017AJ....154..107P, 2017AJ....154..108J, 2017AJ....154..109F}) did the observed radius valley become clear. Now, the radius valley is easily identifiable from the \Kepler{} planet catalog given the updated stellar parameters from \Gaia{} DR2 \citep{2018A&A...616A...1G, 2018A&A...616A...8A}.} in the \Kepler{} planet population is its bimodal distribution of planet radii, with a deficit of planets near $\sim 1.8$-$2 R_\oplus$ separating the mode of the smaller ``super-Earths" from the peak of the larger ``sub-Neptunes" \citep{2017AJ....154..109F, 2018MNRAS.479.4786V}. This feature is commonly described as the planet ``radius valley" and was predicted by photoevaporation theory \citep{2013ApJ...775..105O}. As such, the existence of the radius valley is widely interpreted as a signature of atmospheric mass loss processes that sculpt the observed planet distribution shortly after formation. In the theory of photoevaporation, intense X-ray and extreme ultraviolet (XUV) radiation from the host star drives the hydrodynamic escape of primordial hydrogen-helium envelopes from planets that are not massive enough to gravitationally retain their atmospheres \citep{2013ApJ...776....2L, 2017ApJ...847...29O}. Another mechanism, referred to as the ``core-powered mass loss" (CPML), has also been proposed to explain the same feature, in which the atmospheric mass loss is powered by the thermal energy retained from formation and emitted through the planet's core \citep{2018MNRAS.476..759G, 2019MNRAS.487...24G, 2020MNRAS.493..792G}. Both models make very similar predictions that are consistent with the observed features of the radius valley, including its position as a function of orbital separation \citep{2020ApJ...890...23L, 2021MNRAS.508.5886R, 2023MNRAS.519.4056H, 2023arXiv230200009B}.
In more recent works, other pathways for creating the radius valley have also been proposed, including envelope loss due to giant impacts during the migration phase \citep{2022ApJ...939L..19I} or even formation in a gas-poor disk \citep{2022ApJ...941..186L}.
However, it remains unclear how the radius valley connects with the other architectural properties of multi-planet systems, and whether these processes can explain the other features of the radius distribution.

Another striking pattern in the distribution of planet sizes emerges from analyses of how the planets are distributed among the multi-planet systems. Planets within the same system exhibit a remarkable degree of size similarity, a trend often described as the ``peas-in-a-pod" pattern \citep{2018AJ....155...48W, 2017ApJ...849L..33M, 2020ApJ...893L...1W}. The radii of observed adjacent pairs of planets are highly correlated \citep{2018AJ....155...48W}, and this trend is also seen in the planet masses \citep{2017ApJ...849L..33M, 2022ApJ...933..162G}. These patterns can be expected as an outcome of energy optimization for systems with low mass planets (total mass $\lesssim 40 M_\oplus$; \citealt{2019MNRAS.488.1446A, 2020MNRAS.493.5520A}). Furthermore, the observed preference for size similarity extends beyond simple pair-wise correlations, to a system-level structure in multi-planet systems \citep{2020AJ....159..281G}. However, the detection biases of the \Kepler{} mission have posed a challenge in determining to what extent these patterns are physical in nature \citep{2020AJ....159..188Z, 2020AJ....160..160M}, underscoring the need for forward modeling the \Kepler{} transit survey. As such, forward models of the \Kepler{} mission have enabled us to disentangle the detection biases from the intrinsic patterns, revealing the underlying physical nature of the intra-system size similarity patterns \citep{2019MNRAS.490.4575H, 2020AJ....160..276H, 2021AJ....161...16H, 2021A&A...656A..74M}.

While numerous works have studied the observed radius valley and of the observed intra-system size similarity patterns separately (as described above), a unified model for reproducing both of these features is still lacking. Importantly, these correlations persist across the super-Earth to sub-Neptune transition, implying that the processes that produce the radius valley and those responsible for intra-system size similarity may be physically connected. The size similarity may be even stronger for systems that form planets predominantly on one side of the radius valley \citep{2021ApJ...920L..34M}. A cohesive understanding of these planet size patterns requires a multi-planet population model that can be compared to numerous metrics of the observed \Kepler{} sample, including the planet radius distribution as well as the metrics that quantify the system-level patterns \citep{2020AJ....159..281G}.

In this study, we develop a detailed multi-planet population model for describing the exoplanet ensemble observed by \Kepler{}, with the goal of jointly reproducing the planet radius valley and intra-system size similarity patterns. We develop this model as part of the Exoplanets Systems Simulator (``\SysSim{}"; \citealt{eric_ford_2022_5915004}) codebase which implements a sophisticated forward model for the \Kepler{} detection pipeline that makes use of multiple \Kepler{} data products \citep{2017ksci.rept...18C, 2017ksci.rept...14B, 2017ksci.rept...19B}. The details of how \SysSim{} simulates the \Kepler{} detection efficiency are described in \citet{2018AJ....155..205H, 2019AJ....158..109H, 2020MNRAS.498.2249H} (which compute planet occurrence rates for \Kepler{} FGK and M dwarfs). \SysSim{} has been extensively used to constrain the multi-planet population models in the ``Architectures of Exoplanetary Systems" paper series: \PaperI{} \citep{2019MNRAS.490.4575H} introduced an initial, clustered Poisson point-process model; \PaperII{} \citep{2021AJ....161...16H} extended the model to measure the change in the fraction of stars with planets as a function of host stellar type; finally, \PaperIII{} \citep{2020AJ....160..276H} incorporated the concept of the angular momentum deficit (AMD) stability criterion for describing the orbital architectures of multi-planet systems and resolving the apparent \Kepler{} dichotomy. In this installment (``Paper IV"), we construct a ``hybrid" model that combines features of the clustered, maximum AMD model from \PaperIII{} with a joint mass-radius-period model involving photoevaporation-driven envelope mass loss from \citet{2020ApJ...891...12N} (hereafter \NR20{}).

This paper is organized as follows. In \S\ref{sec:methods}, we describe our Methods, including a summary of the main features of the previous \SysSim{} models (\S\ref{sec:methods:prev_models}), the new model (\S\ref{sec:methods:new_model}), and our procedure for performing inference on the new model (\S\ref{sec:methods:model_inference}, including the \Kepler{} data, summary statistics, and distance function we use). The main results are described in \S\ref{sec:results}, beginning with the constraints on the parameters of the new models and how well they fit the observed distributions of the \Kepler{} catalog (\S\ref{sec:results:new_model_fit}), then focusing on the observed planet radius valley (\S\ref{sec:results:radius_valley}), before describing the model predictions for the underlying distributions of planetary systems (\S\ref{sec:results:model_underlying}). In \S\ref{sec:discussion}, we discuss our choice of the distance function (\S\ref{sec:discussion:distance_function}) and the planet radius ``cliff" (\S\ref{sec:discussion:radius_cliff}). We also compute detailed planet occurrence rates and fractions of stars with planets and compare with previous values in the literature in \S\ref{sec:discussion:occurrence_rates}, and discuss the limitations and avenues for future work in \S\ref{sec:discussion:limitations}. Finally, we summarize our main results and conclusions in \S\ref{sec:summary}.

\section{Methods} \label{sec:methods}

\SysSim{} works by generating large numbers of synthetic planetary systems from a given population model (with tunable model parameters), applying a detailed simulation of the \Kepler{} detection pipeline to generate catalogs of ``observed" systems of transiting planets, and finally comparing these simulated catalogs to the actual \Kepler{} planet catalog using the framework of approximate Bayesian computation (ABC). This procedure constitutes a full forward model for the \Kepler{} primary mission and can be summarized by the following steps:
\begin{enumerate}
 \item Define a model for the intrinsic, underlying distribution of exoplanetary systems. \label{step:model}
 \item Generate a ``physical catalog", by drawing a number of stars from the \Kepler{} target list and a planetary system from the model to assign to each star. \label{step:physical_catalog}
 \item Generate an ``observed catalog" from the physical catalog, by running the \SysSim{} detection pipeline to simulate the \Kepler{} detection efficiency, along with the transit observables and their measurement uncertainties, for each planet.
 \item Compare the simulated observed catalog with the \Kepler{} planet catalog using a number of statistical metrics, which go into computing a ``distance function" for quantifying the goodness of the model.
\end{enumerate}
The ``model" in step \ref{step:model} can be a combination of statistical and/or physical descriptions for the distribution of planetary systems, typically with several free model parameters which we wish to infer, and it must be sufficiently well defined such that the physical properties (planet radii, masses, orbital periods, and other orbital elements) of the planetary systems can be directly drawn from it. In this paper, we will primarily introduce a new, updated model for describing multi-planet systems.

The steps above are then iterated many times in an inference stage, in order to find the best-fit model parameters as well as the credible regions/distributions of the parameters. This inference stage involves applying an ABC algorithm and is implemented as follows:
\begin{enumerate}[label=(\alph*)]
 \item Optimize the distance function using a differential evolution algorithm to find the region of best-fit model parameters and obtain a large sample of ``training points" (sets of model parameters with distances evaluated from the forward model). \label{step:optimizer}
 \item Train a Gaussian Process (GP) emulator for rapidly predicting the distance function for a given draw of parameters, using the training points obtained from running the optimizer. \label{step:emulator}
 \item Compute the ABC posterior distributions for the model parameters by drawing points from a prior distribution, evaluating the distance function using the emulator, and accepting those that are below a distance threshold. \label{step:abc}
\end{enumerate}
In theory, step \ref{step:abc} can be performed with the full forward model, without using the emulator in step \ref{step:emulator}. In practice, the forward model is computationally expensive enough (with most of the time spent on step \ref{step:physical_catalog} to generate the physical catalog) such that the emulator is needed to rapidly assess the distance function in order to obtain a sufficient number of points for characterizing the ABC posterior.

\subsection{Previous \SysSim{} models} \label{sec:methods:prev_models}

Before introducing our new model, we first briefly review and summarize the main features of the previous \SysSim{} models.

\subsubsection{Clusters of planets} \label{sec:methods:prev_models:clusters}

\PaperI{} introduced the first ``clustered" model in which planets are drawn from a Poisson point process, where the planets within the same cluster have correlated orbital periods and planet radii.
The number of clusters ($N_c$), and planets per cluster ($N_p$), are drawn independently from zero-truncated Poisson (ZTP) distributions given by the probability mass function:
\begin{equation}
 p(k; \lambda) = \frac{\lambda^k}{k!(e^\lambda -1)}, \label{eq:ZTP}
\end{equation}
where $k > 0$ is a natural number and $\lambda$ is the rate parameter. Thus we have $N_c \sim {\rm ZTP}(\lambda_c)$ and $N_p \sim {\rm ZTP}(\lambda_p)$ where $\lambda_c$ and $\lambda_p$ are free parameters of the model.
Although \PaperI{} found that most ($\sim 80\%$) of the systems have only one cluster when constrained by the \Kepler{} data, multiple clusters can also exist in some planetary systems allowing for widely spaced and/or different size-classes of planets.

\subsubsection{Occurrence with spectral type}

In \PaperII{}, a change in the occurrence of planetary systems across host stellar type, in the form of a linear dependence on the \Gaia{} $b_p-r_p$ color (a proxy for spectral type) for the fraction of stars with planets, $f_{\rm swpa}$, was added:
\begin{eqnarray}
 && f_{\rm swpa}(c) = \\
 && \quad \max \Big\{ 0,\min \Big[ m \Big(c - c_{\rm med} \Big) + f_{\rm swpa,med}, 1 \Big] \Big\} \nonumber \label{eq:fswp}
\end{eqnarray}
where $c = b_p-r_p-E^*$ is the reddening-corrected \Gaia{} color, $c_{\rm med} \simeq 0.81$ is the median value for our FGK sample, and $m \equiv d{f_{\rm swpa}}/d(c)$ and $f_{\rm swpa,med} \equiv f_{\rm swpa}(c_{\rm med})$ are the slope and normalization fitted as free parameters of the model. This model demonstrated that an increase in planetary system occurrence towards later type dwarfs (redder stars, i.e. higher values of $c$, such that $m > 0$) is necessary to account for the relative numbers of observed transiting planets across the FGK range, while also retaining the other model features.

\subsubsection{AMD-stable systems} \label{sec:methods:prev_models:amd_stability}

Finally, in \PaperIII{} a procedure for drawing the orbital eccentricities and mutual inclinations for multi-planet systems by enforcing the AMD-stability criteria was invented. This procedure replaces the parameterizations of the eccentricity and mutual inclination distributions, previously described by a mixture of Rayleigh distributions.
The total AMD of a planetary system is simply the sum of the AMD of the individual planets:
\begin{equation}
 {\rm AMD}_{\rm tot} = \sum_{k=1}^{N} \Lambda_k \Big(1 - \sqrt{1-e_k^2}\cos{i_{m,k}} \Big), \label{eq:amd_tot}
\end{equation}
where $\Lambda_k = \mu_k\sqrt{a_k}$ and $\mu_k = M_{p,k}/M_\star$ is the planet-star mass ratio (in units of $GM_\star \equiv 1$), and $a_k$, $e_k$, and $i_{m,k}$ are the semi-major axis, orbital eccentricity, and mutual inclination (relative to the system invariable plane) of the $k^{\rm th}$ planet, respectively.
It can be shown that there exists a critical value:
\begin{eqnarray}
 {\rm AMD}_{\rm crit} &=& {\rm min}\bigg[ \Big\{ \Lambda_k {\rm min}\big( \mathcal{C}_{\rm coll}, \mathcal{C}_{\rm mmr} \big) : k=2,\dots,N \Big\} \nonumber \\
 && \quad\quad \cup \Big\{\Lambda_1 \Big\} \bigg], \label{eq:amd_crit}
\end{eqnarray}
below which the planets are prevented from having any overlapping orbits regardless of how this AMD is partitioned between them (as well as preventing a collision between the inner-most planet and the host star, i.e. the $\Lambda_1$ term). Here, $\mathcal{C}_{\rm coll}$ and $\mathcal{C}_{\rm mmr}$ are the ``critical relative AMD" for collision and MMR overlap, respectively, computable from the masses and semi-major axes of the planets alone (see \S2.3.1 of \PaperIII{} for a summary of these equations/derivations, or \citealt{2017A&A...605A..72L, 2017A&A...607A..35P} for more complete derivations).
In addition, the criterion $\alpha < \alpha_{\rm crit} \simeq 1-1.46(\mu_1 + \mu_2)^{2/7}$ where $\alpha = a_1/a_2$ is the semi-major axis ratio and $\mu_1$, $\mu_2$ are the planet-star mass ratios is also enforced for all adjacent planet pairs to ensure stability against MMR overlap for circular orbits (\citealt{2017A&A...607A..35P}; see also \citealt{1980AJ.....85.1122W, 2013ApJ...774..129D}).

The so-called ``maximum AMD model" (hereafter, the H20 model) from \PaperIII{} thus assigns a total amount of AMD equal to the critical AMD, for each system, thereby producing planetary systems that are at the boundary of AMD-stability. This AMD ``budget" is partitioned per unit mass between the individual planets, so as to provide the same degree of dynamical excitation for all planets in a given system. Finally, the AMD given to each planet is randomly distributed among the three excitation components $x = e\sin{\omega}$, $y = e\cos{\omega}$, and $z = \sin{i_m}$ via the constraint:
\begin{equation}
 x^2 + y^2 + z^2 = \frac{{\rm AMD}_k}{\Lambda_k} \bigg(2 - \frac{{\rm AMD}_k}{\Lambda_k} \bigg), \label{eq:amd_partition}
\end{equation}
achieved by drawing $x^2$, $y^2$, and $z^2$ from a symmetric Dirichlet distribution with concentration parameter $\alpha = (1,1,1)$ and then multiplying by the right-hand side of equation \ref{eq:amd_partition} to yield the constraint. Finally, the orbital eccentricity, argument of pericenter, and mutual inclination of the planet relative to the system invariable plane can be easily computed via $e = \sqrt{x^2 + y^2}$, $\omega = {\rm atan}(x,y)$, and $i_m = \sin^{-1}z$, respectively.

In addition to the AMD-stability criteria, a simple criterion for the minimum separation between adjacent planets is also enforced (as was also done in Papers I--II):
\begin{eqnarray}
 \Delta_c < \Delta = \frac{a_{\rm out}(1 - e_{\rm out}) - a_{\rm in}(1 + e_{\rm in})}{R_H}, \label{eq:min_separation} \\
 R_H = \bigg(\frac{a_{\rm in} + a_{\rm out}}{2}\bigg) \bigg[\frac{m_{p,\rm in} + m_{p,\rm out}}{3 M_\star}\bigg]^{1/3}, \label{eq:mutual_hill_radius}
\end{eqnarray}
where $\Delta$ is the separation between the inner planet's apoapsis and the outer planet's periapsis in units of mutual Hill radii $R_H$, and $\Delta_c$ is the minimum separation parameter.

\subsubsection{Underlying distribution of planet radii and masses} \label{sec:methods:prev_models:mass_radius}

A choice common to the clustered models from all three papers is the underlying planet radius distribution and mass-radius (M--R) relationship. The planet radius distribution is described by draws from a lognormal distribution centered around a cluster ``radius scale" ($R_{p,c}$) for the planets in the same cluster, where the cluster radius scale is drawn from a broken power-law:
\begin{eqnarray}
 R_p &\sim& \ln\mathcal{N}(R_{p,c}, \sigma_R), \label{eq:radii_clustered} \\
 f(R_{p,c}) &\propto& \left\{
 \begin{array}{ll}
  {R_{p,c}}^{\alpha_{R1}}, & R_{p,\rm min} \leq R_{p,c} \leq R_{p,\rm break} \label{eq:radii_scales} \\
  {R_{p,c}}^{\alpha_{R2}}, & R_{p,\rm break} < R_{p,c} \leq R_{p,\rm max}
 \end{array} \right..
\end{eqnarray}
Here, $R_p$ is a planet radius, $\alpha_{R1}$ and $\alpha_{R2}$ are the power-law indices below and above the break radius $R_{p,\rm break}$, and $\sigma_R$ is a cluster ``width" that controls the degree of size similarity. The radii bounds were chosen to be $(R_{p,\rm min}, R_{p,\rm max}) = (0.5, 10) R_\oplus$ and the break radius fixed at $R_{p,\rm break} = 3 R_\oplus$.

The planet masses are then drawn from the nonparametric and probabilistic M--R relation from \citet{2018ApJ...869....5N} (hereafter NWG18), with a modification for small planets below $R_p \simeq 1.47 R_\oplus$ introduced in \PaperIII{}. The NWG18 relation uses a series of Bernstein polynomials for fitting the joint M--R distribution with 55 degrees of freedom in each dimension, and was fit to a sample of 127 \Kepler{} exoplanets with measured masses from RVs or transit timing variations. 
The modification adopted in \PaperIII{} involves switching to a lognormal distribution centered around the ``Earth-like rocky" model from \citet{2019PNAS..116.9723Z}:
\begin{eqnarray}
 \log_{10}\bigg(\frac{M_p}{M_\oplus}\bigg) &\sim& \mathcal{N}(\mu, \sigma), \nonumber \\
 \mu &=& \log_{10}\Big(\frac{M_{p,\rm ELR}(R_p)}{M_\oplus}\Big), \nonumber \\
 \sigma &=& m(R_p - R_{p,\rm trans}) + \sigma(R_{p,\rm trans}) \label{eq:MRsmall}
\end{eqnarray}
where $M_{p,\rm ELR}(R_p)$ is the M--R relation from interpolating the Earth-like rocky model \citep{2019PNAS..116.9723Z} and $R_{p,\rm trans} = 1.472 R_\oplus$ is the transition radius, chosen to be where $M_{p,\rm ELR}(R_p)$ intersects the mean prediction of the NWG18 model. The scatter in mass, $\sigma$, is parameterized as a linear function of $R_p$,  where the slope $m = (\sigma(R_{p,\rm trans}) - \sigma(R_{p,\rm min}))/(R_{p,\rm trans} - R_{p,\rm min})$ takes the scatter from $\sim 10\%$ at the smallest radii ($\sigma(R_{p,\rm min}) = 0.04$) to a scatter similar to that of the NWG18 model at the transition radius ($\sigma(R_{p,\rm trans}) = 0.3$). Thus, the M--R relation is approximately continuous across the transition radius both in terms of the mean prediction and the scatter, and it remains probabilistic at small planet sizes.

It is worth emphasizing that the previous models involved an M--R relation simply for the purposes of assigning reasonable planet masses, and not for performing inference on the details of the M--R model, as is a goal of this paper. In Papers I--III, the planet radius distribution was observationally constrained in two main ways: (1) by the observed distribution of transit depths, which is a function of the ratio of planet radii to stellar radii and is directly fitted (i.e. included as a term in the distance function), and (2) by the detectability of each transiting planet, which is a strong function of the transit depths and thus affects all of the other observable distributions. On the other hand, the distribution of planet masses is not well constrained by the data. The planet masses only indirectly influence the observed properties of the transiting planets by their orbital separations (via their spacings in mutual Hill radii) and excitations (via their eccentricities and mutual inclinations drawn from the AMD in the H20 model, which is a function of planet mass). Further, the M--R relation described above was kept fixed (there are no free parameters), and thus any of these influences would propagate to constrain the parameters of the radius distribution and other model parameters when fitting to the \Kepler{} catalog via ABC.

The full procedure for drawing from the H20 model is described in Section 2.3.4 of \PaperIII{}. We will replace the underlying model from which the planet radii and masses are drawn, and describe a new, step-by-step procedure for drawing planetary systems from the new model in \S\ref{sec:methods:new_model}.

\subsection{New model} \label{sec:methods:new_model}

\begin{figure}
\centering
\includegraphics[scale=0.52,trim={0.5cm 0.5cm 0.5cm 0.2cm},clip]{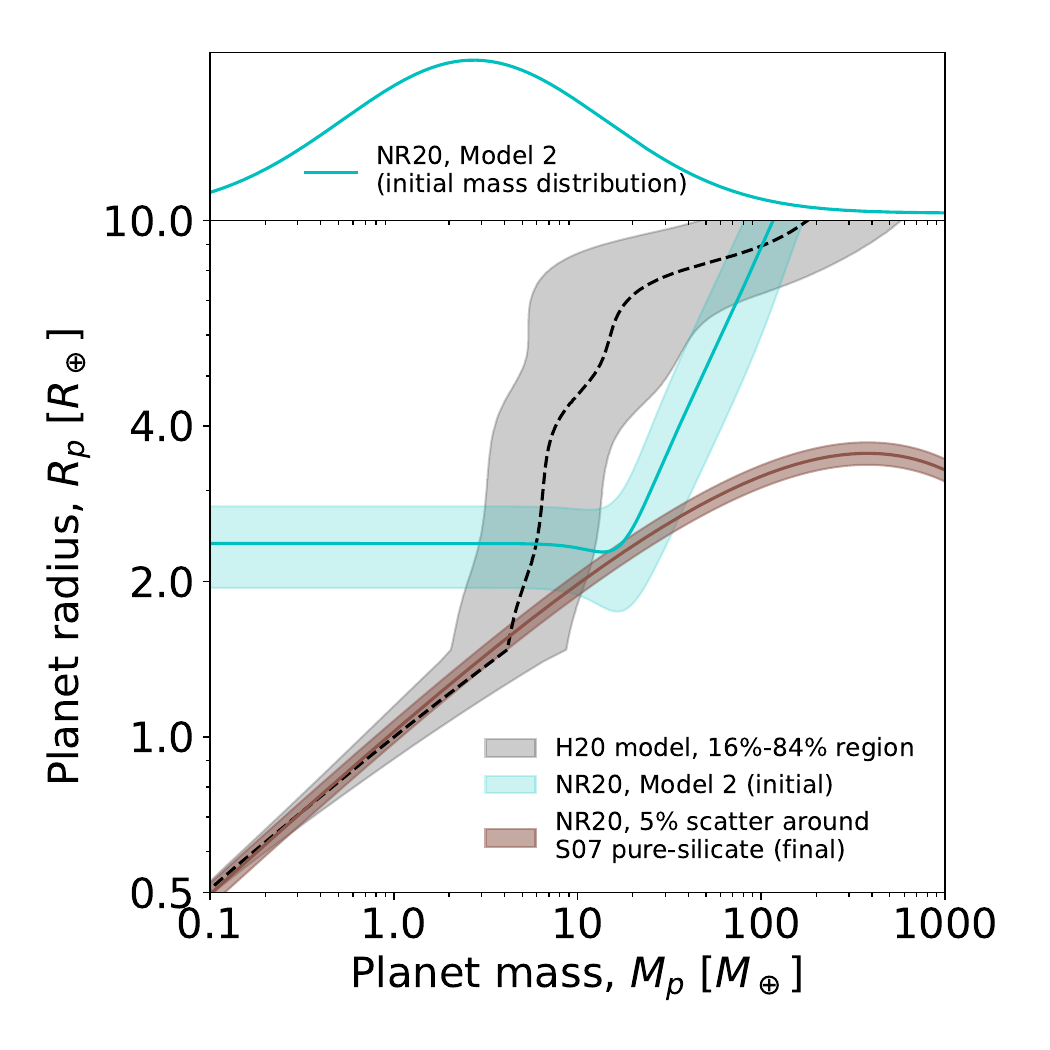} 
\caption{Radius-mass relationships of various models. The H20 model is shown as the black dashed line (median prediction) and the gray shaded region (16-84\% quantiles), for reference. The initial radius-mass relation from the \NR20{} models is shown in cyan (solid line for the median prediction and shaded region for the 16-84\% quantiles). It is a power-law with two break points, smoothed by logistic functions, as also described in \S\ref{sec:methods:new_model:initialMR}; for illustrative purposes, here we have adopted the best-fit parameters from ``Model 2" of \NR20{} (the upper break occurs at $>10 R_\oplus$ and is thus off the axes). The brown curve shows the pure-silicate relation from \citet{2007ApJ...669.1279S} (equation \ref{eq:radius_mass_pure_silicate}), where the shaded region denotes a 5\% scatter as adopted by \NR20{} for drawing final planet radii from their core masses for the planets that have lost their envelopes.
The top panel shows the initial planet mass distribution from \NR20{} model 2.}
\label{fig:radius_mass}
\end{figure}

\subsubsection{Distribution of initial planet masses and radii} \label{sec:methods:new_model:initialMR}

Our new model is strongly inspired by the models from \NR20{}. In that paper, a series of four models, each with increasing complexity and numbers of model parameters over the previous, were introduced with the goal of reproducing the observed radius valley. Each of those models assumed an underlying, \textit{initial} planet mass distribution characterized by a lognormal distribution that is independent of all other factors, and an \textit{initial} planet radius distribution that is conditioned on the planet masses. It is important to emphasize that these are the initial masses and radii, which can be interpreted as the planet properties after formation but before any subsequent photoevaporation or other sculpting processes. This is in contrast to the previous \SysSim{} models, which parameterized the \textit{final} (i.e. ``present day") underlying distributions of planet masses and radii. For our new model, we adopt the same initial planet mass distribution as \NR20{}:
\begin{equation}
 M_{p,\rm init} \sim \ln\mathcal{N}(\mu_M, \sigma_M), \label{eq:mass_init}
\end{equation}
where $M_{p,\rm init}$ is an (initial) planet mass, and $\mu_M$ and $\sigma_M$ are the mean and standard deviation of the log-planet masses (i.e. of $\ln{M_{p,\rm init}}$) to be inferred.

We also adopt a similar radius-mass relationship, where the planet radii are drawn from a probabilistic model characterized by a normal distribution around the mean prediction with a fractional scatter:
\begin{equation}
 R_{p,\rm init} | M_{p,\rm init} \sim \mathcal{N}(\mu_R, \sigma_R \mu_R), \label{eq:radius_init}
\end{equation}
where $R_{p,\rm init}$ is the (initial) planet radius, $\mu_R = \mu_R(M_{p,\rm init})$ is the mean radius at a given mass, and $\sigma_R = \sigma_R(M_{p,\rm init})$ is a scatter term in the radius (it is scaled by the mean $\mu_R$ to account for the greater scatter towards larger masses). While \NR20{} used a double broken power-law (i.e. three regimes, following \citealt{2017ApJ...834...17C}) for the mean mass-radius relation, smoothed by a logistic function, we use a single broken power-law:
\begin{eqnarray}
 \mu_{R,0} &=& R_{p,\rm norm} M_{p,\rm init}^{\gamma_0}, \\ 
 \mu_{R,1} &=& R_{p,\rm norm} M_{p,\rm break}^{\gamma_0-\gamma_1} M_{p,\rm init}^{\gamma_1}, \label{eq:radius_mass} 
\end{eqnarray}
where $M_{p,\rm break}$ is the break mass, $\gamma_0$ and $\gamma_1$ are the power-law indices below and above the break, respectively, and $R_{p,\rm norm}$ is a normalization constant (units of planet radius, in $R_\oplus$). We choose $M_{p,\rm min} = 0.1 M_\oplus$ and $M_{p,\rm max} = 1000 M_\oplus$, approximately corresponding to the range between the estimated masses of the smallest \Kepler{} planets and about three Jupiter masses; the break mass $M_{p,\rm break}$ is a free parameter. This broken power-law is also smoothed by:
\begin{eqnarray}
 \mu_R &=& (1-S)\mu_{R,0} + S \mu_{R,1}, \\ \label{eq:radius_mass_smoothed}
 \sigma_R &=& (1-S)\sigma_0 + S \sigma_1, \\
 S &=& \frac{1}{1 + {\rm exp}(-5(\ln{M_{p,\rm init}} - \ln{M_{p,\rm break}))}}. \label{eq:logistic_smoothed}
\end{eqnarray}
This is a simplification of equations (8)-(10) in \NR20{}, as we have a single break in the radius-mass relationship instead of their double breaks. They find the upper break to be $\sim 170 M_\oplus$ in their models 2--4; given that most of the planets drawn in our models will be less massive than that, our results will not be strongly affected by the choice of a simpler functional form, and this has the advantage of reducing the number of model parameters by three (the upper break mass, as well as an additional power-law index and scatter).

In Figure \ref{fig:radius_mass}, we plot the radius-mass model from \NR20{} along with the mass-radius relationship from the H20 model. The mass-radius relationship used in the H20 model, described in \S\ref{sec:methods:prev_models:mass_radius}, is shown in black (the dashed line for the median prediction and the gray shaded region for the 16\%-84\% quantiles). The cyan curve and shaded region displays the median and 16\%-84\% quantiles of the radius-mass relation given by equations \ref{eq:radius_init}-\ref{eq:logistic_smoothed}, where we have adopted the best-fit ``Model 2" parameters from \NR20{} for illustrative purposes ($\mu_M = 1$, $\sigma_M = 1.65$, $R_{p,\rm norm} = 2.37 R_\oplus$\footnote{This parameter is denoted by the symbol ``$C$" in \NR20{}.}, $M_{p,\rm break} = 17.4 M_\oplus$, $\gamma_0 = 0$, $\gamma_1 = 0.74$, $\sigma_0 = 0.18$, and $\sigma_1 = 0.34$).

\subsubsection{Clustered initial masses}
\label{sec:methods:new_model:clustered}

As discussed in \S\ref{sec:intro}, there is observational evidence for intra-system size similarity among \Kepler{}'s multi-planet systems, both in terms of the planet radii \citep{2018AJ....155...48W, 2020ApJ...893L...1W, 2022arXiv220310076W} and the planet masses \citep{2017ApJ...849L..33M, 2017RNAAS...1...26W, 2020MNRAS.493.5520A, 2022ApJ...933..162G}, although the detection biases of the \Kepler{} mission plays a confounding role \citep{2020AJ....159..188Z, 2020AJ....160..160M}. In Papers I--III, we showed via forward modeling that a clustering in the physical planet radius distribution is necessary to generate the observed ``peas-in-a-pod" patterns. Developing a new model in \SysSim{} to better reproduce the observed planet radius distribution provides an ideal opportunity to revisit the extent of these size similarity patterns.

The model described above can produce populations of planets with similar masses and radii, depending on the values of the parameters defining the initial mass distribution and radius-mass relation. In particular, a smaller value of $\sigma_M$ will produce more similar draws of the initial masses, and a flatter radius-mass relation (e.g. $\gamma_0 \sim 0$) with a tighter spread around the mean relation (i.e. smaller $\sigma_0$ and $\sigma_1$) will lead to similar draws of the initial radii. For example, \NR20{} find a very flat radius-mass relation below the break mass; their $\gamma_0$ is consistent with zero for their ``Model 2" and ``Model 4", producing similar initial radii for all planets below the break mass. 

However, this model cannot produce \textit{intra-system} size similarity, which is a preference for the sizes (either radii or masses) of the planets in the same system to be more similar than compared to the overall population. In other words, the size similarity of planets within the same system is not enhanced compared to planets drawn randomly from different systems. In order to allow for the possibility of intra-system size similarity, we also consider an alternative model in which the initial planet masses are clustered, replacing Equation \ref{eq:mass_init} with:
\begin{eqnarray}
 \mu_{M,c} &\sim& \ln\mathcal{N}(\mu_M, \sigma_{M}), \nonumber \\
 M_{p,\rm init} &\sim& \ln\mathcal{N}(\mu_{M,c}, \sigma_{M,c}). \label{eq:mass_init_clustered}
\end{eqnarray}
Here, $\mu_{M,c}$ is the mean of the log-planet masses of a given cluster (drawn from the lognormal distribution for the overall population), and $\sigma_{M,c}$ is the cluster width of the lognormal distribution for drawing the (initial) masses within the cluster. This is an additional free parameter to be inferred, with smaller values of $\sigma_{M,c}$ indicating stronger clustering (i.e., more correlated initial masses for planets within the same cluster).

We choose to parameterize the pattern of intra-system similarity in terms of the planet masses instead of the planet radii (as was done in \PaperI{}-\PaperIII{}), because the new models start with drawing planet masses instead of radii. Furthermore, mass is the more fundamental quantity in terms of the energy optimization of a planetary system, which may explain the transition from the peas-in-a-pod patterns to runaway growth of a single planet \citep{2019MNRAS.488.1446A, 2020MNRAS.493.5520A}. Nevertheless, the clustering in planet masses propagates into also producing clustered planet radii.
Instead of assuming that intra-system size similarity exists in this study, we will also test the need for clustered planet sizes by considering two versions of the new model, one without and one with the clustered initial masses.

\subsubsection{Envelope mass loss via photoevaporation}
\label{sec:methods:new_model:photoevap}

Previous studies have already demonstrated that the process of photoevaporation due to incident X-ray and extreme ultraviolet (collectively, XUV) radiation is able to reproduce the planet radius valley reported in \citet{2017AJ....154..109F, 2018MNRAS.479.4786V}. 
\NR20{} adopted the analytic prescription for XUV-driven mass loss from \citet{2012ApJ...761...59L} for their models, which we also restate and use here. First, each planet is prescribed an amount of mass in its gaseous envelope, $M_{\rm env}$, which can potentially be lost due to photoevaporation, chosen as a simple, two-regime function of the total planet mass that is also smoothed by a logistic function:
\begin{eqnarray}
 M_{\rm env} &=& (1-S)M_{\rm env,0} + S M_{\rm env,1}, \label{eq:envelope_mass} \\
 M_{\rm env,0} &=& 0.1 M_{p,\rm init}, \label{eq:envelope_mass_low} \\
 M_{\rm env,1} &=& M_{p,\rm init} - \sqrt{(M_{p,\rm init})}, \label{eq:envelope_mass_high} \\
 S &=& \frac{1}{1 + {\rm exp}(-5(\ln{M_{p,\rm init}} - \ln{M_{\rm trans}}))}.
\end{eqnarray}
Here, \NR20{} adopted the scaling of planet mass in heavy elements from \citet{2016ApJ...831...64T} and a transition mass of $M_{\rm trans} = 20 M_\oplus$, which we also use. The mass-loss timescale (\citealt{2012ApJ...761...59L}, \NR20{}) can then be computed by:
\begin{equation}
 t_{\rm loss} = \frac{GM_{\rm env}^2}{\pi \epsilon R_{p,\rm init}^3 F_{\rm XUV,E100}} \cdot \frac{F_\oplus}{F_p}, \label{eq:mass_loss_timescale}
\end{equation}
where $G$ is the gravitational constant, $\epsilon$ is the mass-loss efficiency, $F_{\rm XUV,E100}$ is the XUV flux on Earth at 100 Myr,  and $F_p$ is the bolometric incident flux on the planet. The mass-loss timescale of each planet is then compared to the age of the system $\tau$ to model the probability of envelope mass-loss as:
\begin{eqnarray}
 p_{\rm loss} &=& 1 - p_{\rm ret} \\
 &=& 1 - {\rm min}\Big(\alpha_{\rm ret}\frac{t_{\rm loss}}{\tau}, 1\Big), \label{eq:prob_mass_loss}
\end{eqnarray}
where $p_{\rm ret}$ is the probability of the planet \textit{retaining} its envelope and $\alpha_{\rm ret}$ is a normalization factor for capturing our uncertainties in the adopted nominal values of $\epsilon = 0.1$, $F_{\rm XUV,E100} = 504$ erg s$^{-1}$ cm$^{-2}$, and $\tau = 5$ Gyr for each star.

For each planet, we draw a random variable $x \sim {\rm Unif}(0,1)$ to determine whether it will lose its envelope. If $x < p_{\rm loss}$ (implying envelope mass-loss has occurred due to photoevaporation), we subtract the planet's envelope from its total initial mass to compute its final mass:
\begin{equation}
 M_{p,\rm final} = M_{p,\rm init} - M_{\rm env}. \label{eq:mass_final}
\end{equation}
Following \NR20{}, we then treat the resulting planet mass as a rocky core of pure silicate by drawing a radius from a normal distribution centered around the analytical radius-mass relation from (\citealt{2007ApJ...669.1279S}; $R_{\rm si}$) with a fixed fractional scatter of 5\%:
\begin{eqnarray}
  R_{p,\rm final} &\sim& \mathcal{N}(R_{\rm si}, 0.05 R_{\rm si}), \label{eq:radius_final} \\
  R_{\rm si} &=& R_1 \cdot 10^{k_1 + \frac{1}{3}\log_{10}(M_p/M_1) - k_2 (M_p/M_1)^{k_3}}, \label{eq:radius_mass_pure_silicate}
\end{eqnarray}
where $R_1 = 3.90$, $M_1 = 10.55$, $(k_1, k_2, k_3) = (-0.209594, 0.0799, 0.413)$ \citep{2007ApJ...669.1279S}, and $M_p$ is taken to be $M_{p,\rm final}$. This relation is shown as the brown curve (along with the 5\% fractional scatter as the shaded region) in Figure \ref{fig:radius_mass}. We note that this is very similar to the ``Earth-like rocky" model from \citet{2019PNAS..116.9723Z} adopted for small planets ($R_p \lesssim 1.47 R_\oplus$) in the H20 model and described in \S\ref{sec:methods:prev_models:mass_radius}. The planets that have lost their envelopes thus scatter around this curve.

\begin{figure}
\centering
\includegraphics[scale=0.24,trim={0.5cm 0cm 0cm 0cm},clip]{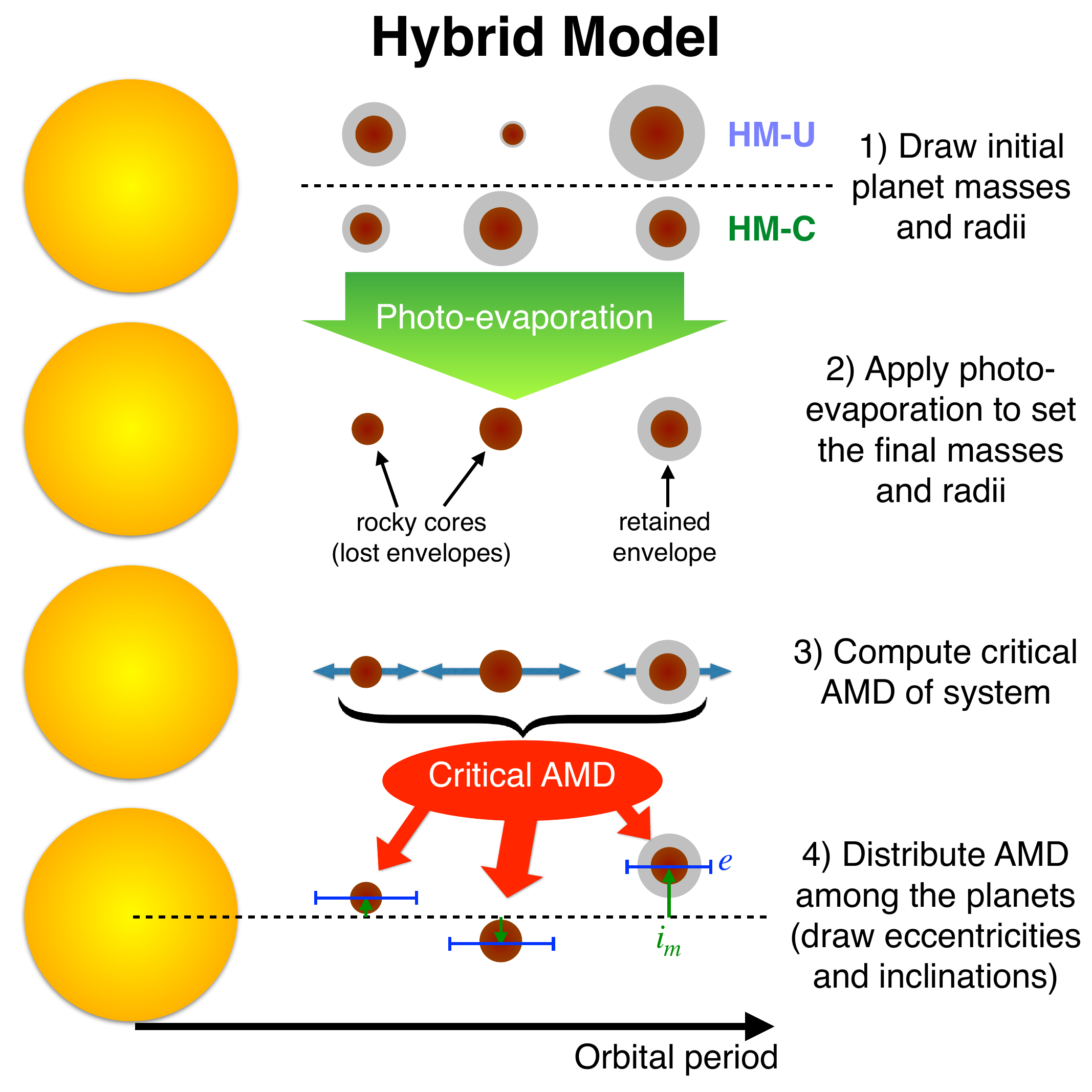} 
\caption{Cartoon illustration of the hybrid models. The difference between \HMU{} and \HMC{} is in the first step, where the initial planet masses and radii are drawn (\HMC{} draws clustered initial masses as in \S\ref{sec:methods:new_model:clustered}, while \HMU{} does not). The planets are also assigned envelope masses in this step. The planets are then subject to photoevaporation and some lose their gaseous envelopes, resulting in the final masses and radii (\S\ref{sec:methods:new_model:photoevap}). Finally, the critical AMD of the system is computed and distributed among the individual planets to draw their orbital eccentricities and mutual inclinations, as in the H20 model.}
\label{fig:cartoon}
\end{figure}

\subsubsection{Putting it all together: a hybrid model} \label{sec:methods:new_model:procedure}

We now describe the new, complete procedure for drawing a physical catalog from the new model, which is a ``hybrid" between the H20 model and the \NR20{} model. This ``hybrid model" effectively combines the distributions of initial planet masses and radii (\S\ref{sec:methods:new_model:initialMR}-\ref{sec:methods:new_model:clustered}), subject to envelope mass loss via the photoevaporation model (\S\ref{sec:methods:new_model:photoevap}), with the remaining features of the H20 model (everything described in \S\ref{sec:methods:prev_models} except \S\ref{sec:methods:prev_models:mass_radius}). As discussed in \S\ref{sec:methods:new_model:clustered}, we also consider two versions of the new model: a Hybrid Model with \textbf{Unclustered} initial masses (hereafter ``\textbf{HM-U}") and a Hybrid Model with \textbf{Clustered} initial masses (hereafter ``\textbf{HM-C}").
Figure \ref{fig:cartoon} shows a cartoon depicting the main idea of the hybrid models.
For a randomly drawn star in the \Kepler{} catalog (described in \S\ref{sec:methods:model_inference:data}):
\begin{enumerate}[leftmargin=*]
 \item Draw a number of planets and planet-clusters in the system:
 \begin{enumerate}
  \item First, determine whether the star will host a planetary system at all by computing the probability that the star will host at least one planet (in the parameter space we are considering, i.e. between $R_p = 0.5$-$10 R_\oplus$, $M_p = 0.1$-$1000 M_\oplus$, and $P = 3$-300 days), given as the fraction of stars with planets $f_{\rm swpa}(c)$ from equation \ref{eq:fswp}.\footnote{We fix the slope $m = 0.9$ and normalization $f_{\rm swpa,med} = 0.88$ to the best-fit values found from \PaperIII{}, in order to reduce the number of free parameters. In theory, these parameters may also vary given the new model.} Then draw a number $u \sim {\rm Unif}(0,1)$. If $u > f_{\rm swpa}$, return the star with no planets; otherwise, continue.
  \item Draw a number of clusters in the system to attempt, $N_c \sim {\rm ZTP}(\lambda_c)$ (resampled such that $N_c \leq N_{c,\rm max}$), where ${\rm ZTP}$ is the zero-truncated Poisson distribution.
  \item For each cluster, draw a number of planets in the cluster, $N_p \sim {\rm ZTP}(\lambda_p)$ (resampled such that $N_p \leq N_{p,\rm max}$).
 \end{enumerate}
 \item Draw the physical properties (initial masses, radii, and orbital periods) of the planets. For each cluster:
 \begin{enumerate}
  \item Draw a set of initial planet masses and radii for the $N_p$ planets following \S\ref{sec:methods:new_model:initialMR}. For \HMU{}, this involves drawing $M_{p,\rm init}$ from equation \ref{eq:mass_init} and $R_{p,\rm init} | M_{p,\rm init}$ from equations \ref{eq:radius_init}-\ref{eq:logistic_smoothed}; for \HMC{}, $M_{p,\rm init}$ are drawn from equation \ref{eq:mass_init_clustered} instead of equation \ref{eq:mass_init}. The entire set is resampled until all the masses and radii are between the bounds (0.1-$1000 M_\oplus$ and 0.5-$10 R_\oplus$), up to 10 max attempts per cluster.
  \item Draw unscaled periods for the planets in the cluster. If $N_p = 1$, assign an unscaled period of $P' = 1$. If $N_p > 1$, draw their unscaled periods $P' \sim \ln\mathcal{N}(0, N_p\sigma_P)$. Check if the conditions for mutual Hill stability ($\Delta \geq \Delta_c$) and stability against MMR overlap ($\alpha < \alpha_{\rm crit}$), assuming circular orbits, are satisfied for all adjacent pairs in the cluster. Resample the unscaled periods $P'$ until this condition is satisfied or until 1000 max attempts is reached (in which case the entire cluster is discarded).
  \item Draw a period scale factor $P_c$ (days) for the cluster and multiply each planet's unscaled period by the period scale: 
  \begin{eqnarray}
   f(P_c) \propto P_c^{\alpha_P}, \\
   P = {P_c}P',
  \end{eqnarray}
  such that all the planet periods are in the range $P = 3$-300 days.\footnote{This can be achieved by drawing $P_c$ in the range $\Big[\frac{P_{\rm min}}{{\rm min}(P'_k)}, \frac{P_{\rm max}}{{\rm max}(P'_k)}\Big]$, where $k = 1,\dots,N_p$.} Check if $\Delta \geq \Delta_c$ and $\alpha < \alpha_{\rm crit}$ are satisfied for all adjacent planet pairs in the entire system, including planets from previously drawn clusters. Resample $P_c$ for the current cluster until this condition is satisfied or until 100 max attempts is reached (in which case the current cluster is discarded).
 \end{enumerate}
 \item For the remaining (successfully drawn) planets, assign envelope masses and apply the photoevaporation model to draw their resulting physical properties (final masses and radii):
 \begin{enumerate}
  \item Compute the envelope masses using equation \ref{eq:envelope_mass}.
  \item Compute the mass-loss timescales $t_{\rm loss}$ and the probabilities of envelope mass-loss $p_{\rm loss}$ from equations \ref{eq:mass_loss_timescale}-\ref{eq:prob_mass_loss}. Draw a number $u \sim {\rm Unif}(0,1)$ for each planet. If $u < p_{\rm loss}$, subtract the planet's envelope mass from its total initial mass (equation \ref{eq:mass_final}).
  \item For the planets that have lost their envelopes, compute their final radii from their final masses, $R_{p,\rm final} | M_{p,\rm final}$, using equations \ref{eq:radius_final}-\ref{eq:radius_mass_pure_silicate}.
 \end{enumerate}
 \item Draw the orbital excitations (eccentricities and mutual inclinations) of the planets using the procedure of distributing AMD:
 \begin{enumerate}
  \item If the total number of (successfully attempted) planets in the system is $N = 1$, draw an eccentricity $e \sim {\rm Rayleigh}(\sigma_{e,1})$ and argument of pericenter $\omega \sim {\rm Unif}(0,2\pi)$, and skip to step \ref{step_orbit}.
  \item If there are two or more planets in the system, compute the critical AMD of the system, ${\rm AMD}_{\rm crit}$ (equation \ref{eq:amd_crit}).
  \item Draw the eccentricities ($e$), arguments of pericenter ($\omega$), and mutual inclinations ($i_m$) of the planets by distributing ${\rm AMD}_{\rm crit}$ amongst the individual planets and their excitation components subject to equation \ref{eq:amd_partition}.
  \item Draw an angle of ascending node, $\Omega \sim {\rm Unif}(0,2\pi)$, and mean anomaly, $M \sim {\rm Unif}(0,2\pi)$, (relative to the system invariant plane) for each planet. \label{step_orbit}
 \end{enumerate}
 \item Specify the system invariant plane by drawing a random normal vector relative to the observer sky ($z$) axis.
 \item Compute the inclination angle $i$ (relative to the plane of the sky) for each planet's orbit, using rotations and dot products relative to the system invariant plane.
\end{enumerate}

The hybrid models contain a large number of parameters. Not all of them are constrainable by our forward modeling procedure, nor are we interested in constraining every possible parameter in each model. Thus, we will fix some parameters and attempt to constrain the rest, and describe the various sets of model inference runs in \S\ref{sec:methods:model_inference:optimization}.


\subsection{Fitting the new model via ABC} \label{sec:methods:model_inference}

The method of Approximate Bayesian computation provides a way to estimate the posterior distribution of the model parameters when there is no explicit likelihood function \citep{Rubin1984, Beaumont2009}. The basic steps of the ABC algorithm can be summarized as follows:
\begin{enumerate}
 \item Sample a set of model parameters from the prior distribution.
 \item Simulate a dataset drawn from the model with those model parameters and calculate a set of summary statistics.
 \item Compare the simulated data with the real data by evaluating a distance function, which quantifies how close the summary statistics are between the simulated and real data.
 \item Accept the model parameters if the distance is less than a set distance threshold.
\end{enumerate}
These steps are repeated until enough points (sets of model parameters) are accepted to form the posterior distribution.
Applied to our problem, the ``simulated data'' is an \textit{observed} catalog generated from applying the \SysSim{} pipeline to a physical catalog drawn from the model as described in \S\ref{sec:methods:new_model:procedure}, and the ``real" data is the \Kepler{}-observed catalog (the sample we use is summarized in \S\ref{sec:methods:model_inference:data}). The summary statistics refer to a set of properties that can be computed from both the simulated and \Kepler{} catalogs, which we describe in \S\ref{sec:methods:model_inference:summary_stats}. The distance function is defined in \S\ref{sec:methods:model_inference:distance_function}. Finally, we discuss the choice of the prior distribution and distance threshold, as well as the implementation details of the ABC algorithm, in \S\ref{sec:methods:model_inference:optimization}.

\subsubsection{The \Kepler{} stellar and planet catalogs} \label{sec:methods:model_inference:data}

In order to robustly infer the population parameters of planets around stars from the \Kepler{} mission, a clean, uniform sample of stars and planets must be defined. We adopt the same sample as the one defined in \PaperIII{}, which primarily includes FGK dwarfs selected from the \Kepler{} DR25 target list \citep{2018ApJS..235...38T} and with stellar properties (radii and masses) updated from the \Gaia{}-\Kepler{} Stellar Properties Catalog (GKSPC; \citealt{2020AJ....159..280B}). These target stars were selected using a series of cuts (following \citealt{2019AJ....158..109H} and \PaperIII{}), in order to filter out close-in binaries and stars with poorly measured radii, by requiring: (i) a consistent \Kepler{} magnitude and \Gaia{} G magnitude, (ii) a good astrometric fit (\Gaia{} GOF\_AL $\leq 20$ and astrometric excess noise $\leq 5$), and (iii) a precise parallax (fractional error within $10\%$). The sample was then reduced to main sequence FGK dwarfs via their \Gaia{} DR2 $b_p-r_p$ colors, by requiring $0.5 \leq b_p-r_p - E^* \leq 1.7$ and $L \leq 1.75 L_{\rm MS}(b_p-r_p-E^*)$ where $E^*$ is a differential reddening amount estimated from smoothly interpolating the measured values of $E(b_p-r_p)/d$ (and $d=1/\pi$ is the distance to the star from the parallax $\pi$), and $L_{\rm MS}(b_p-r_p-E^*)$ is the luminosity of the main sequence derived from iterative fitting (see Section 2.1 of \PaperIII{} for more details). The final stellar catalog contains $N_\star = 86,760$ target stars.

For this sample of stars, we considered the planet candidates reported in the \Kepler{} DR25 KOI table. For each planet, (i) the transit depths and durations were updated with the median values from the posteriors in \citet{2015ApJS..217...16R}, and (ii) the planet radii were recomputed from the transit depths and the stellar radii (which have been updated from the \citealt{2020AJ....159..280B} catalog, as described above). Finally, the planet sample was restricted to only include planets with periods in the range $P \in [3,300]$ days and planet radii in the range $R_p \in [0.5,10] R_\oplus$. This sample consists of $N_{p,\rm obs} = 2169$ planet candidates, 964 of which are in observed multi-transiting systems. This catalog of \Kepler{} transiting planets around FGK dwarfs was used to constrain the H20 model from \PaperIII{}, and will serve to also constrain the models in this paper.

\subsubsection{Summary statistics} \label{sec:methods:model_inference:summary_stats}

We define a set of summary statistics which captures the population-level properties that we wish to fit. We maintain most of the summary statistics adopted in \PaperIII{}, with a few modifications designed to capture more of the patterns in the observed planet radii (and less of the orbital eccentricities and mutual inclinations as was the primary focus of that paper). For the \Kepler{} catalog and each simulated observed catalog, we compute the following observables:
\begin{enumerate}
 \item $f = N_{p,\rm obs}/N_\star$: the total number of observed planets relative to the total number of target stars,
 \item $\{N_m\}$: the observed multiplicity distribution, where $N_m$ is the number of systems with $m$ observed planets and $m=1,2,3,\dots$,
 \item $\{P\}$: the observed orbital period distribution,
 \item $\{\mathcal{P} = P_{i+1}/P_i\}$: the observed period ratio distribution,
 \item $\{t_{\rm dur}\}$: the observed transit duration distribution,
 \item $\{\xi = (t_{\rm dur,in}/t_{\rm dur,out})(P_{\rm out}/P_{\rm in})^{1/3}\}$: the observed period-normalized transit duration ratio distribution,
 \item $\{\delta\}$: the observed transit depth distribution,
 \item $\{\delta_{i+1}/\delta_i\}$: the observed transit depth ratio distribution,
 \item $\{R_p\}$: the observed planet radius distribution,
 \item $\{R_p - R_{\rm gap}(P)\}$: the observed ``gap-subtracted" planet radius distribution,
 \item $\{\mathcal{Q}_R\}$: the observed radii partitioning distribution,
 \item $\{\mathcal{M}_R\}$: the observed radii monotonicity distribution,
\end{enumerate}
where the ratio observables ($\mathcal{P}$, $\xi$, and $\delta_{i+1}/\delta_i$) are computed for adjacent observed planet pairs.
This collection of summary statistics enables detailed constraints on the overall distribution of planets across and within planetary systems. Summary statistics 1 and 2 are crucial for constraining the overall rate of observed planets and how they are distributed among the stars. More broadly, the distributions of 3-6 encode the overall orbital properties of the planetary systems, while 7-12 capture the distributions and patterns in the planet sizes.

The observed ``gap-subtracted" planet radius distribution is simply the distribution of the differences in the observed planet radii from the location of the radius gap (i.e. valley) at each planet's orbital period. It is well observed that the location of the radius valley is a function of orbital period, often parameterized as a power-law \citep[e.g.,][]{2020ApJ...890...23L, 2021MNRAS.508.5886R, 2023MNRAS.519.4056H, 2023arXiv230200009B}:
\begin{equation}
 R_{\rm gap}(P) = R_{\rm gap,0} P^\gamma, \label{eq:radius_gap}
\end{equation}
where $\gamma$ is the power-law index (i.e. slope in $\log(R_{\rm gap})$-$\log(P)$ space) and $R_{\rm gap,0}$ is the normalization (which we adopt at $P = 1$ day). We adopt values of $\gamma = -0.10$ and $R_{\rm gap,0} = 2.40 R_\oplus$ (the median best-fit values from \citealt{2018MNRAS.479.4786V}).\footnote{We also performed a fit of Equation \ref{eq:radius_gap} to our sample of the \Kepler{} catalog defined in \S\ref{sec:methods:model_inference:data}, using the \texttt{gapfit} package \citep{2020ApJ...890...23L}, and found similar values.}
The distribution of gap-substracted radii provides a clearer view of the observed radius valley than just the observed radius distribution \citep{2020ApJ...890...23L}, the latter of which marginalizes over the orbital periods and thus dilutes the true extent of the valley in period-radius space. Since a primary goal of this paper is to reproduce the observed radius valley, we fit both distributions in our distance function defined in \S\ref{sec:methods:model_inference:distance_function}. We note that we also subtract the same power-law for the location of the radius valley (i.e. equation \ref{eq:radius_gap} with $\gamma = -0.10$ and $R_{\rm gap,0} = 2.40 R_\oplus$, which was fit to the \Kepler{} data) for each simulated observed catalog, since we are interested in reproducing the \Kepler{}-observed radius valley, and not simply \textit{any} observed radius valley.

The observables $\mathcal{Q}_R$ and $\mathcal{M_R}$ are the system-level metrics inspired by \citet{2020AJ....159..281G}, adapted to be computed from the observed planet radii instead of their masses (which have not been measured for most \Kepler{} planets), as also in \PaperIII{}. For each observed system, these metrics can be computed as:
\begin{eqnarray}
 \mathcal{Q}_R &\equiv& \bigg(\frac{m}{m - 1}\bigg) \Bigg(\sum_{k=1}^{m}\Big(R_{p,k}^{*} - \frac{1}{m}\Big)^2 \Bigg), \label{eq:radius_partitioning} \\
 \mathcal{M}_R &\equiv& \rho_S{\mathcal{Q}_R}^{1/m}, \label{eq:radius_monotonicity}
\end{eqnarray}
where $R_{p,k}^{*}$ is the radius of the $k^{\rm th}$ planet normalized by the sum of all the planet radii, $\rho_S$ is the Spearman's rank correlation coefficient of the planet radii versus their index order in orbital period, and $m$ is the observed multiplicity. $\mathcal{Q}_R$ is designed to capture the size similarity of all the planets in the system as a whole (its value ranges from 0 for identically sized planets, to 1 for one planet being much larger than the rest), while $\mathcal{M_R}$ quantifies the size ordering of the planets (positive values imply that the planets get larger towards longer periods while negative values imply the opposite).

We note that unlike in Papers II--III, we no longer divide the stellar sample into two halves (and thus compute all of the summary statistics for each half) based on their \Gaia{} $b_p-r_p-E^*$ colors. Dividing the sample was necessary to constrain the change in the occurrence of planetary systems as a function of host star spectral type, namely the slope ($m$) and normalization ($f_{\rm swpa,med}$) parameters in equation \ref{eq:fswp}. However, in order to reduce the number of free model parameters in the hybrid models (as well as the complexity of the distance function defined in \S\ref{sec:methods:model_inference:distance_function}) while still retaining the rise in occurrence towards later spectral types, we simply adopt fixed values of $m = 0.9$ and $f_{\rm swpa,med} = 0.88$ based on the best-fit H20 model. This produces a lower rate of planetary systems ($f_{\rm swpa} \lesssim 0.4$) for the bluest stars in our catalog, and rises sharply towards unity ($f_{\rm swpa} \simeq 1$) for K-dwarfs and redder stars.



\subsubsection{Distance function} \label{sec:methods:model_inference:distance_function}

The ABC algorithm involves evaluating a distance function, which captures the goodness of fit of a simulated catalog to the \Kepler{} catalog. We adopt a similar formulation as defined in \PaperIII{}, by calculating a distance metric for each marginal distribution (summary statistic) described in \S\ref{sec:methods:model_inference:summary_stats} and computing a weighted sum of the individual distance terms:
\begin{equation}
 \mathcal{D}_W = \sum_i w_i \mathcal{D}_i, \label{eq:dist}
\end{equation}
where $w_i$ and $\mathcal{D}_i,$ are the weight and distance metric, respectively, of the $i^{\rm th}$ summary statistic.

The first term simply measures the absolute difference in the overall rate of observed planets per star:
\begin{equation}
 \mathcal{D}_1 = \mathcal{D}_f \equiv | f_{\rm sim} - f_{\rm Kepler} |, \label{eq:dist_rate}
\end{equation}
where $f_{\rm sim}$ and $f_{\rm Kepler}$ are the total number of observed planets divided by the total number of target stars, i.e. the first summary statistic for the simulated and \Kepler{} catalogs, respectively.

For the second term, we continue to use the Cressie-Read power divergence \citep{cr1984} which is a measure of how close two distributions of discrete bins are, as is the case of the planet multiplicity distribution:
\begin{equation}
 \mathcal{D}_2 = \rho_{\rm CRPD} \equiv \frac{9}{5}\sum_j O_j \bigg[{\bigg(\frac{O_j}{E_j}\bigg)}^{2/3} - 1 \bigg], \label{eq:dist_mult}
\end{equation}
where $O_j$ and $E_j$ are the numbers of ``observed" systems (from our models) and ``expected" systems (from the \Kepler{} data), for each multiplicity bin $j = 1,2,3,4,5+$.

The remaining summary statistics are continuous distributions, for which we continue to adopt the two-sample Kolmogorov-Smirnov (KS; \citealt{kolmogorov1933, smirnov1948}) statistic,
\begin{equation}
 \mathcal{D}_{\rm KS} = {\rm max}| F(x) - G(x) |, \label{eq:KS_dist}
\end{equation}
which is simply the maximum absolute difference in the empirical cumulative distribution functions of the two samples, $F(x)$ and $G(x)$.

The distance terms must be weighted when summing them, as in equation \ref{eq:dist}, because each distance metric has a different characteristic scale. In theory, a catalog that is identical to the \Kepler{} catalog (in terms of the summary statistics we have chosen) would result in distances that are zero for each metric. However, planet formation is a stochastic process, and we expect that even a ``perfect" model would result in slightly varying simulated catalogs and thus non-zero distance terms, given the Monte Carlo noise. The typical distance for each metric arising from stochastic noise depends on the noisiness of each summary statistic; in general, the summary statistics with fewer data points (such as the system-level metrics) are noisier than those with more data points (such as the marginal distributions of the observed planets' properties). We compute the weights $w_i$ in the same manner as was done in Papers II--III, by taking the reciprocal of the root mean square of each distance term, $\hat{\sigma}(\mathcal{D}_i)$, computed over many catalog generations from the same model compared to each other:
\begin{equation}
 w_i = 1/\hat{\sigma}(\mathcal{D}_i) = \Bigg( {\frac{1}{n} \sum_{k \neq l} \mathcal{D}_i (k,l)^2 } \Bigg)^{-0.5}, \label{eq:weights}
\end{equation}
where $\mathcal{D}_i (k,l)$ is simply the $i^{\rm th}$ distance metric taken between two simulated catalogs labeled by $k$ and $l$, and $n$ is the total number of unique catalog combinations. By simulating catalogs with the same number of targets as in the \Kepler{} catalog and from the exact same model (i.e. with identical model parameters), we are quantifying the variations in each distance metric resulting purely from stochastic noise, given the sample size of the \Kepler{} catalog. In this manner, a ``perfect" model would tend to produce weighted distances $w_i \mathcal{D}_i \to 1$ for each summary statistic. Therefore, a perfect model would result in a total distance $\mathcal{D}_W$ close to the number of individual distance terms included. For calculating the weights used in this paper, we simulated 100 catalogs drawn from a nominal, best-fit ``maximum AMD model" from \PaperIII{}, resulting in $n = {100 \choose 2} = 4950$ unique pairs.

\subsubsection{Model optimization} \label{sec:methods:model_inference:optimization}

With our distance function defined in \S\ref{sec:methods:model_inference:distance_function}, we just need to define a prior distribution and set a distance threshold in order to perform the ABC rejection algorithm for inferring the posterior distributions of our models.

To inform our choice of the prior distribution and the distance threshold, we employ a model optimization scheme.
We use a Differential Evolution (DE) algorithm to estimate the minimum value of the distance function achievable for the given model, and then approximately locate the corresponding region of parameter space and sample a large number data points (model evaluations) near that region. We adopt the DE optimizer given by the ``BlackBoxOptim.jl" Julia package, which implements a population-based genetic algorithm beginning with a number of initial points in the multi-dimensional parameter space which are then evolved such that those with better ``fitness" (i.e. lower values of the distance function) are more likely to pass on their properties to future generations. We choose a population size equal to four times the number of free model parameters and evolve for 5000 model evaluations. This optimization algorithm is then repeated 50 times with different seeds of the initial points, resulting in a collective pool of $2.5 \times 10^5$ points in the parameter space (at which we have evaluated the full forward model).

We choose a bounded uniform prior for each of our free model parameters. The bounds are chosen based on visual inspection of the distribution of model parameters resulting from the optimization pool, and are somewhat narrower than the optimizer bounds listed in Table \ref{tab:param_fits} (varying slightly between different runs). The distance threshold is set to a value that is moderately larger than the smallest distance found during the optimization stage, $\mathcal{D}_{W,\rm thres} = 25$. While a smaller threshold could have been chosen, it would result in a significantly reduced rate of acceptance. Our choice of the distance threshold improves the computational efficiency at the cost of resulting in a slightly wider posterior distribution (i.e., larger credible intervals).

Finally, we perform ABC by repeatedly drawing model parameters from the prior distribution and accepting those that pass our distance threshold. A complete treatment of the ABC process would involve simulating a catalog using the full forward model for each set of drawn model parameters, and evaluating the distance function. However, this procedure becomes computationally infeasible when drawing the very large numbers of sets of model parameters needed to collect a sufficient number of points for constructing the posterior distribution. For reference, we aim to generate $10^5$ accepted points when the rate of acceptance is roughly $10^{-5}$ for our choice of distance threshold; therefore, the distance must be evaluated on the order of $\sim 10^{10}$ points, and it takes on the order of $\sim 10$s to simulate a single pair of physical and observed catalogs from the forward model. Thus, we train a Gaussian process (GP) emulator for rapidly predicting the distance function given a proposed set of model parameters, following the implementation described in Papers I--III. In summary, we use a sample of the points found by the DE optimization algorithm to train the emulator, which provides an estimate of the distance for acceptance or rejection at any given new point in parameter space without having to run the full forward model.

Given the large number of model parameters and summary statistics we wish to fit, we performed several sets of model optimization ``runs" by repeating the entire procedure described above for a few different combinations of free parameters and summary statistics to be included in the distance function, as described below. The free and fixed parameters for each of these runs are summarized in Table \ref{tab:param_fits}.

\textit{Run 0 (initial run) of \HMU{}.} We initially attempted to fit \textit{most} of the free parameters (13): $\{\lambda_c, \lambda_p, \alpha_P, \sigma_P, R_{p,\rm norm}, M_{p,\rm break}, \gamma_0, \gamma_1, \sigma_0, \sigma_1, \alpha_{\rm ret}, \mu_M,$ $\sigma_M\}$. However, we found that many of the parameters were poorly constrained. In particular, the break mass ($M_{p,\rm break}$) and the radius-mass relation above the break mass ($\gamma_1$ and $\sigma_1$) were unconstrained over the wide ranges of values we allowed. In retrospect, this is unsurprising because our summary statistics do not involve any direct constraints on the planet masses (besides indirectly through the planet radii, and weakly through the stability criteria), and there are also relatively fewer planets larger than $\sim 5 R_\oplus$ in the \Kepler{} data. Simulations of catalogs with parameters drawn from this run also rarely showed an observed radius valley. Therefore, we will not rely on the results of this run, and instead, we consider simplifying the model inference problem by fixing some parameters and varying the rest, as in the following runs.

\textit{Run 1 of \HMU{}.} We performed another run with fixed values for the break mass and upper radius-mass relation ($M_{p,\rm break} = 20 M_\oplus$, $\gamma_1 = 0.5$, and $\sigma_1 = 0.3$; these values were chosen as guided by the best-fit values from \NR20{}, Table 1 therein), as well as the period cluster scale ($\sigma_P = 0.25$; based on the H20 model), while allowing the remaining parameters (9) to vary. In order to focus on reproducing the distributions involving planet radii, we also reduced the number of terms in the distance function by excluding the period ratios, transit durations, and period-normalized transit duration ratios. Given that several parameters were also fixed to the best-fit values from the H20 model (e.g. $\Delta_c = 10$, $\sigma_{e,1} = 0.25$, and $\sigma_P = 0.25$), and the implementation of AMD-stability remains the same, we do not expect these excluded distributions to deviate substantially from the \Kepler{} distributions. Nevertheless, we will examine these distributions when assessing the resulting catalogs from the hybrid models.

\textit{Run 1 of \HMC{}.} Finally, we repeated Run 1 for the hybrid model with clustered initial masses, which has an additional free parameter ($\sigma_{M,c}$) for a total of 10 free parameters. All other settings (values of the fixed parameters, terms in the distance functions, etc.) were kept the same as in Run 1 of \HMU{}. This is the primary run we will focus on for the remainder of the paper.

\begin{deluxetable*}{lcccccc}
\centering
\tablecaption{Best-fit credible intervals (16-84$\%$) for each parameter of each hybrid model, from various inference runs.}
\tablehead{
 \colhead{Parameter} & \colhead{Units} & \multicolumn2c{\HMU{} (Unclustered masses)} & \multicolumn2c{\HMC{} (Clustered masses)} & \colhead{Optimizer Bounds} \\
 \cmidrule(r){3-4} \cmidrule(r){5-6}
 & & \colhead{Run 0 (initial)} & \colhead{Run 1} & \colhead{Run 1} & \colhead{Run 1, $\Delta_{\rm valley} \geq 0.29$} & 
}
\decimalcolnumbers
\startdata
 $\ln{(\lambda_c)}$            & -- & $-0.40_{-0.68}^{+0.59}$ & $-0.49_{-0.60}^{+0.65}$ & $-0.82_{-0.40}^{+0.48}$ & $-0.91_{-0.35}^{+0.49}$ & $(\ln{0.2},\ln{5})$ \\[5pt]
 $\ln{(\lambda_p)}$            & -- & $0.37_{-0.49}^{+0.54}$ & $0.56_{-0.40}^{+0.33}$ & $0.81_{-0.20}^{+0.18}$ & $0.75_{-0.18}^{+0.14}$ & $(\ln{0.2},\ln{5})$ \\[5pt]
 $\ln{(\lambda_c\lambda_p)}^{\dagger}$   & -- & $-0.05_{-0.56}^{+0.56}$ & $0.06_{-0.49}^{+0.51}$ & $0.00_{-0.36}^{+0.38}$ & $-0.14_{-0.31}^{+0.35}$ & -- \\[5pt]
 $\ln{(\lambda_p/\lambda_c)}^{\dagger}$  & -- & $0.77_{-1.00}^{+1.17}$ & $1.07_{-0.97}^{+0.79}$ & $1.65_{-0.63}^{+0.49}$ & $1.66_{-0.64}^{+0.45}$ & -- \\[5pt]
 $\alpha_P$                    & -- & $-0.05_{-0.45}^{+0.72}$ & $0.15_{-0.33}^{+0.46}$ & $0.17_{-0.35}^{+0.54}$ & $-0.09_{-0.21}^{+0.30}$ & $(-2,2)$ \\[5pt]
 $\sigma_P$                    & -- & $0.18_{-0.11}^{+0.14}$ & 0.25 & 0.25 & 0.25 & $(0,0.5)$ \\[5pt]
 $R_{p,\rm norm}$              & $R_\oplus$ & $1.55_{-0.38}^{+0.59}$ & $1.81_{-0.32}^{+0.32}$ & $1.80_{-0.30}^{+0.29}$ & $2.07_{-0.21}^{+0.17}$ & $(1,5)$ \\[5pt]
 $M_{p,\rm break}$               & $M_\oplus$ & $65_{-35}^{+23}$ & 20 & 20 & 20 & $(1,100)$ \\[5pt]
 $\gamma_0$                    & -- & $0.16_{-0.11}^{+0.18}$ & $0.10_{-0.06}^{+0.10}$ & $0.09_{-0.05}^{+0.07}$ & $0.07_{-0.04}^{+0.06}$ & $(0,1)$ \\[5pt]
 $\gamma_1$                    & -- & $0.47_{-0.29}^{+0.29}$ & 0.5 & 0.5 & 0.5 & $(0,1)$ \\[5pt]
 $\sigma_0$                    & -- & $0.39_{-0.14}^{+0.08}$ & $0.34_{-0.14}^{+0.10}$ & $0.37_{-0.12}^{+0.08}$ & $0.23_{-0.08}^{+0.10}$ & $(0,0.5)$ \\[5pt]
 $\sigma_1$                    & -- & $0.28_{-0.18}^{+0.16}$ & 0.3 & 0.3 & 0.3 & $(0,0.5)$ \\[5pt]
 $\ln{(\alpha_{\rm ret})}$ & -- & $3.75_{-0.96}^{+0.64}*$ & $2.35_{-0.95}^{+0.89}$ & $2.27_{-0.64}^{+0.65}$ & $2.07_{-0.59}^{+0.52}$ & $(\ln{0.01},\ln{1000})$ \\[5pt]
 $\mu_M$                       & $\ln{(M/M_\oplus)}$ & $1.18_{-0.73}^{+0.80}$ & $0.58_{-0.34}^{+0.41}$ & $0.48_{-0.27}^{+0.33}$ & $0.35_{-0.20}^{+0.26}$ & $(\ln{1},\ln{100})$ \\[5pt]
 $\sigma_M$                    & $\ln{(M/M_\oplus)}$ & $1.37_{-0.50}^{+0.65}$ & $1.52_{-0.37}^{+0.36}$ & $1.77_{-0.33}^{+0.40}$ & $1.57_{-0.23}^{+0.27}$ & $(0,3)$ \\[5pt]
 $\sigma_{M,c}$      & $\ln{(M/M_\oplus)}$ & -- & -- & $0.37_{-0.20}^{+0.24}$ & $0.48_{-0.21}^{+0.23}$ & $(0,3)$ \\[5pt]
\enddata
\tablecomments{The values without uncertainties denote the parameters that were fixed for that run.}
\tablenotetext{^\dagger}{These values are directly computed from $\ln{(\lambda_c)}$ and $\ln{(\lambda_p)}$, and are provided for reference only.}
\tablenotetext{*}{For this run, the parameter $\alpha_{\rm ret}$ was directly varied instead of its logged value, and thus the prior was uniform in $\alpha_{\rm ret}$ instead of $\ln{(\alpha_{\rm ret})}$.}
\label{tab:param_fits}
\end{deluxetable*}

\begin{figure*}
\centering
\includegraphics[scale=0.32,trim={0.1cm 0.1cm 0.2cm 0.1cm},clip]{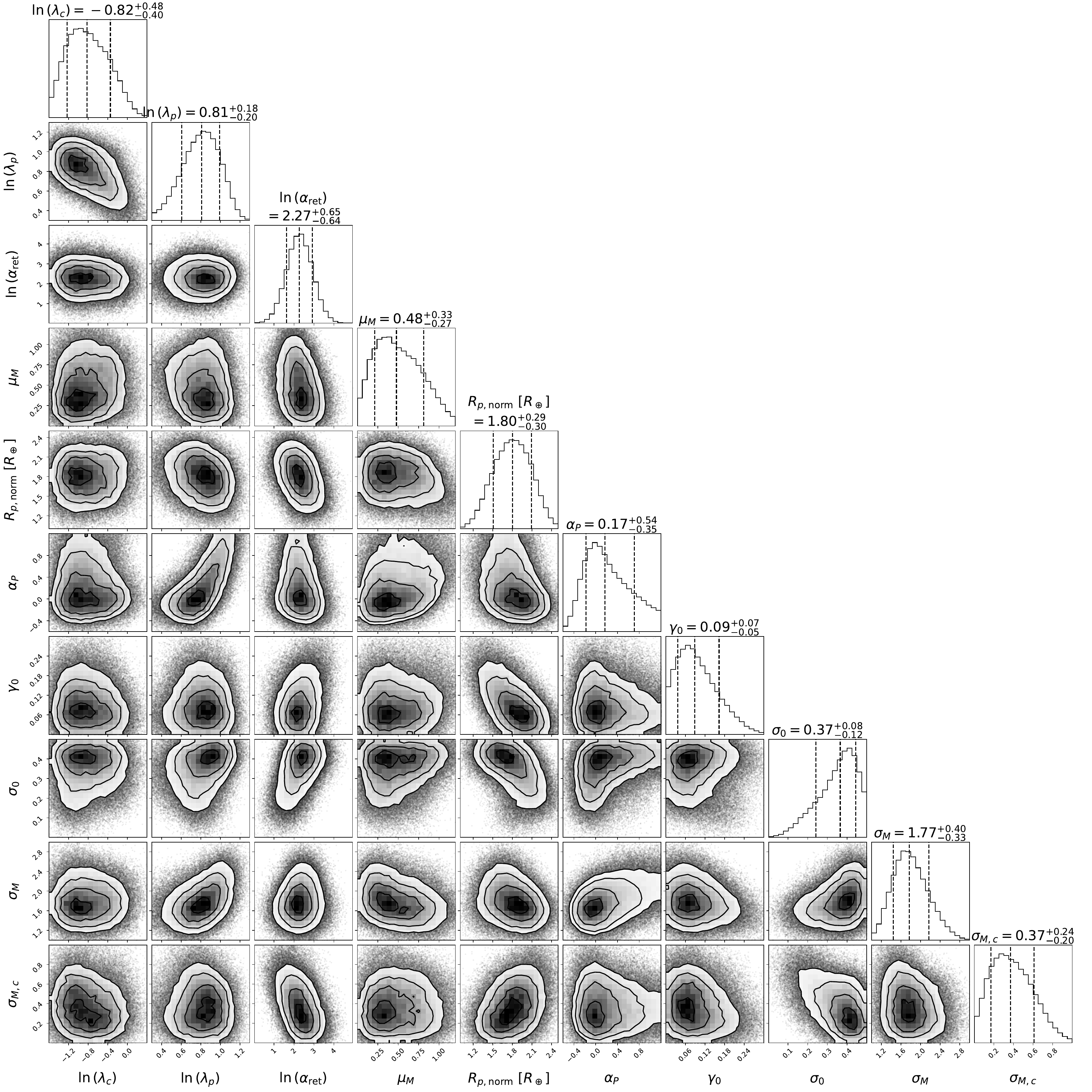} 
\caption{ABC posterior distribution of the free parameters from Run 1 of \HMC{}. A distance threshold of $\mathcal{D}_W \leq 20$ is used and $10^5$ points passing this threshold as evaluated using the GP emulator are plotted. The median and central 68.3\% credible intervals for each parameter are labeled above each histogram, and are also listed in Table \ref{tab:param_fits}.}
\label{fig:hm2_posterior_params10}
\end{figure*}

\section{Results} \label{sec:results}

We organize the main results as follows. First, we report the results of the hybrid models as fit to the \Kepler{} catalog in \S\ref{sec:results:new_model_fit}, in terms of the best-fit model parameters (\S\ref{sec:results:new_model_fit:hm1_params} and \S\ref{sec:results:new_model_fit:hm2_params} for \HMU{} and \HMC{}, respectively) and how the marginal distributions of the observed statistics compare (\S\ref{sec:results:new_model_fit:observed}-\S\ref{sec:results:new_model_fit:observed_size_clustering}).
We then focus on the observed planet radius valley, a central motivation for constructing the hybrid models, in \S\ref{sec:results:radius_valley}. We then examine the inferred underlying distributions from the hybrid models and how they compare to the H20 model in \S\ref{sec:results:model_underlying}, including the intrinsic planet radius distribution (\S\ref{sec:results:model_underlying:radius}), the intrinsic degree of intra-system planet mass and radius similarity (\S\ref{sec:results:models_underlying:size_similarity}), and other properties of the orbital architectures (\S\ref{sec:results:model_underlying:model_comparison}), as well as a brief discussion of the initial core mass distribution (\S\ref{sec:results:model_underlying:core_masses}).

\begin{figure*}
\centering
\includegraphics[scale=0.6,trim={0.1cm 0.5cm 0.2cm 0.5cm},clip]{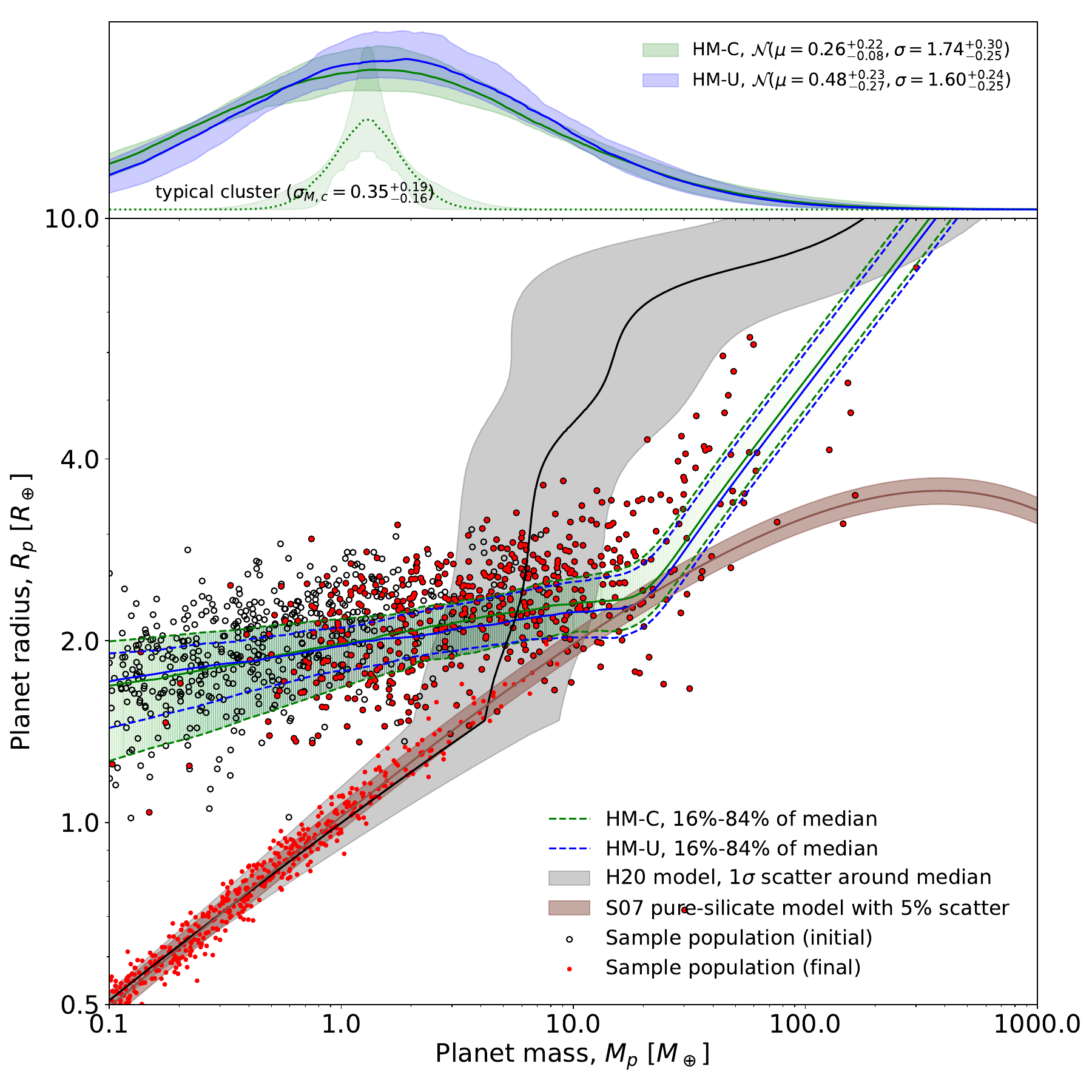} 
\caption{Same as Figure \ref{fig:radius_mass}, but for the radius-mass distributions inferred from the hybrid models. \textbf{Main panel:} As in Figure \ref{fig:radius_mass}, the black and brown curves show the H20 model and the pure silicate relation (equation \ref{eq:radius_mass_pure_silicate}), respectively, for reference. The blue and green curves now denote the initial radius-mass distributions from \HMU{} and \HMC{}, respectively (for each model, the solid line represents the median prediction and the dashed lines denote the 16-84\% credible regions of the median). For \HMC{}, the green shading indicates the relative occurrence at that initial mass (i.e., the darkness is scaled to the lognormal distribution shown in the top panel). We note that for both models, the break mass and radius-mass relation above it were set to fixed values instead of being inferred (see Table \ref{tab:param_fits}). A sample of planets drawn from \HMC{} are plotted as scatter points; their initial masses and radii are shown as black circles, while their final values (after photoevaporation) are shown as red dots. \textbf{Top panel:} the credible regions of the initial planet mass distributions (lognormal distributions with parameters $\mu = \mu_M$ and $\sigma = \sigma_M$, as labeled). The parameters listed here differ slightly from those in Table \ref{tab:param_fits}, because they are computed from 100 simulated catalogs drawn from the posterior distribution that also pass the distance threshold whereas the table values are directly computed from the posterior based on the emulator-predicted distances. In \HMC{} (green), the narrow lognormal distribution (green dotted line) illustrates a typical cluster; its location is arbitrary and is drawn from the broader lognormal distribution.}
\label{fig:radius_mass_credible}
\end{figure*}

\subsection{Constraining the hybrid models} \label{sec:results:new_model_fit}

We report the distributions of best-fit model parameters of the hybrid models. The median and 16\%-84\% quantiles of each parameter, of each run, are given in Table \ref{tab:param_fits}.

\subsubsection{\HMU{} parameters} \label{sec:results:new_model_fit:hm1_params}

First, we briefly describe the results from the initial run for the Hybrid Model with Unclustered masses (\HMU{}), in which we varied all of the free parameters we wish to infer (``Run 0" in Table \ref{tab:param_fits}). We find that several of the model parameters are not well constrained. This includes the break mass for which we find $M_{p,\rm break} = 65_{-35}^{+23} M_\oplus$ (note the large uncertainties). We note that our inferred value is between the two break masses that were included in the \NR20{} model ($\sim 10$-$20 M_\oplus$ and $\sim 170 M_\oplus$, depending on which model is chosen between their models 2--4; see Table 1 of \NR20{}). While the radius-mass power-law below the break mass is reasonably well constrained ($\gamma_0 = 0.16_{-0.11}^{+0.18}$ and $\sigma_0 = 0.39_{-0.14}^{+0.08}$), the power-law above the break mass is not, with $\gamma_1 = 0.47_{-0.29}^{+0.29}$ and $\sigma_1 = 0.28_{-0.18}^{+0.16}$ (note the significantly larger uncertainties). This can be explained by the fact that the model often does not draw many planets with masses greater than $\sim 100 M_\oplus$, as evidenced by the best-fit parameters of the initial planet mass (lognormal) distribution: $\mu_M = 1.18_{-0.73}^{+0.80}$ and $\sigma_M = 1.37_{-0.50}^{+0.65}$ (units of $\ln(M_p/M_\oplus)$; for example, the median values of $\mu_M = 1.18$ and $\sigma_M = 1.37$ are equivalent to a normal distribution in $\ln{M_p}$ in which $2\sigma$ above the mean is $e^{3.92} \simeq 50 M_\oplus$, i.e. less than 2.5\% of the planets are more massive than this). However, these values are comparable to those found by \NR20{} in their ``Model 2" ($\mu_M \simeq 1.00$ and $\sigma_M \simeq 1.65$). 
We find that the normalization constant for the radius-mass relation is $R_{p,\rm norm} = 1.55_{-0.38}^{+0.59} R_\oplus$ (normalized at an initial mass of $1 M_\oplus$). This is somewhat lower than the value found by \NR20{} ($C \sim 2.1-2.4 R_\oplus$ for their models 2-4).

The remaining free parameters control the number of planets and the period distribution. These parameters carry the same meaning as in the H20 model: we find $\ln{(\lambda_c)} = -0.40_{-0.68}^{+0.59}$ and $\ln{(\lambda_p)} = 0.37_{-0.49}^{+0.54}$, and the period power-law is given by $\alpha_P = -0.05_{-0.45}^{+0.72}$ with a cluster width of $\sigma_P = 0.18_{-0.11}^{+0.14}$. Given the large uncertainties, the values of these parameters are generally consistent with those found in the H20 model (see Table 4 of \PaperIII{}).

While the results of Run 0 for \HMU{} serve as a good initial point of comparison, the poor constraints on several of the model parameters motivated ``Run 1", in which we fix the parameters that exhibited large uncertainties: namely, we set $M_{p,\rm break} = 20 M_\oplus$, $\gamma_1 = 0.5$, and $\sigma_1 = 0.3$. We note that these chosen values are close to the median values found from Run 0, except for $M_{p,\rm break}$ which was guided by the results of \NR20{} (who found a break mass of $\sim 10$-$20 M_\oplus$ across their models 2--4). The period cluster scale is also fixed to $\sigma_P = 0.25$ based on the H20 model in order to further reduce the number of free parameters. The full ABC posterior distribution for this run is shown in Appendix Figure \ref{fig:hm1_posterior_params9}.
The free parameters inferred from this run are mostly consistent with those from Run 0, with some notable differences. The median value of $R_{p,\rm norm} = 1.81_{-0.32}^{+0.32}$ is shifted to a modestly higher value; the slope of the radius-mass power-law below the break mass is decreased to $\gamma_0 = 0.10_{-0.06}^{+0.10}$ (note that this is similar to the value found from \NR20{} Model 3: $\gamma_0 = 0.08_{-0.07}^{+0.06}$). The normalization factor for the mass-loss timescale, $\ln{(\alpha_{\rm ret})} = 2.35_{-0.95}^{+0.89}$, is also comparable to the values found by \NR20{} ($\alpha \simeq 8 \Rightarrow \ln{(\alpha)} \simeq 2.08$ in all of their models). Interestingly, the location of the initial planet mass distribution is shifted to lower masses, $\mu_M = 0.58_{-0.34}^{+0.41}$, while the width remains similar, $\sigma_M = 1.52_{-0.37}^{+0.36}$. We also note that many of the free parameters in this run have smaller uncertainties than in Run 0; by fixing some of the model parameters, the remaining free parameters are more easily constrained.

\subsubsection{\HMC{} parameters} \label{sec:results:new_model_fit:hm2_params}

Next, we discuss the results of Run 1 for the Hybrid Model with Clustered masses (\HMC{}). This is identical to Run 1 for \HMU{}, but the model now has an additional parameter for the cluster width of the initial planet masses, $\sigma_{M,c}$. The full ABC posterior distribution for this run is shown in Figure \ref{fig:hm2_posterior_params10}. We find that all of the free parameters essentially remain the same as in \HMU{} (i.e., they are all consistent within the error bars/central 68.3\% uncertainties). However, we find that the clustering parameter for the initial planet masses is $\sigma_{M,c} = 0.37_{-0.20}^{+0.24}$. Importantly, this is much smaller than the width of the overall initial planet mass distribution, $\sigma_M = 1.77_{-0.33}^{+0.40}$, implying that the initial planet masses are significantly clustered.

In Figure \ref{fig:radius_mass_credible}, we plot the radius-mass relations and initial mass distributions from the hybrid models. This is the same as Figure \ref{fig:radius_mass}, except now the best-fit models from \HMU{} (blue) and \HMC{} (green) are shown instead of an \NR20{} model. Nevertheless, the hybrid models are very similar to the \NR20{} model (and to each other) given their shared parameterizations and comparable values for the parameters. The initial planet mass distributions are also highly alike between the two hybrid models. 
For \HMC{}, a typical cluster is shown as the green dotted curve, illustrating the extent of the similarity in (initial) planet masses between planets in the same cluster, in contrast to the overall population (green solid curve). In \S\ref{sec:results:new_model_fit:observed_size_clustering}, we will show that \HMC{} produces a substantially better fit to several of the \Kepler{}-observed distributions than \HMU{}, due to this clustering in planet masses.

\begin{figure*}
\centering
\begin{tabular}{cc}
 \includegraphics[scale=0.425,trim={0 0.5cm 0 0.2cm},clip]{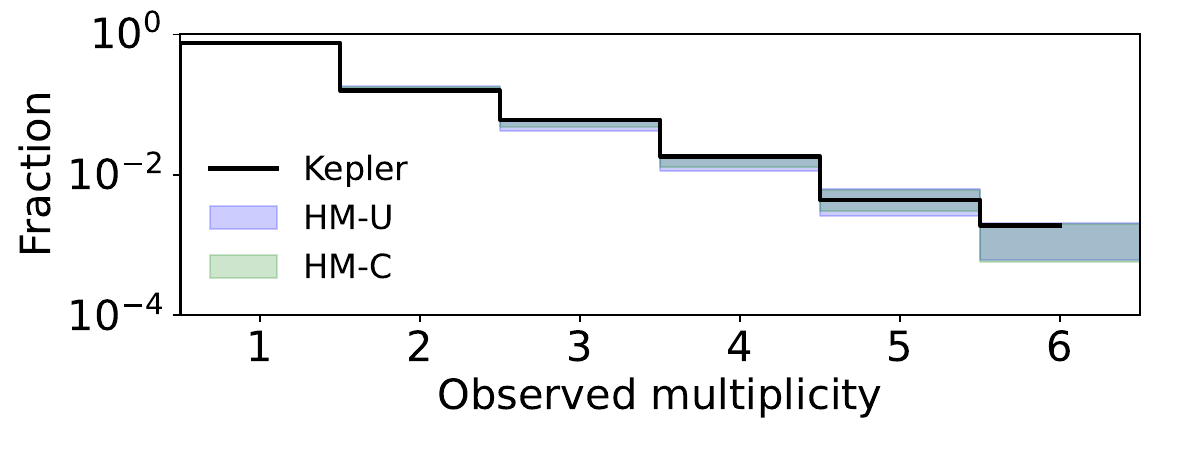} & 
 \includegraphics[scale=0.425,trim={0 0.5cm 0 0.2cm},clip]{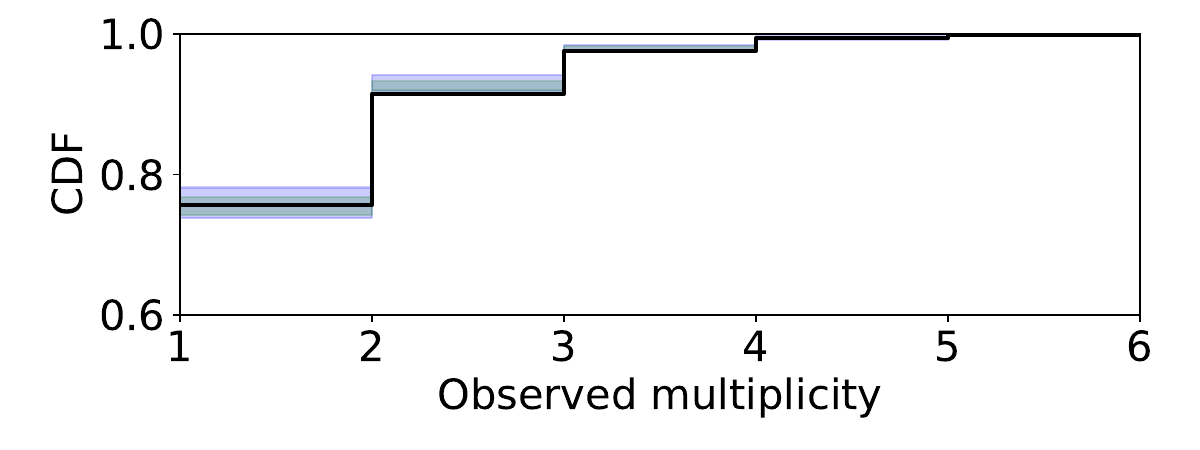} \\
 \includegraphics[scale=0.425,trim={0 0.5cm 0 0.2cm},clip]{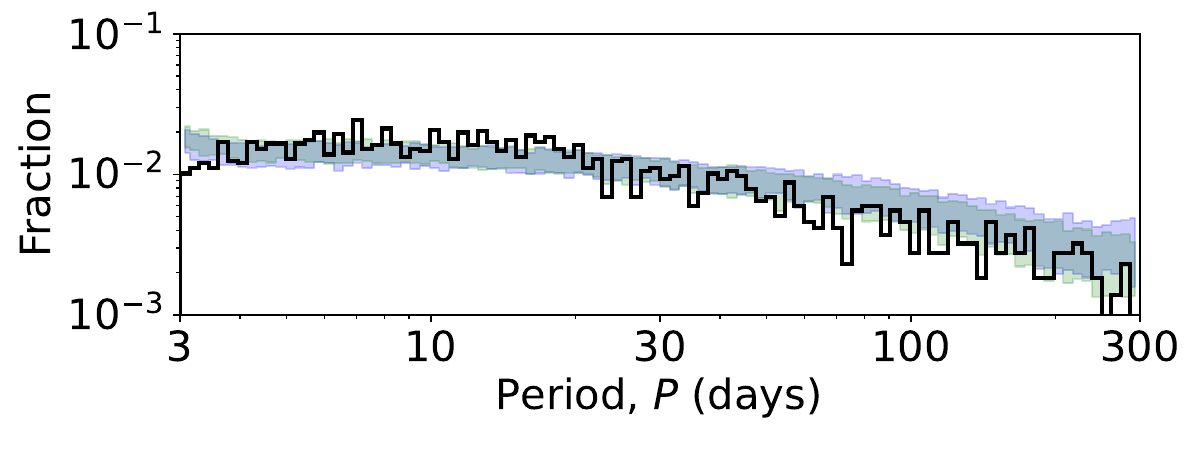} &
 \includegraphics[scale=0.425,trim={0 0.5cm 0 0.2cm},clip]{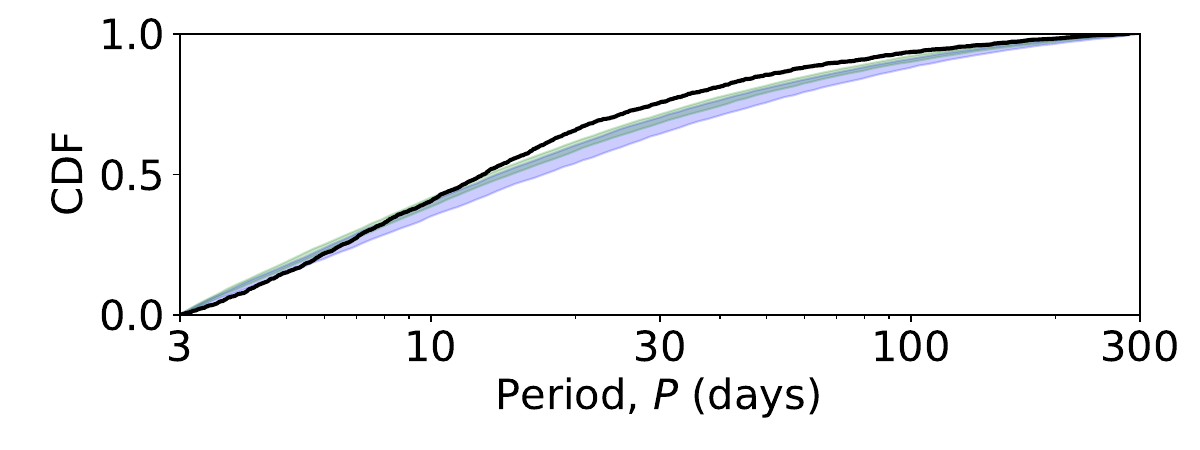} \\
 \includegraphics[scale=0.425,trim={0 0.5cm 0 0.2cm},clip]{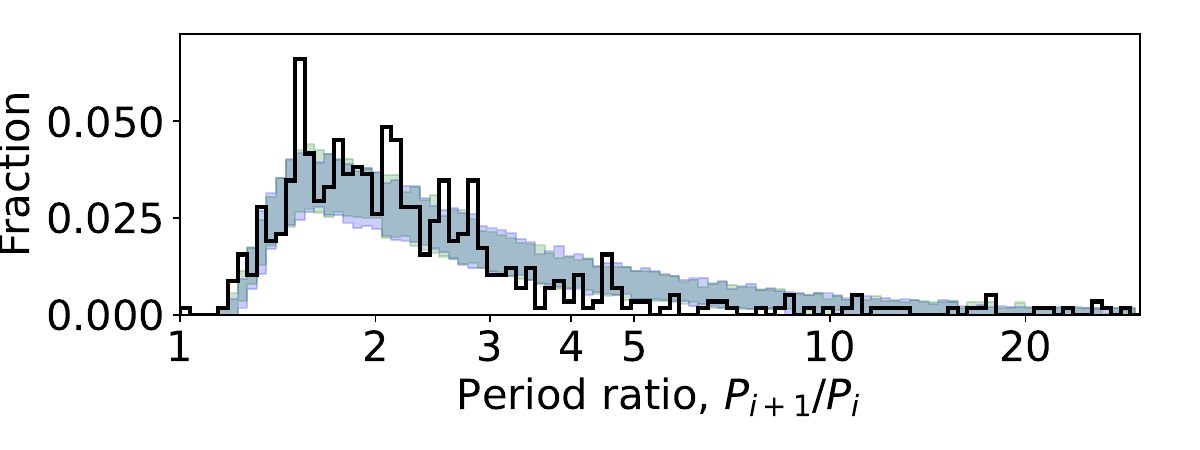} &
 \includegraphics[scale=0.425,trim={0 0.5cm 0 0.2cm},clip]{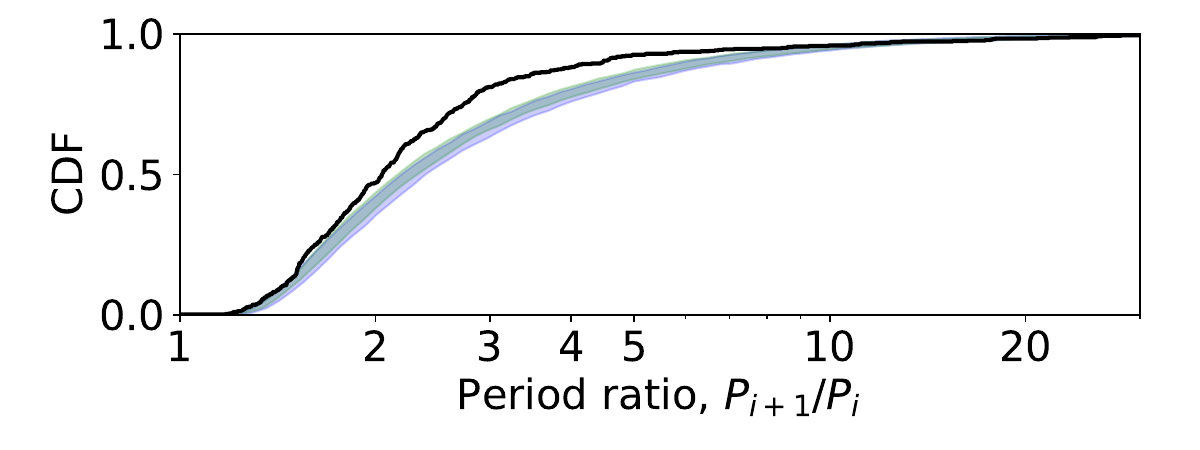} \\
 \includegraphics[scale=0.425,trim={0 0.5cm 0 0.2cm},clip]{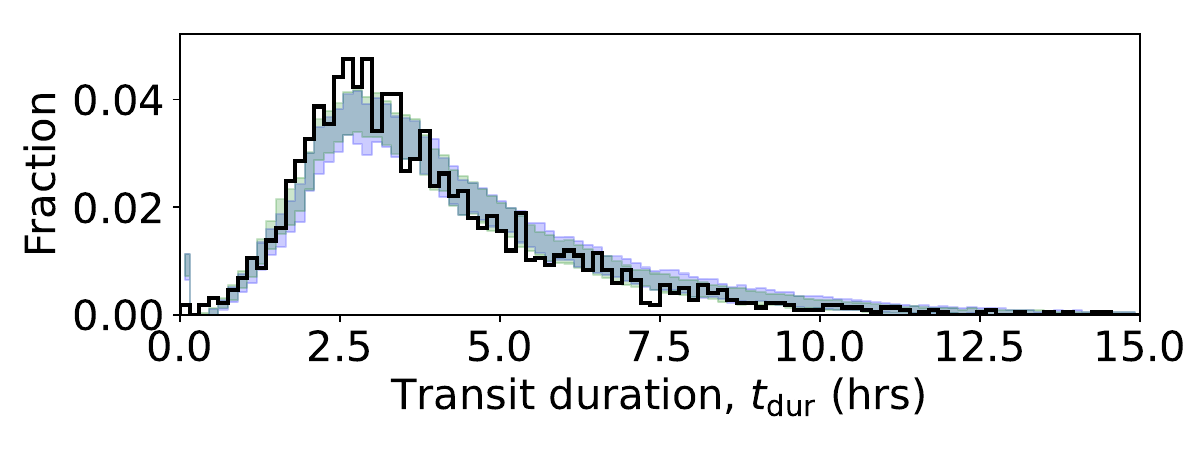} &
 \includegraphics[scale=0.425,trim={0 0.5cm 0 0.2cm},clip]{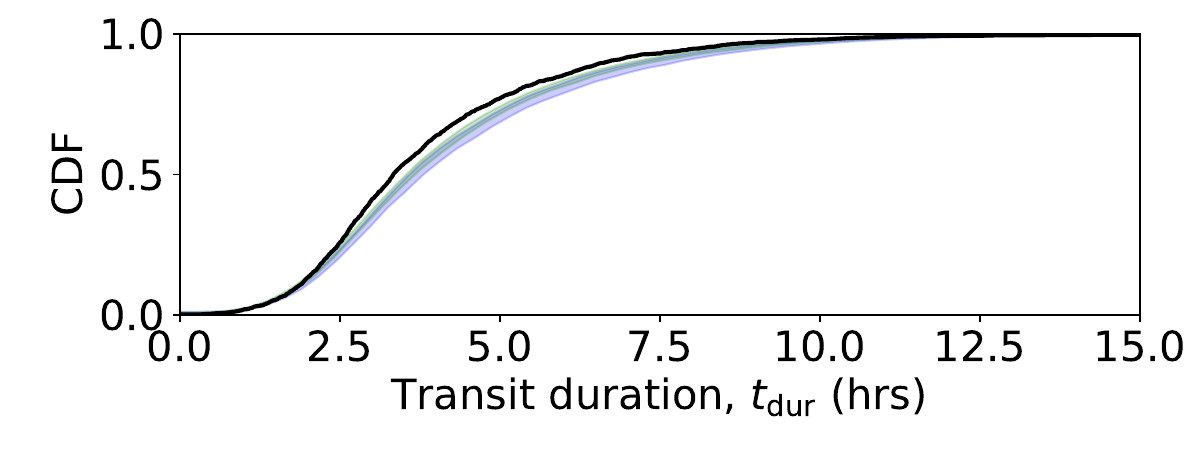} \\
 \includegraphics[scale=0.425,trim={0 0.5cm 0 0.2cm},clip]{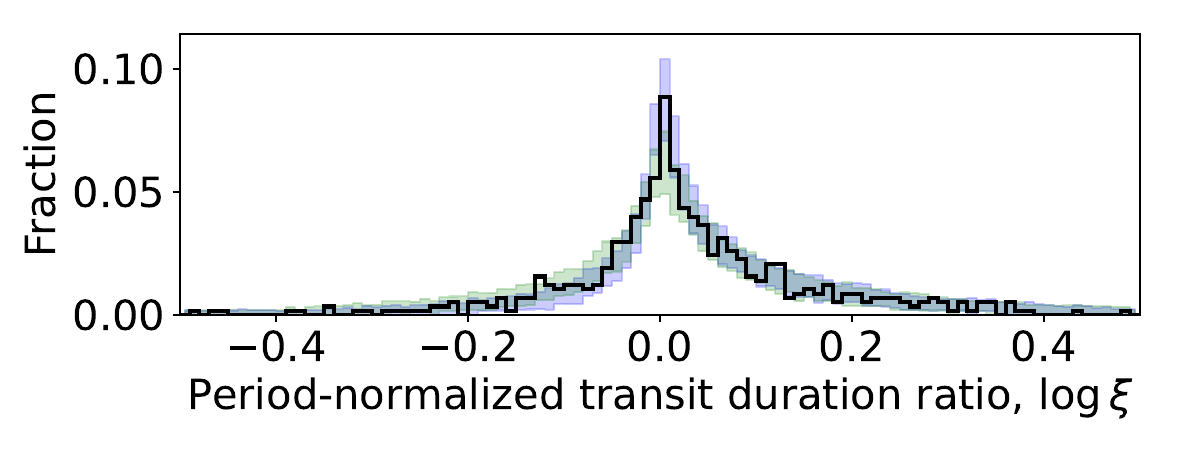} &
 \includegraphics[scale=0.425,trim={0 0.5cm 0 0.2cm},clip]{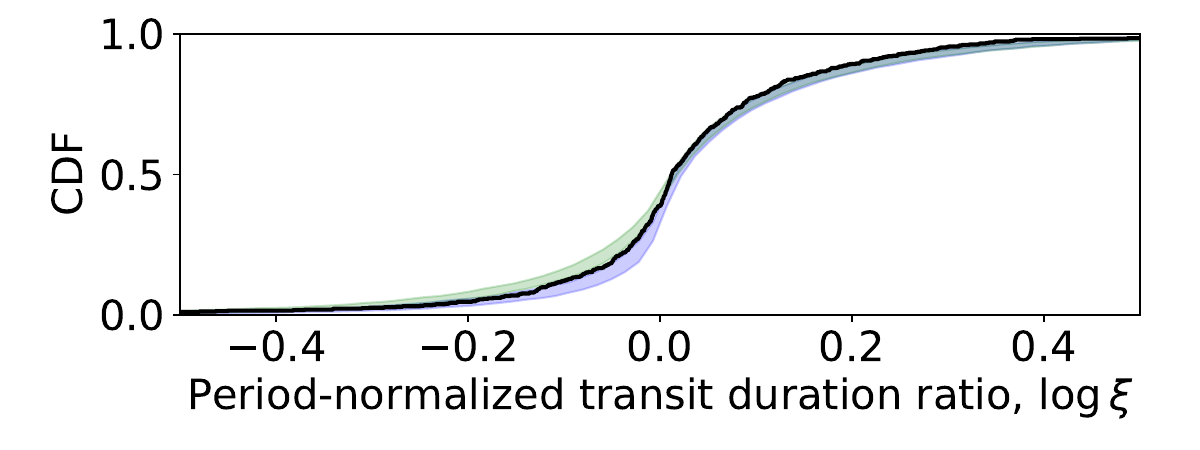} \\
 \includegraphics[scale=0.425,trim={0 0.5cm 0 0.2cm},clip]{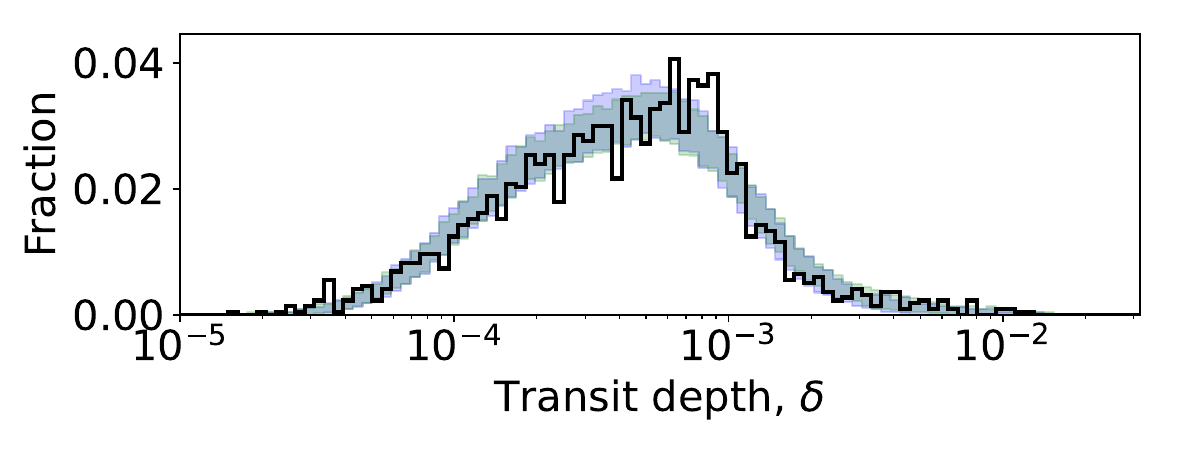} &
 \includegraphics[scale=0.425,trim={0 0.5cm 0 0.2cm},clip]{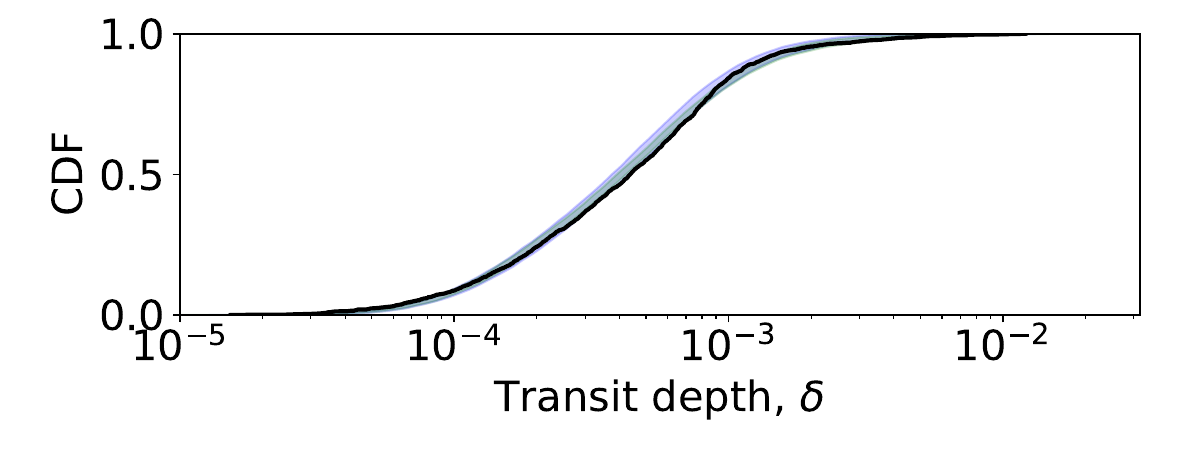} \\
\end{tabular}
\caption{Marginal distributions of the observed planetary properties, for both hybrid models as compared to the \Kepler{} catalog (black histogram). The left panels show histograms of the observed distributions, while the right panels show the cumulative distributions of the same distributions. In each panel, the colored shaded regions denote the 16-84\% credible region of each bin computed over 100 simulated catalogs from each hybrid model. Given that the only difference in the two hybrid models is the drawing of clustered (\HMC{}; green) vs. unclustered (\HMU{}; blue) initial planet masses, these observed distributions are very similar between the two models.}
\label{fig:observed_marginals_compare}
\end{figure*}

\subsubsection{Comparison to the observed \Kepler{} catalog} \label{sec:results:new_model_fit:observed}

\begin{figure*}
\centering
\begin{tabular}{cc}
 \includegraphics[scale=0.425,trim={0 0.5cm 0 0.2cm},clip]{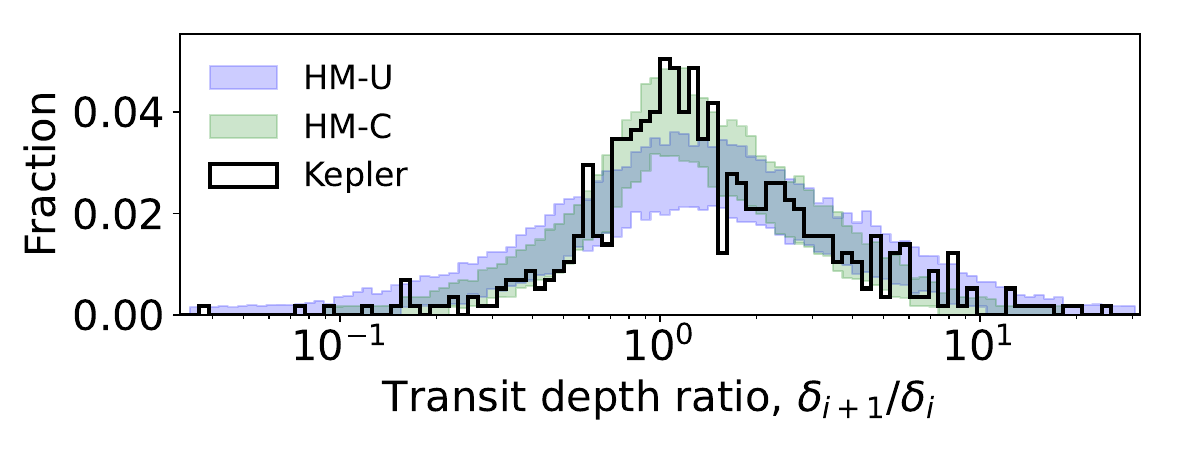} &
 \includegraphics[scale=0.425,trim={0 0.5cm 0 0.2cm},clip]{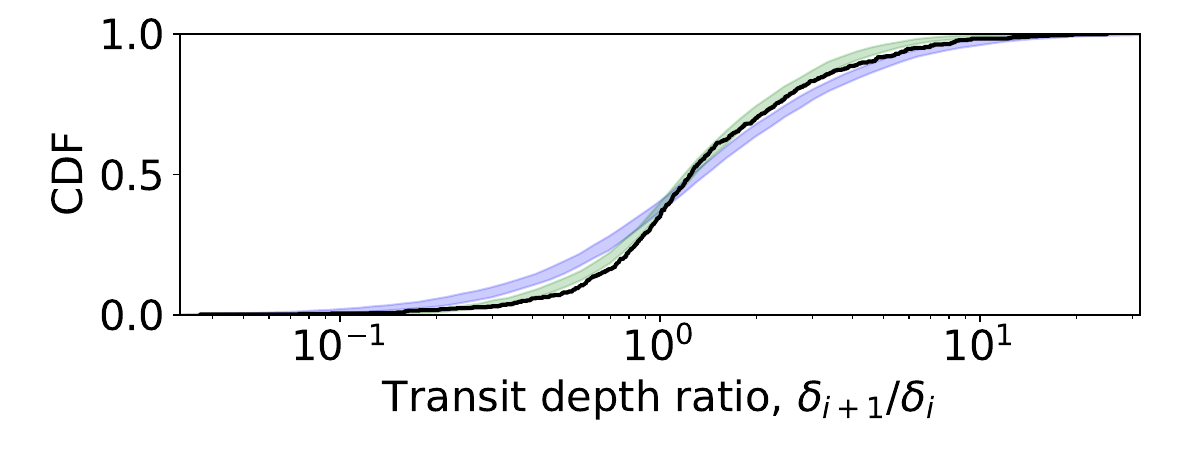} \\
 \includegraphics[scale=0.425,trim={0 0.5cm 0 0.2cm},clip]{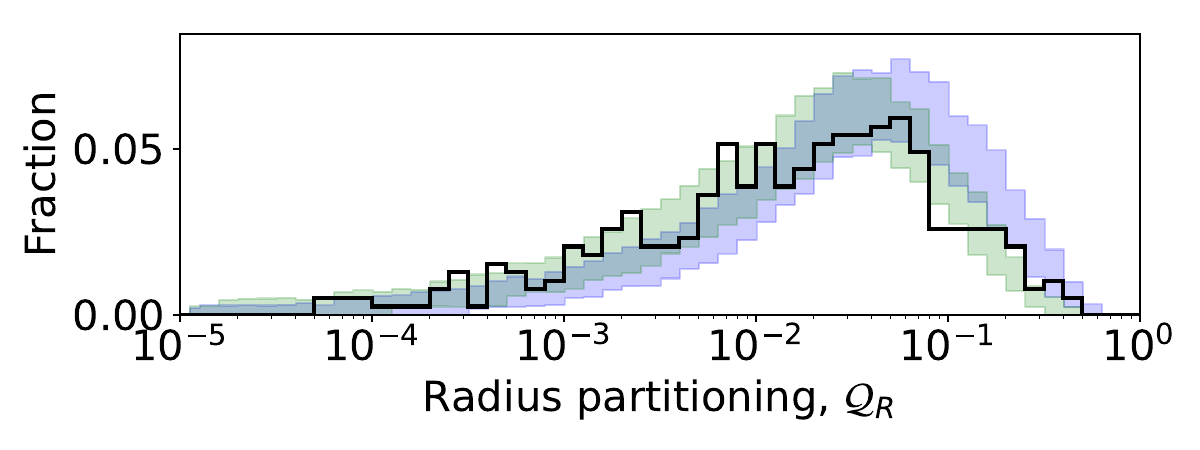} &
 \includegraphics[scale=0.425,trim={0 0.5cm 0 0.2cm},clip]{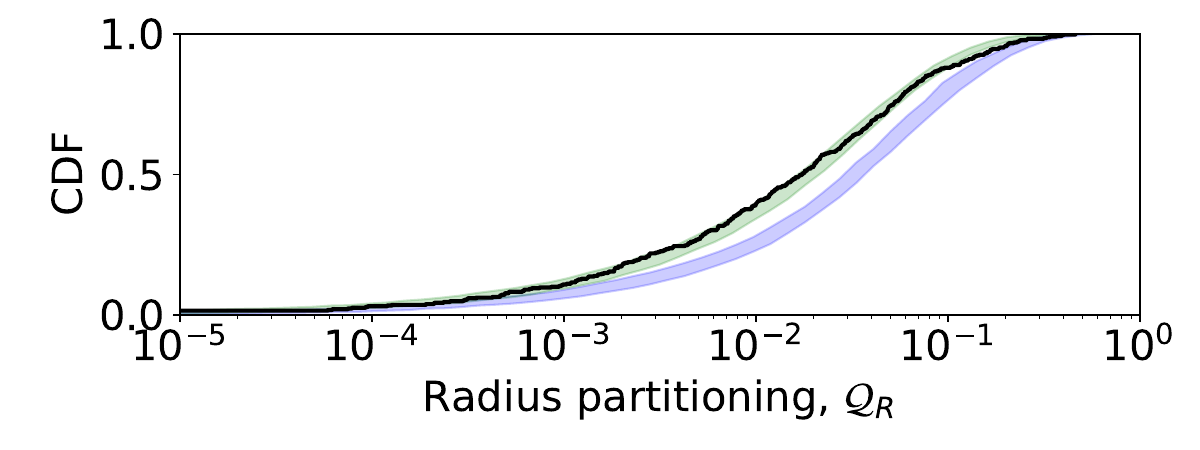} \\
 \includegraphics[scale=0.425,trim={0 0.5cm 0 0.2cm},clip]{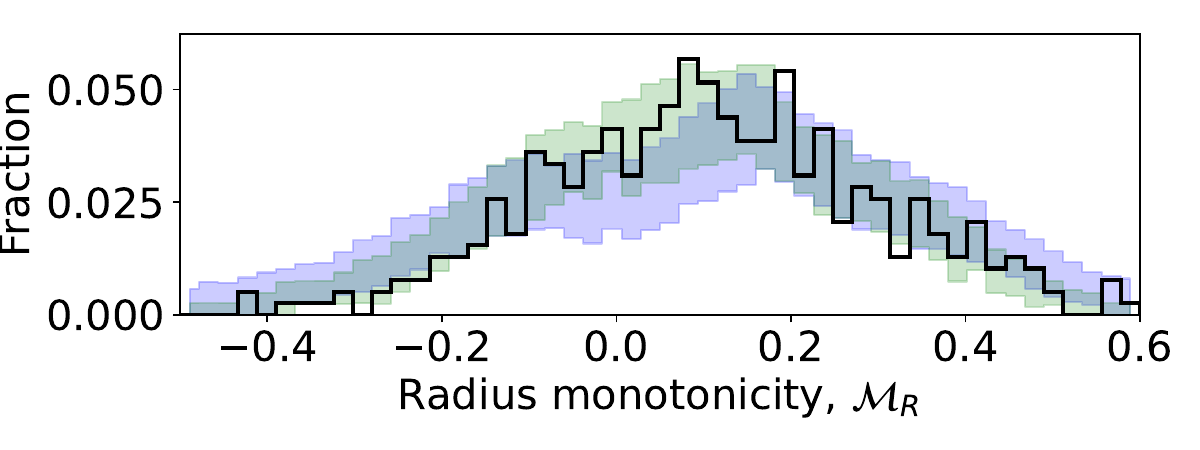} &
 \includegraphics[scale=0.425,trim={0 0.5cm 0 0.2cm},clip]{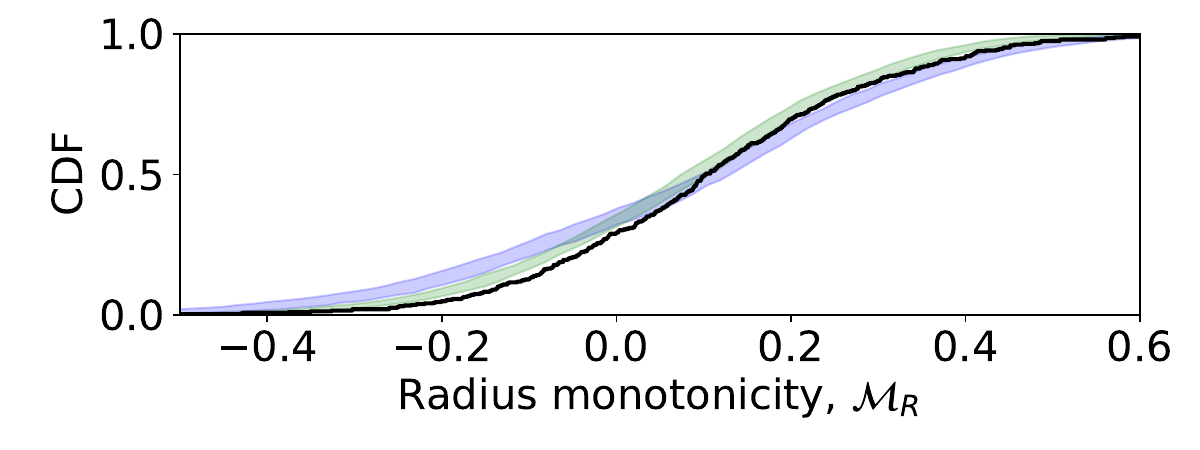} \\
\end{tabular}
\caption{Marginal distributions of the observed planet radius similarity metrics, for both hybrid models (colored regions showing the 16-84\% quantiles of each model) as compared to the \Kepler{} catalog (black histogram). From top to bottom: (1) transit depth ratios (equivalent to the radius ratios squared) of adjacent planet pairs; (2) radius partitioning ($\mathcal{Q}_R$), which measures the degree of planet size similarity, with lower (higher) values denoting more (less) similar planet radii; and (3) radius monotonicity ($\mathcal{M}_R$), which measures the degree of planet size ordering, with positive (negative) values denoting a preference for increasing (decreasing) planet sizes towards longer periods. \HMC{} (green) provides a significantly better fit to the \Kepler{} data than \HMU{} (blue), for all three distributions, due to the clustering in initial planet masses which produces more similar planet sizes within the same system.}
\label{fig:observed_marginals_radii_metrics_compare}
\end{figure*}

Here, we show how the hybrid models compare to the observed \Kepler{} catalog in terms of the marginal distributions (summary statistics described in \S\ref{sec:methods:model_inference:summary_stats}). In Figures \ref{fig:observed_marginals_compare} and \ref{fig:observed_marginals_radii_metrics_compare}, we plot all of the fitted marginal distributions, except the observed planet radius and gap-subtracted radius distributions, for the \Kepler{} catalog (solid black line in each panel) as well as the central 16\%-84\% quantile regions of the simulated catalogs drawn from Run 1 of each hybrid model (blue and green shaded region in each panel, for \HMU{} and \HMC{}, respectively). The left and right panels show the histograms and the cumulative distribution functions (CDFs) of the same data, respectively. The observed radius and gap-subtracted radius distributions will be shown and discussed in \S\ref{sec:results:radius_valley}.

The panels in Figure \ref{fig:observed_marginals_compare}, from top to bottom, show the marginal observed distributions of: (1) planet multiplicities; (2) orbital periods; (3) period ratios of adjacent planet pairs; (4) transit durations; (5) log period-normalized transit duration ratios; and (6) transit depths.
In terms of these marginal distributions, the hybrid models are almost indistinguishable from each other, and both produce very similar fits to the \Kepler{} catalog. It is unsurprising that \HMU{} and \HMC{} produce nearly identical distributions of these observables because they are essentially the same model (with largely overlapping values for the inferred model parameters), except that one draws clustered initial masses and the other does not.
We note that these distributions are also very similar to those generated by the H20 model (see \PaperIII{}), given that they share many of the same architectural elements.

The observed multiplicity distribution is fit very well by both hybrid models (as by the H20 model), given that they both consist of similar numbers of planets and have orbital mutual inclinations sculpted by AMD stability (see \S4.1 of \PaperIII{} for a detailed discussion of the so-called ``\Kepler{} dichotomy" and the challenges of fitting the observed multiplicity distribution with a single population model).

The period and period ratio distributions are acceptable, although there are appreciable differences and the same two caveats as with the H20 model: (1) the single power-law over-produces the number of planets at the innermost edge, and (2) there are features near the first-order mean motion resonances (MMRs) that cannot be captured by power-law models. We note that a broken power-law in orbital period is likely to provide an even better fit, as was implemented by \NR20{} (who find a break period of $P_{\rm break} \simeq 7$-8 days; see also \citealt{2012ApJS..201...15H, 2018AJ....155...89P, 2018AJ....156...24M, 2021MNRAS.503.1526R}), though a simple power-law reduces the number of model parameters and provides an adequate fit given our minimum period bound of $P_{\rm min} = 3$ days. Also, we note that in Run 1 of either model, the period ratio distribution was excluded from the distance function, and yet the best-fit models found from these runs still reproduce this distribution reasonably well. It is likely that the fit to the period ratio distribution could be improved if we had included this in the distance function and allowed $\sigma_P$ to vary.

The transit duration and period-normalized transit duration ratio distributions were also not directly fit in Run 1, and yet are extremely well reproduced. The transit depth distribution is also recovered well by the hybrid models (though there appears to be a peak at $\delta \simeq 10^{-3}$ that is difficult to fully capture), and is an improvement over the H20 model, a promising sign given that it is directly related to the planet radius distribution.

\subsubsection{The need for clustered planet sizes} \label{sec:results:new_model_fit:observed_size_clustering}

Here, we directly assess the need for clustered planet sizes (i.e., the improvement of \HMC{} over \HMU{}) by examining the fits to the three summary statistics designed to capture the intra-system size similarity patterns.

Figure \ref{fig:observed_marginals_radii_metrics_compare} shows how the hybrid models compare to the \Kepler{} catalog in terms of the observed distributions of: (1) transit depth ratios of adjacent planets, $\delta_{i+1}/\delta_i$; (2) radius partitioning, $\mathcal{Q}_R$ (equation \ref{eq:radius_partitioning}); and (3) radius monotonicity, $\mathcal{M}_R$ (equation \ref{eq:radius_monotonicity}). \HMC{} (green) provides a marked improvement for all three distributions compared to \HMU{} (blue), for reproducing the \Kepler{} data (black). \HMU{} produces a distribution of $\delta_{i+1}/\delta_i$ that is broader and does not capture the strong peak around 1 (equal planet sizes) seen in the data, while \HMC{} is able to capture this feature and the overall distribution very well. \HMU{} also produces a distribution of $\mathcal{Q}_R$ that is shifted to higher values (implying less uniform planet sizes), while \HMC{} provides a very close fit. Lastly, while both models produce distributions of $\mathcal{M}_R$ that are more heavily weighted towards positive values (implying a preference for increasing planet sizes towards longer periods), \HMC{} clearly does a better job of matching the \Kepler{} distribution.

These results overwhelmingly demonstrate the need for clustered planet sizes in the underlying population of multi-planet systems. \HMC{} provides a significantly better fit to the observed distributions that are designed to measure the degree of intra-system similarity in the planet radii. The distances to the \Kepler{} data for these distributions are also significantly improved for \HMC{} compared to \HMU{} (see Appendix Figure \ref{fig:dists}, right-bottom three panels). Moreover, \HMC{} often achieves a weighted distance of $w\mathcal{D} \simeq 1$ for each of distributions, implying we have found a model that is nearly indistinguishable from the \Kepler{} data when accounting for the stochastic noise that arises from the repeated catalog simulations from the same model.
The ability of \HMC{} to fully capture the observed preference for similar planet radii within the same system implies that these patterns can be understood as an outcome of planet mass uniformity, a point that we will revisit in \S\ref{sec:results:models_underlying:size_similarity} when examining the underlying distributions of the models.


\begin{figure*}
\centering
\begin{tabular}{cc}
 \includegraphics[scale=0.47,trim={1.2cm 0.5cm 0.5cm 0.1cm},clip]{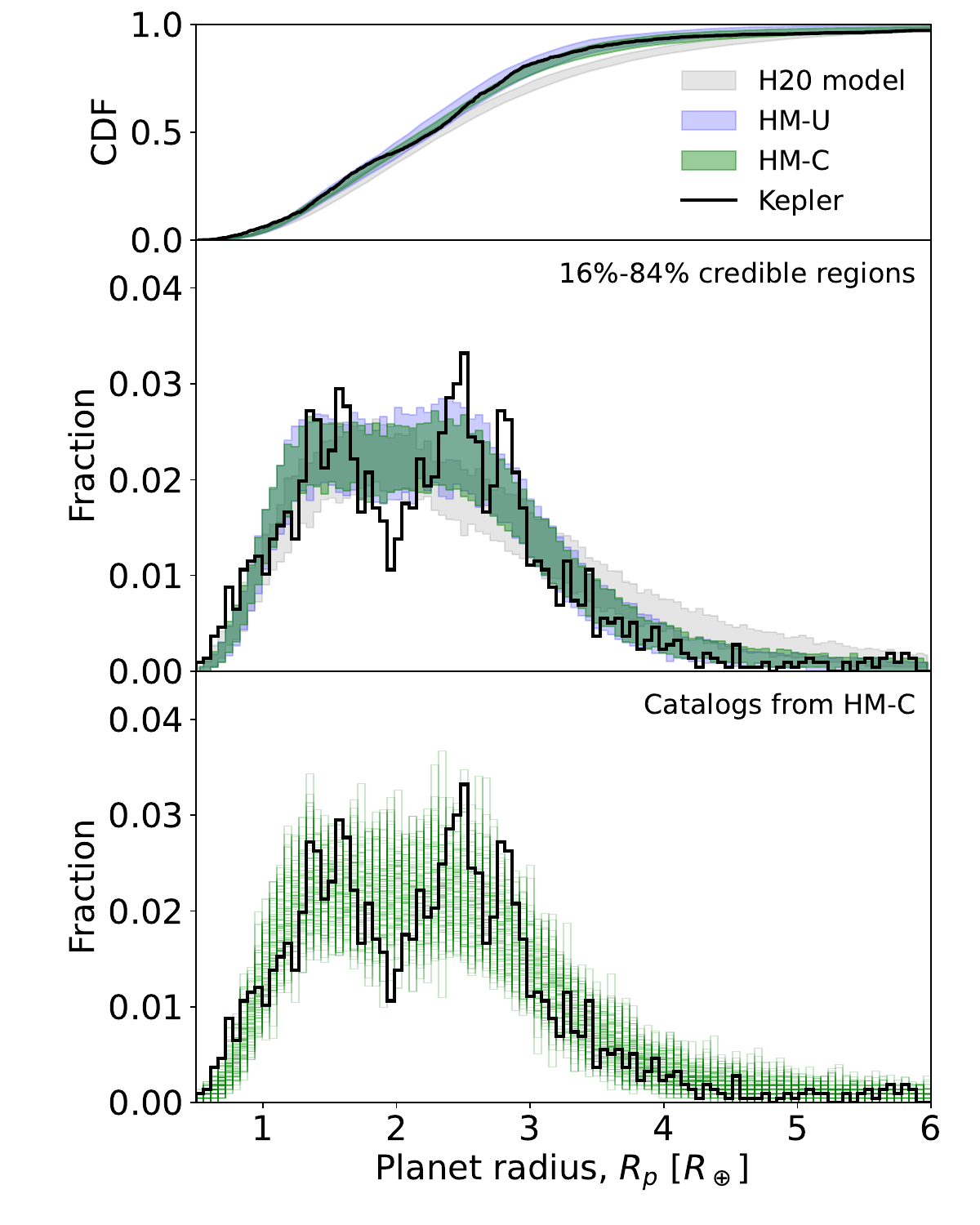} &
 \includegraphics[scale=0.47,trim={1.2cm 0.5cm 0.5cm 0.1cm},clip]{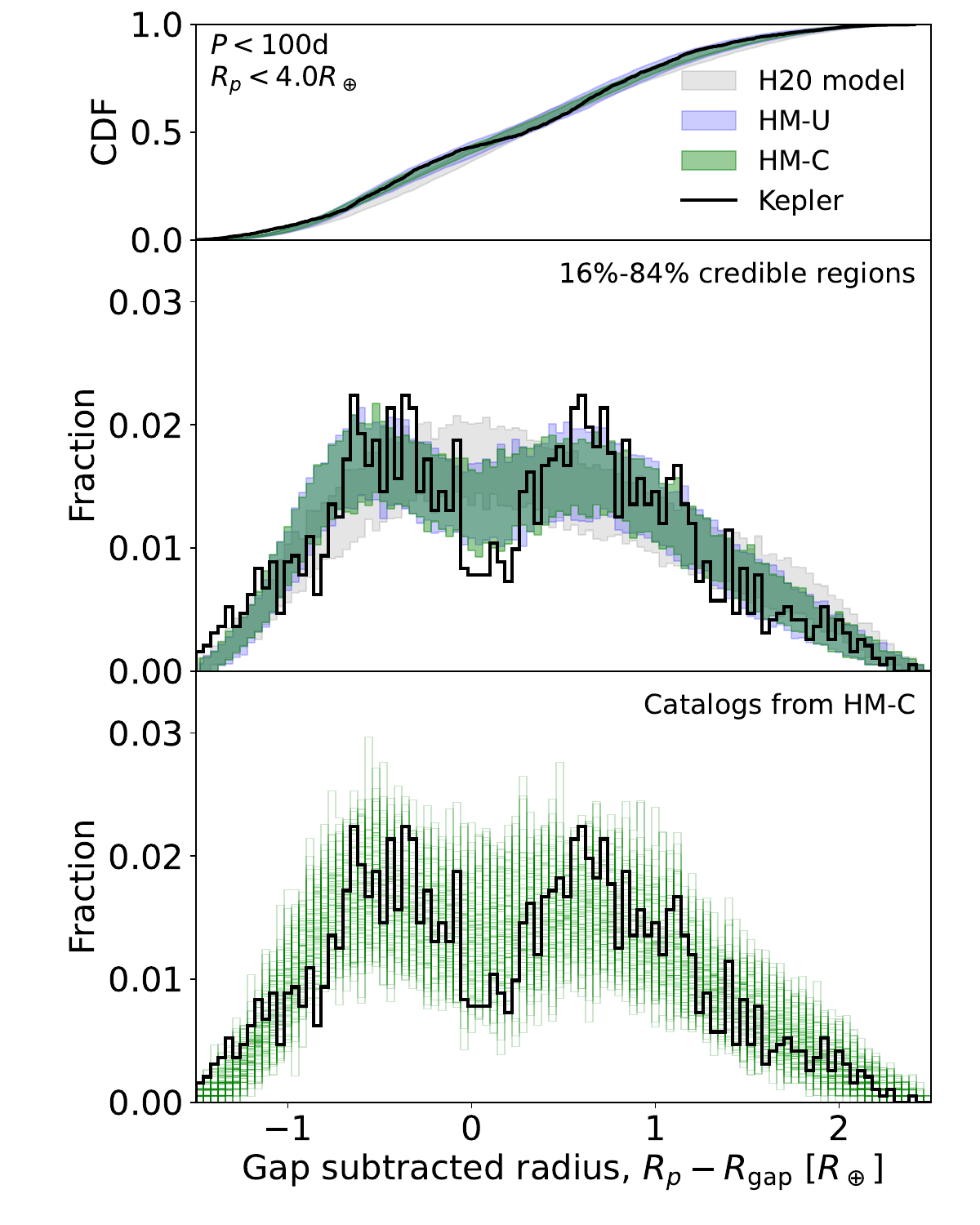} \\
\end{tabular}
\caption{The observed distributions of planet radii (\textbf{left panels}) and gap-subtracted radii (\textbf{right panels}) for the \Kepler{} catalog and simulated catalogs drawn from the posteriors of the hybrid and H20 models. The histograms (\textbf{middle panels}) and the cumulative distribution functions (\textbf{top panels}) for the \Kepler{} catalog and each model, as labeled. In each panel, the black curve shows the \Kepler{} catalog. In the top and middle panels, the blue and green shaded regions denote the 16-84\% credible regions of each bin for \HMU{} and \HMC{}, respectively, while the gray shaded region denotes the 16-84\% credible region of the H20 model, for comparison. Each shaded region is computed from 100 simulated catalogs from the respective model.
\textbf{Bottom panels:} the individual histograms drawn from the \HMC{}. For the gap-substracted radii (right panels), the catalogs are also restricted to planets with $P < 100$ days and $R_p < 4 R_\oplus$.}
\label{fig:observed_radii_models}
\end{figure*}

\subsection{The planet radius valley} \label{sec:results:radius_valley}

A primary motivation for constructing the ``hybrid models" between the H20 and \NR20{} models is to determine whether the \Kepler{} observed planet radius valley can be reproduced simultaneously with clustered planets in a multi-planet model. Here, we show that the hybrid models are capable of producing an observed radius valley, even as strong as that of the \Kepler{} data. However, random draws from the posterior distribution (for either \HMU{} or \HMC{}) do not reliably generate a radius valley. Nevertheless, first we show that the hybrid models provide a significantly better fit to the \Kepler{}-observed radius distribution than the H20 model, in \S\ref{sec:results:radius_valley:observed}. We then present a simple method for measuring the ``depth" of the radius valley in \S\ref{sec:results:radius_valley:depth} and use it to show that the hybrid models can produce a radius valley as deep as that of the \Kepler{} data for specific regions of parameter space, in \S\ref{sec:results:radius_valley:depth_large}. Lastly, we show how the hybrid models compare to the \Kepler{} catalog in terms of the radius valley in the period-radius distribution in \S\ref{sec:results:radius_valley:radius_period}.

\subsubsection{A better model for the observed radius distribution} \label{sec:results:radius_valley:observed}

In Figure \ref{fig:observed_radii_models}, we first plot the distribution of observed planet radii (left panel) and gap-subtracted planet radii (right panel) from the simulated catalogs drawn from each model compared to the \Kepler{} catalog. The top panels show the cumulative distributions while the middle and bottom panels show histograms. In each panel, the \Kepler{} distribution is shown as the solid black curve/histogram. In the top and middle panels, all three models are shown: (1) the H20 model (gray), (2) \HMU{} (blue), and (3) \HMC{} (green), where the shaded regions denote the 16\%-84\% quantiles of the histograms for 100 simulated catalogs with parameters drawn from the posterior distributions of each model. In the bottom panel, the 100 simulated catalogs from \HMC{} are plotted as individual histograms.

First, we focus on the observed radius distribution (i.e. left panel of Figure \ref{fig:observed_radii_models}). All models produce planets up to $10 R_\oplus$ but we restrict the axes to $6 R_\oplus$ to focus on the radius valley. The \Kepler{} data clearly exhibits the well known radius valley in the form of a local minimum around $2 R_\oplus$. For all three models, the 16\%-84\% credible regions do not fully capture this feature. However, the hybrid models provide improvements over the H20 model. The H20 model (gray) produces a distribution that completely smooths over this feature (in fact, instead it peaks at where the \Kepler{}-observed valley is). It also does not fall as rapidly towards larger radii ($R_p \gtrsim 3 R_\oplus$) as the \Kepler{} distribution, producing roughly twice as many planets in this regime (this has been referred to as the ``radius cliff" and we will return to discuss this feature in \S\ref{sec:discussion:radius_cliff}). The hybrid models produce very similar distributions; while on first glance it appears that they still struggle to generate a substantial radius valley (based on the credible regions), this is due to the averaging over many simulated catalogs, some of which have a radius valley and some of which do not (in \S\ref{sec:results:radius_valley:depth}, we will focus on the catalogs that produce strong radius valleys). Even from these distributions, there is a hint of two peaks close to where the super-Earth ($\sim 1.5 R_\oplus$) and sub-Neptune ($\sim 2.5 R_\oplus$) peaks of the \Kepler{} data are located. The uncertainties of the hybrid models in this region are large, especially compared to the H20 model. The individual catalogs (bottom panel) from \HMC{} illustrates the high degree of stochastic noise. Finally, the distributions fall off beyond $\sim 3 R_\oplus$ very closely to that of the \Kepler{} data, a clear improvement over the H20 model. Overall, the distances for the radius distribution are improved for the hybrid models compared to the H20 model (see the panel labeled ``$\{R_p\}$" in Appendix Figure \ref{fig:dists}).

We see similar results in the observed distributions of gap-subtracted radii, $R_p - R_{\rm gap}$ (right panel of Figure \ref{fig:observed_radii_models}). As described in \S\ref{sec:methods:model_inference:summary_stats}, $R_{\rm gap} = 2.4R_\oplus (P/{\rm days})^{-0.1}$ (Equation \ref{eq:radius_gap}) is taken to be the period-dependent location of the radius valley for the \Kepler{} catalog. In these units, the \Kepler{}-observed valley is more pronounced (since the period-dependence is aligned) and is centered around zero, by construction. The valley that is occasionally generated by the hybrid models (green and blue shaded regions) is also seen more clearly. This is a significant improvement over the H20 model (gray shaded region), which is incapable of generating any radius valley and instead still produces a unimodal distribution that peaks near $R_p - R_{\rm gap} \simeq 0$. However, as is the case for the marginal radius distribution, not all simulated catalogs from the hybrid models produce a noticeable valley, and the valley seen in the \Kepler{} data appears deeper than those of the catalogs in the central 16\%-84\% quantiles of either hybrid model. We require a method of quantifying the ``strength" of the observed radius valley in order to reliably distinguish between models that produce a valley and those that do not.


\begin{figure}
\centering
\includegraphics[scale=0.44,trim={0.3cm 0cm 0.3cm 0.3cm},clip]{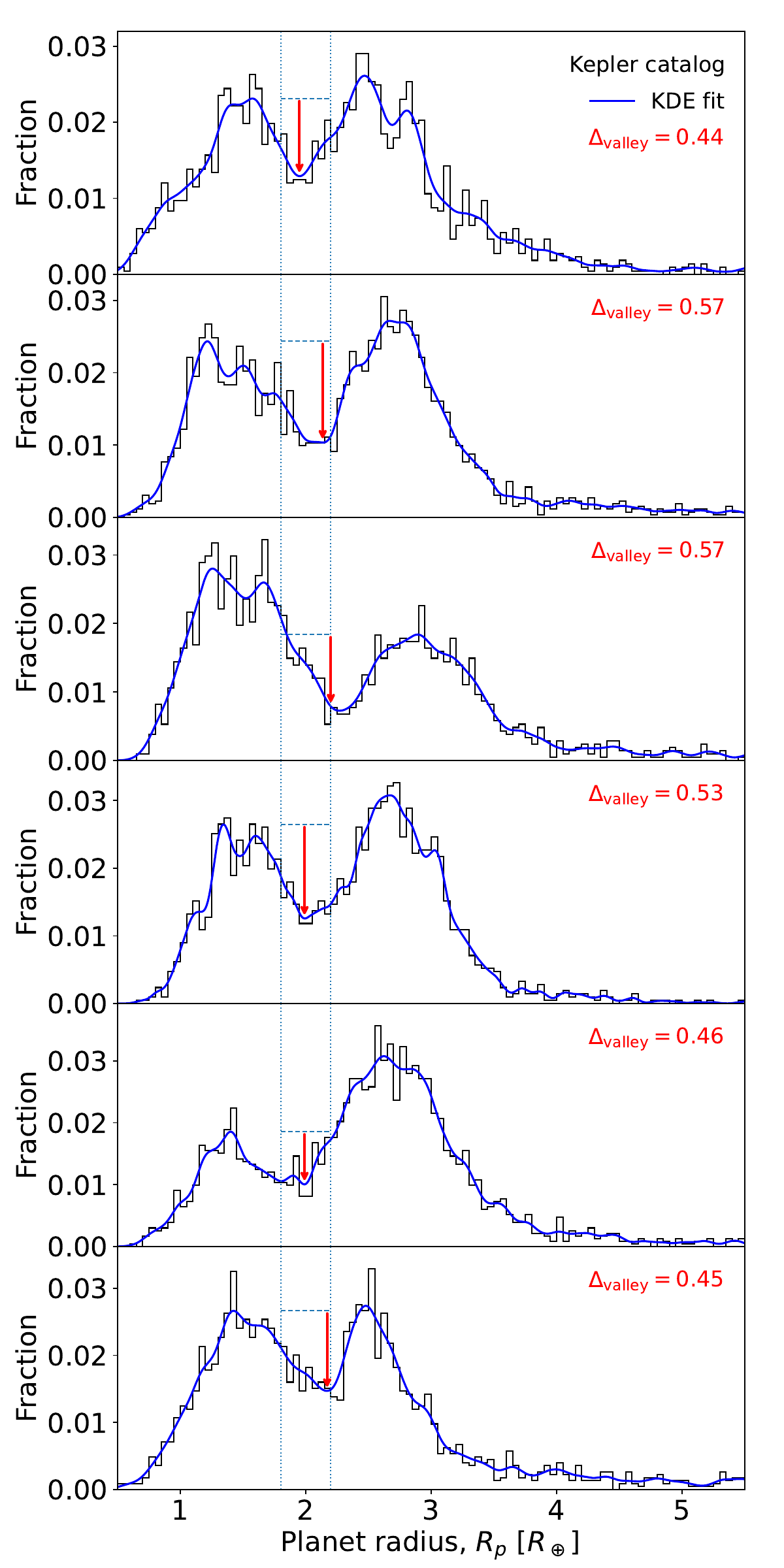} 
\caption{The observed distributions of planet radii and the ``depth" of the observed radius valley ($\Delta_{\rm valley}$), as measured for the \Kepler{} catalog (top-most panel) and simulated catalogs from \HMU{} (all panels but the top-most). In each panel, the black histogram shows the observed distribution of planet radii, while the blue curve is a KDE fit (with bandwidth $h = 0.25n^{-1/5} \simeq 0.05$ where $n$ is the number of data points). The vertical dotted lines mark the bounds ($[1.8, 2.2] R_\oplus$) for the location of the valley. The measured depth, $\Delta_{\rm valley}$, is displayed by the red arrow and labeled.
The five simulated catalogs shown here each exhibit a significant radius valley, with depths larger than or comparable to that seen in the \Kepler{} data.}
\label{fig:observed_radii_depth_models}
\end{figure}

\begin{figure}
\centering
\includegraphics[scale=0.44,trim={0.8cm 0cm 0.3cm 0.1cm},clip]{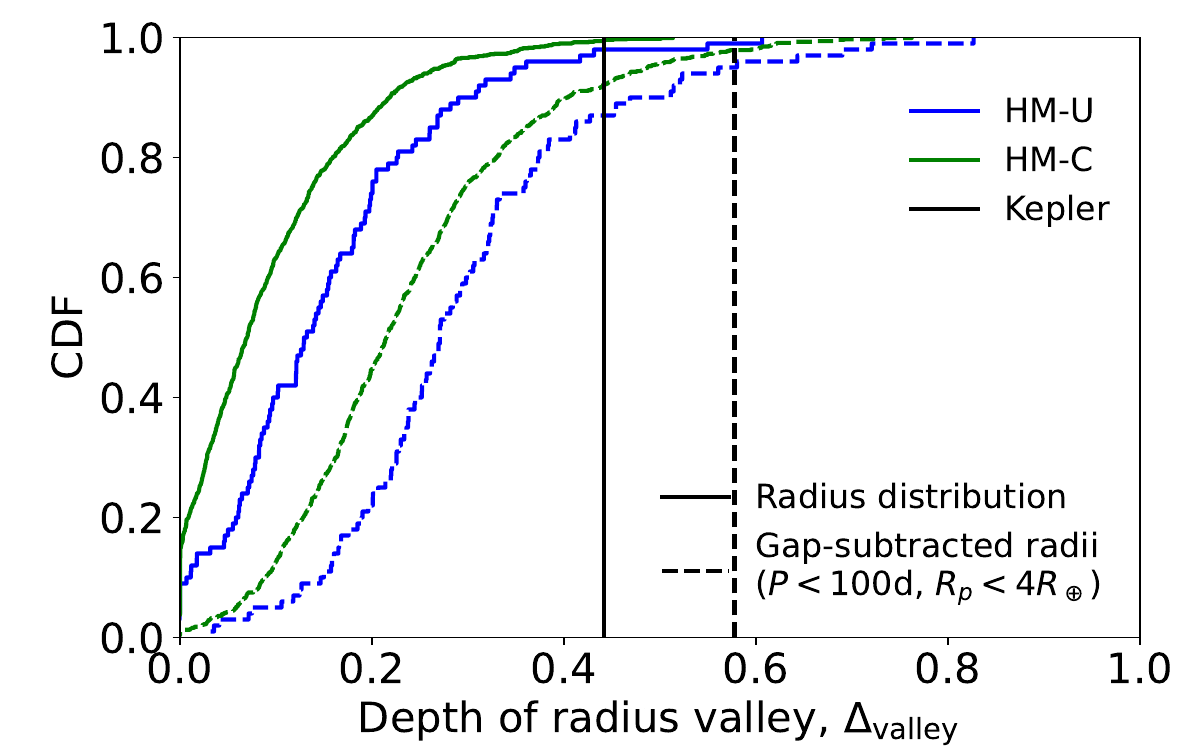} 
\caption{The distribution of measured radius valley ``depths", $\Delta_{\rm valley}$, from each model. The solid lines denote the depths as measured from the observed radius distribution (e.g. as depicted in Figure \ref{fig:observed_radii_depth_models}), while the dashed lines denote the depths as measured from the observed gap-subtracted radius distribution (which is also restricted to planets with $P < 100$ days and $R_p < 4 R_\oplus$). The vertical lines denote the values for the \Kepler{} catalog; $\Delta_{\rm valley} = 0.44$ and 0.58 for the radius and gap-subtracted radius distributions, respectively. For the radius distribution, the \Kepler{} value is higher than 98\% (99.5\%) of the catalogs from \HMU{} (\HMC{}). For the gap-subtracted radius distribution, the \Kepler{} value is higher than 95\% (97.9\%) of the catalogs from \HMU{} (\HMC{}).}
\label{fig:observed_radii_depths_cdfs}
\end{figure}

\begin{figure*}
\centering
\begin{tabular}{cc}
 \includegraphics[scale=0.47,trim={1.2cm 0.5cm 0.5cm 0.1cm},clip]{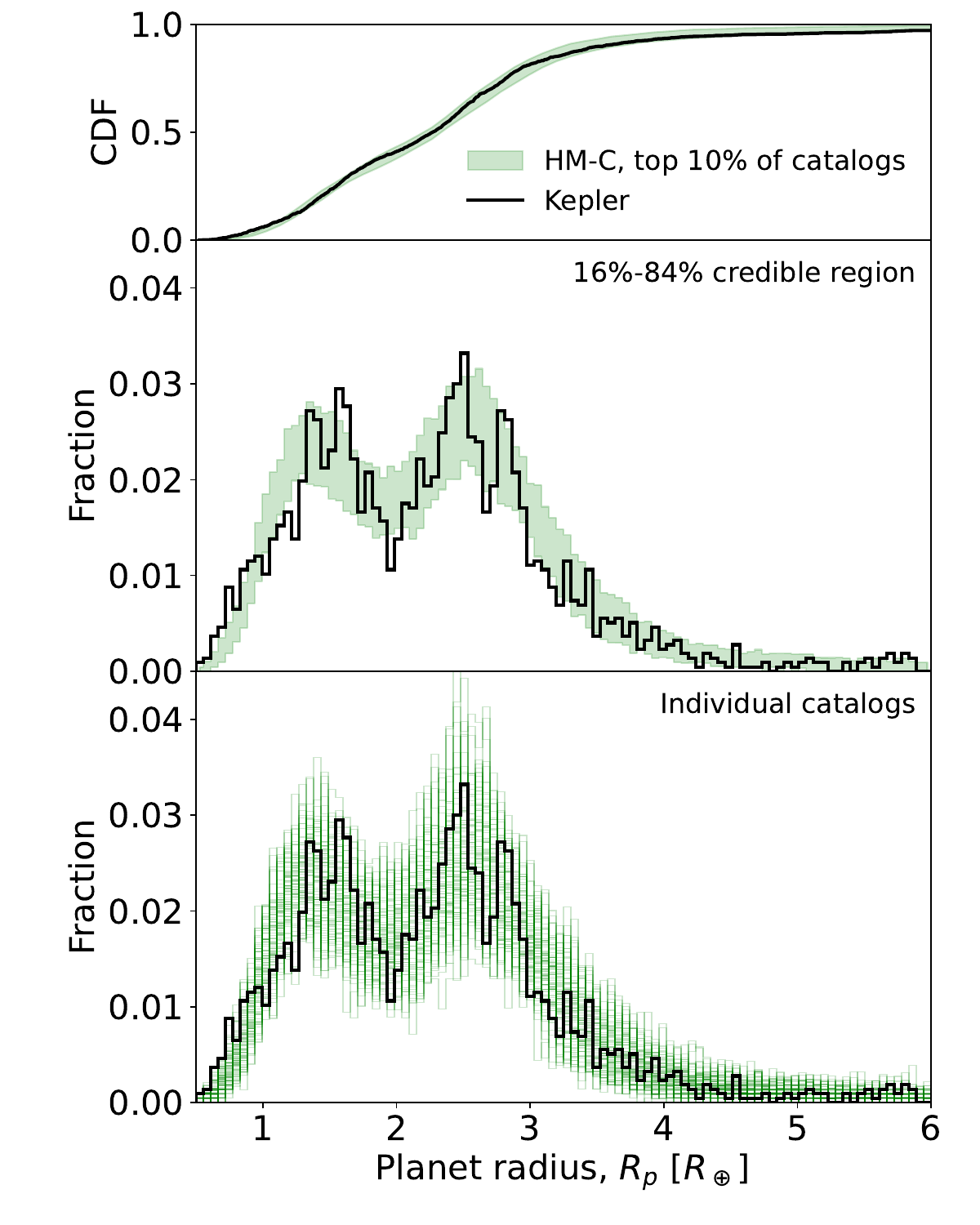} &
 \includegraphics[scale=0.47,trim={1.2cm 0.5cm 0.5cm 0.1cm},clip]{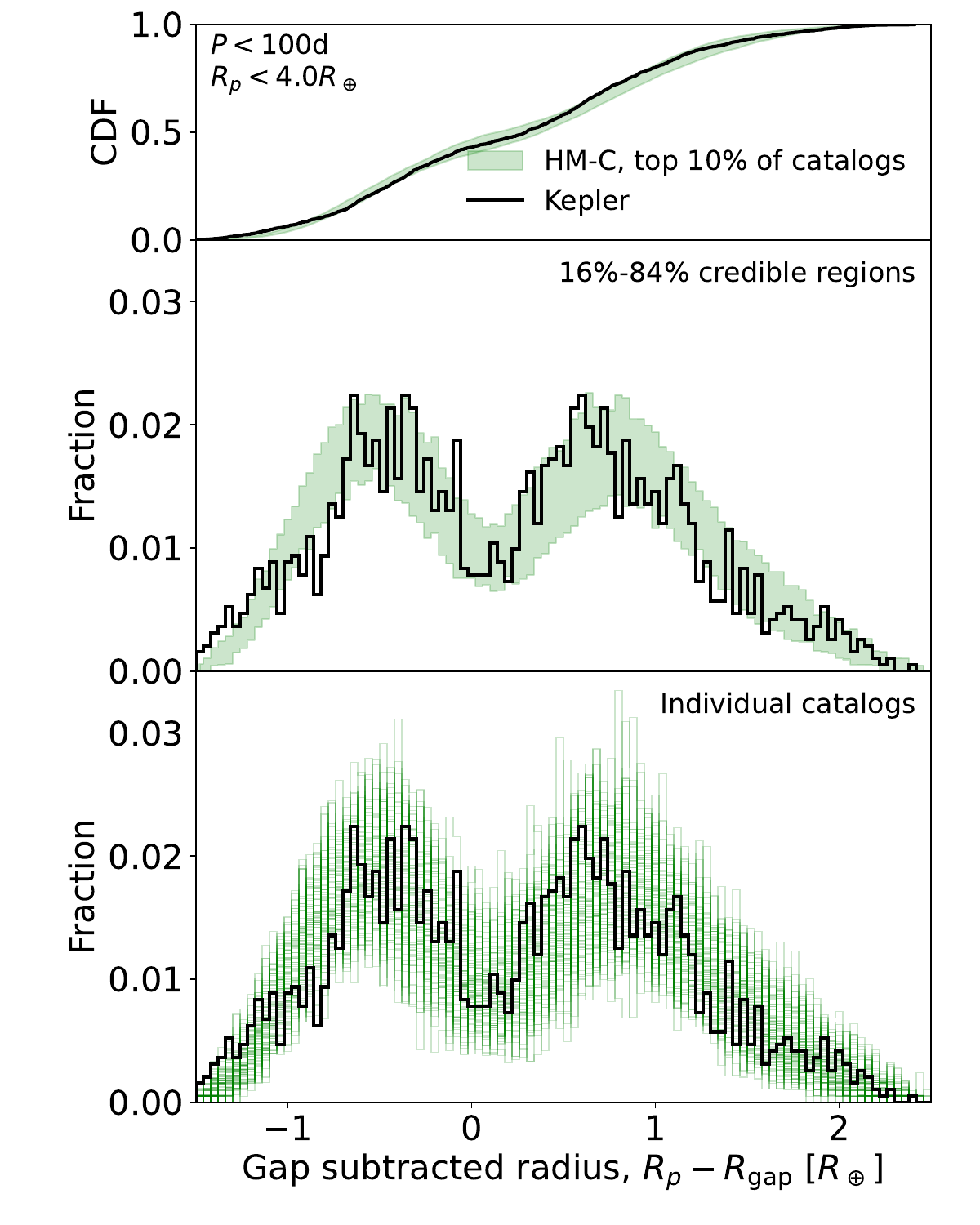} \\
\end{tabular}
    \caption{Same as Figure \ref{fig:observed_radii_models}, but for the top 10\% of the simulated catalogs in terms of their measured radius valley ``depths", from \HMC{}. For the planet radius distribution (\textbf{left panels}), this corresponds to catalogs with measured depths of $\Delta_{\rm valley} \gtrsim 0.26$; the \Kepler{} catalog exhibits $\Delta_{\rm valley} = 0.44$ (as depicted in Figure \ref{fig:observed_radii_depth_models}). For the gap-subtracted radius distribution (\textbf{right panels}; also restricted to planets with $P < 100$ days and $R_p < 4 R_\oplus$), the top 10\% corresponds to catalogs with measured depths of $\Delta_{\rm valley} \gtrsim 0.46$, compared to the \Kepler{} value of $\Delta_{\rm valley} = 0.58$.}
\label{fig:observed_radii_models_top_depths}
\end{figure*}

\subsubsection{How strong is the observed radius valley?} \label{sec:results:radius_valley:depth}

In the previous section, we showed that although the hybrid models are capable of generating a radius valley, such a feature is not consistently reproduced via random draws from the posterior distributions of the models.
This is not due to an inability of the model, but rather due to the difficulty in discerning varying ``strengths" of the radius valley given our distance function; we discuss this issue further in \S\ref{sec:discussion:distance_function}.

Here, we seek a better way to quantify whether a given model produces a feature that is \textit{like} the radius valley seen in the \Kepler{} data. We describe a simple method below for measuring the ``depth" of any valley-like feature that occurs close to the location seen in the \Kepler{} distribution. First, we perform a kernel density estimate (KDE) of the marginal radius distribution. This involves choosing a bandwidth ($h$) for the KDE, a free parameter that controls the extent of smoothing of the distribution. In order to capture a local feature like the observed valley, we must set a relatively small bandwidth so that the KDE does not smooth over the valley; we find that taking a quarter of the bandwidth value from Scott's rule \citep{scott1992multivariate}, $h = 0.25 h_{\rm SR} = 0.25 n^{-1/5}$ (where $n$ is the number of data points) is sufficient for our purposes. We find the local minimum of the KDE within in the range $[1.8, 2.2] R_\oplus$, (which bounds the location of the observed valley), as well as the maxima on either side of the bounded region (i.e. the largest peak below $1.8 R_\oplus$ and the largest peak above $2.2 R_\oplus$). Finally, we compute the ``depth" of the valley, $\Delta_{\rm valley}$, by taking the difference between the local minimum and the lesser height ($H$) of the two peaks, and dividing by the lesser peak height:
\begin{equation}
 \Delta_{\rm valley} \equiv \frac{H - \min\{f(R_p) : R_p \in [1.8,2.2] R_\oplus\}}{H}, \label{eq:depth_valley}
\end{equation}
where
\begin{equation}
 H = \min\Big\{ \max\limits_{R_p < 1.8 R_\oplus} f(R_p), \max\limits_{R_p > 2.2 R_\oplus} f(R_p) \Big\} \label{eq:min_peak}
\end{equation}
and $f(R_p)$ is the KDE function evaluated at $R_p$. This gives a normalized value of the depth ranging from 0 (no valley) to 1 (the valley is completely cleared out, i.e. the distribution drops to zero at some point within the valley).
We note that in Equation \ref{eq:min_peak}, one could have reasonably used the mean of the two peaks instead of the minimum. However, we choose to normalize by the smaller peak in order to avoid assigning especially large depths to models that produce highly asymmetrical peaks on either side of the valley.\footnote{For example, consider a distribution where the ``valley" is actually a flat plateau, and where the distribution drops on one side and increases to a very high peak on the other side. Our measure of $\Delta_{\rm valley}$ would be zero, whereas taking the mean of the peaks on either side of the valley would give a rather large value for the ``depth".} In this way, our measure of the valley depth is also relatively conservative.
We note that a similar approach was adopted by \citet{2023MNRAS.519.4056H}, who fit the planet radius distribution using a mixture of two Gaussians and measured the sub-Neptune and super-Earth peaks relative to the minimum of the valley.

In Figure \ref{fig:observed_radii_depth_models}, we show the application of our method to the \Kepler{} catalog (top-most panel) and several simulated catalogs from \HMU{} (panels below the top-most). For the \Kepler{} catalog, the KDE closely captures the overall shape of the distribution, including the valley at $\sim 2 R_\oplus$, and its depth (relative to the smaller of the two peaks, at $\sim 1.6 R_\oplus$) is measured to be $\Delta_{\rm valley} \simeq 0.44$.

We apply this method to each of the 100 simulated catalogs drawn from each hybrid model (the same catalogs used to compute the credible regions in the top panel of Figure \ref{fig:observed_radii_models}) to measure their observed radius valley depths. This method is also repeated for the gap-subtracted radius distribution, for which the bounds for the location of the valley are adjusted to $[-0.2, 0.2] R_\oplus$. In Figure \ref{fig:observed_radii_depths_cdfs}, we plot the cumulative distribution of $\Delta_{\rm valley}$ from each hybrid model. The distributions of $\Delta_{\rm valley}$ for the radius and gap-subtracted radius distributions are shown as solid and dashed lines, respectively. The measured values for the \Kepler{} catalog are denoted by the vertical lines.
For either model and radius metric, there is a wide range of depths. For the observed radius distributions (solid lines), the values range from $\Delta_{\rm valley} \simeq 0$ to 0.6; thus, some catalogs exhibit little to no radius valley, while others produce valleys as deep as (or even deeper than) that of the \Kepler{} data. As expected, the values of $\Delta_{\rm valley}$ are systematically shifted to somewhat larger values for the gap-subtracted radius distributions (dashed lines) compared to just the marginal radius distributions (solid lines), given that subtracting the location of the gap as a function of orbital period naturally aligns the period-dependence of the radius valley. The shift in the median $\Delta_{\rm valley}$ is consistent between both hybrid models and the \Kepler{} data (an increase in $\Delta_{\rm valley}$ of $\sim 0.14$).

While both hybrid models produce a wide range of $\Delta_{\rm valley}$, the catalogs drawn from each models' posterior distribution still rarely generate a radius valley that is as deep as that of the \Kepler{} data. For the radius distribution, the \Kepler{} value ($\Delta_{\rm valley} = 0.44$) is higher than 98\% (99.5\%) of the catalogs from \HMU{} (\HMC{}). The results are similar for the gap-subtracted radius distribution, where the \Kepler{} measured $\Delta_{\rm valley} = 0.58$ is higher than 95\% (97.9\%) of the catalogs from \HMU{} (\HMC{}). In other words, only the top few percent of catalogs drawn from the hybrid models exhibit an observed radius valley as substantial as that of the \Kepler{} catalog, based on our simple measure of ``depth".

To illustrate the strongest radius valleys generated from the hybrid models, we show five simulated catalogs with values of $\Delta_{\rm valley}$ from \HMU{} at least as large as that of the \Kepler{} data in Figure \ref{fig:observed_radii_depth_models} (panels below the top-most). Each of these catalogs exhibit a prominent radius valley. While some have a larger peak below versus above the valley (i.e. more super-Earths compared to sub-Neptunes) or vice versa, others produce peaks of comparable height, which are remarkably similar to the \Kepler{} data. Furthermore, repeated simulations of catalogs using the exact same model parameters as in these catalogs consistently produce an observed radius valley, demonstrating that the feature is robust given the appropriate parameters. 

We simulate and plot a larger number of catalogs with significant radius valleys in Figure \ref{fig:observed_radii_models_top_depths}, for \HMC{}. The results for \HMU{} are similar. This figure is the same as Figure \ref{fig:observed_radii_models}, except the simulated catalogs with the top 10\% largest $\Delta_{\rm valley}$ are shown instead of all simulated catalogs from the model posterior. For the radius distribution (left panel), this corresponds to simulated catalogs with $\Delta_{\rm valley} \gtrsim 0.26$; for the gap-subtracted radius distribution (right panel), $\Delta_{\rm valley} \gtrsim 0.44$. A total of 100 catalogs are shown in each panel. These catalogs provide a very good match to the \Kepler{} catalog in terms of reproducing the observed radius valley.

The existence of simulated catalogs (from both \HMU{} and \HMC{}) with strong radius valleys closely resembling that of the \Kepler{} data, combined with the wide distribution of $\Delta_{\rm valley}$, implies that the posterior distributions of the models are likely broader than the region of parameter space that can reliably produce a radius valley. Our measure of $\Delta_{\rm valley}$ does a good job of enabling us to rank and identify catalogs that exhibit a strong radius valley, and thus enables us to examine which parameters the radius valley is most sensitive to.

\begin{figure*}
\centering
\includegraphics[scale=0.45,trim={0.1cm 0.5cm 0.2cm 0.1cm},clip]{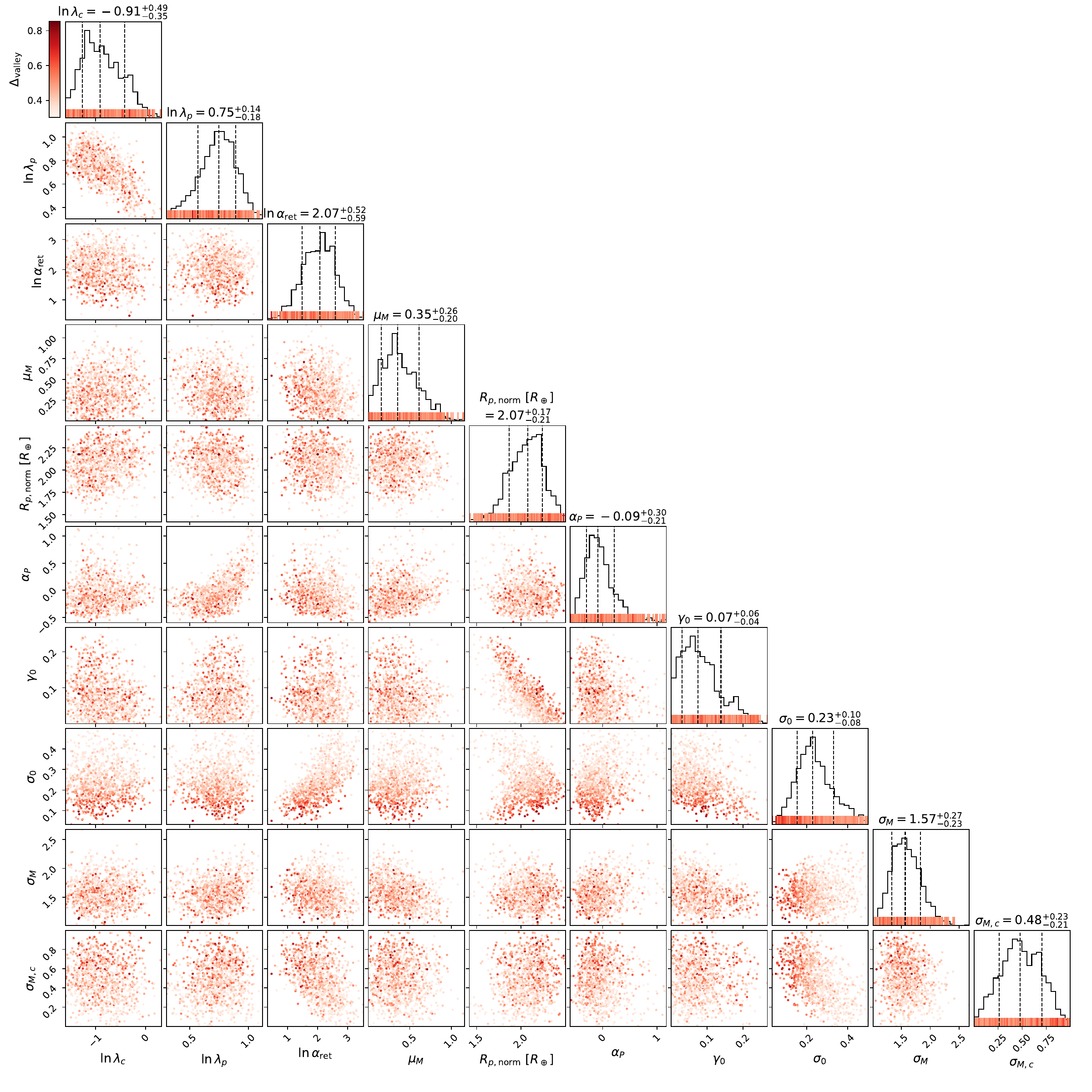} 
\caption{Distribution of the free parameters for simulated catalogs from \HMC{} that produce a strong observed radius valley. The catalogs are simulated with sets of parameters drawn from Run 1, and those passing two criteria are kept: (1) $\mathcal{D}_W \leq 20$, and (2) $\Delta_{\rm valley} \geq 0.29$ (measured from the distribution of gap-subtracted radii for planets with $P < 100$ days and $R_p < 4 R_\oplus$; this threshold corresponds to about half of the value as measured from the corresponding distribution of the \Kepler{} catalog, $\Delta_{\rm valley} = 0.58$). A total of 1000 simulated catalogs are collected and plotted. The color of each point represents the value of $\Delta_{\rm valley}$ (as indicated by the color bar); redder points denote stronger radius valleys. The median and central 68.3\% credible intervals for each parameter are labeled above each histogram, and are also listed in Table \ref{tab:param_fits} (column 6).}
\label{fig:hm2_posterior_large_valleys_params10}
\end{figure*}

\subsubsection{Which parameters lead to strong radius valleys?} \label{sec:results:radius_valley:depth_large}

To better understand which regions of parameter space produce significant valleys in the observed radius distribution, we simulate a much larger number of catalogs with sets of parameters drawn from the posterior distribution of \HMC{}. In addition to our distance criterion of $\mathcal{D}_W \leq 20$, we also require $\Delta_{\rm valley} \geq 0.29$ (as measured from the gap-subtracted radius distribution, which corresponds to roughly half of the value for the \Kepler{} catalog). We apply rejection-sampling to collect 1000 simulated catalogs passing both criteria and plot the distribution of their model parameters in Figure \ref{fig:hm2_posterior_large_valleys_params10}. Each scatter point represents a single catalog's set of model parameters, with the color denoting the value of $\Delta_{\rm valley}$ (as labeled by the color-bar to the left of the top-left panel; redder points indicate stronger radius valleys). The values above the histograms report the median and central 68.3\% credible intervals for each parameter computed from these 1000 catalogs, and are also listed in column (6) of Table \ref{tab:param_fits}.

Compared to the posterior distribution (i.e. Figure \ref{fig:hm2_posterior_params10} and column (5) of Table \ref{tab:param_fits}), there are several notable differences. The most significant differences appear to be a shift towards higher values of $R_{p,\rm norm}$ ($1.80_{-0.30}^{+0.29} R_\oplus$ to $2.07_{-0.21}^{+0.17} R_\oplus$) and lower values of $\sigma_0$ ($0.37_{-0.12}^{+0.08}$ to $0.23_{-0.08}^{+0.10}$). These can be understood to enhance the radius valley because the hybrid model simulates atmospheric mass-loss by moving planets from the initial radius-mass relation to the pure-silicate relation (e.g. from the green curve to the brown curve in Figure \ref{fig:radius_mass_credible}); thus, (1) a higher value of $R_{p,\rm norm}$ implies a greater reduction in the radius of a planet undergoing photoevaporation, and (2) a lower value of $\sigma_0$ implies less scatter around the initial radius-mass relation (the lower tail of which would fill in the region of the radius valley). The value of $\alpha_P$ is also slightly reduced ($0.17_{-0.35}^{+0.54}$ to $-0.09_{-0.21}^{+0.30}$), implying a somewhat slower rise in planet occurrence towards longer periods (thus, more planets are at shorter periods, making them more susceptible to the effects of photoevaporation). The normalization factor for the probability of envelope retention, $\ln{(\alpha_{\rm ret})}$ is also slightly decreased, as would be expected for promoting envelope mass-loss. Finally, $\mu_M$ and $\sigma_M$ are both shifted to moderately smaller values, concentrating the bulk of the simulated planets to the region around a few Earth masses which maps to the super-Earth to sub-Neptune regime. From Figure \ref{fig:hm2_posterior_large_valleys_params10}, it further appears that the reddest points are concentrated towards the lower end of the distribution for $\sigma_0$. This implies that if an even higher threshold for $\Delta_{\rm valley}$ is chosen, the distribution of $\sigma_0$ would shift to even smaller values, but the other parameters would remain largely unaffected.

\begin{figure*}
\centering
\includegraphics[scale=0.45,trim={0.8cm 0.2cm 0.1cm 0.1cm},clip]{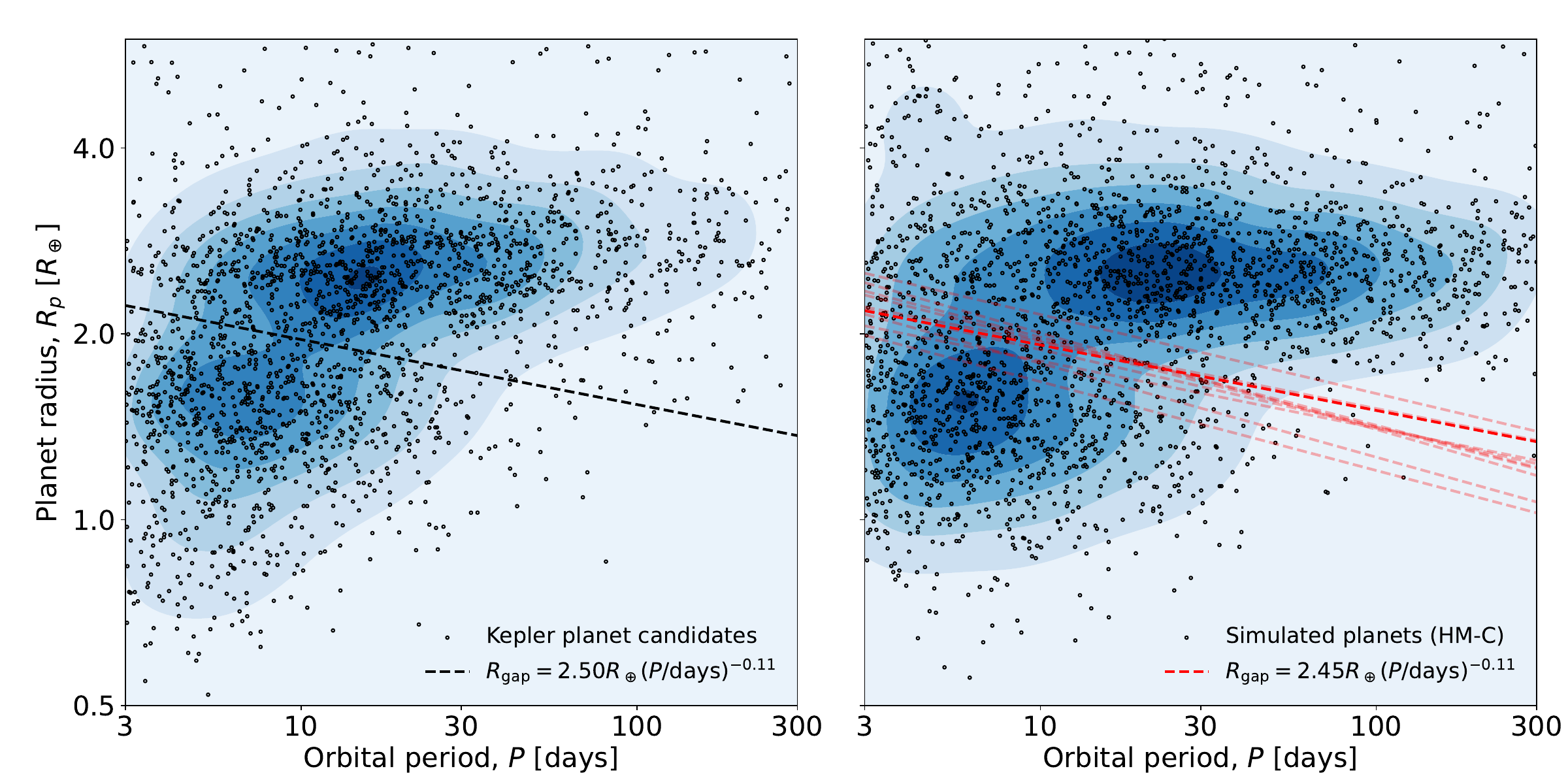} 
\caption{Observed distributions of planet radius vs. orbital period for the \Kepler{} catalog (left panel) and for a simulated catalog from \HMC{} (right panel). The simulated catalog is one of the top 10\% of catalogs sorted by $\Delta_{\rm valley}$, and is representative of the period-radius distribution for catalogs with a strong observed radius valley. Each dashed line denotes the best-fit location of the radius valley as a function of orbital period (Equation \ref{eq:radius_gap}) for a given catalog as determined using the \texttt{gapfit} package; in the right panel, the bright red dashed line represents the fit to the simulated catalog shown, while the dull red dashed lines denote fits to other simulated catalogs in the top 10\% of largest $\Delta_{\rm valley}$ from \HMC{}. The simulated catalog exhibits a bimodal period-radius distribution that highly resembles the \Kepler{} catalog and the location of the valley is also extremely similar.}
\label{fig:observed_period_radius}
\end{figure*}

\subsubsection{The observed radius valley as a function of orbital period} \label{sec:results:radius_valley:radius_period}

The observed radius valley is a function of stellar insolation/orbital separation. In particular, the location of the radius valley increases towards smaller orbital separations (i.e. shorter periods or higher stellar insolation), as larger planets can be stripped of their gaseous envelopes the closer they are to their host star. In Figure \ref{fig:observed_period_radius}, we plot the observed radius-period distribution for the \Kepler{} catalog (left panel) and a simulated catalog from \HMC{} (right panel). The simulated catalog is representative of (drawn from) the top 10\% of catalogs with the largest $\Delta_{\rm valley}$. In both panels, the scatter points denote individual planets while the shaded contours show a 2-D KDE fit. There is a qualitative agreement between the distributions from the simulated and the \Kepler{} catalogs. To compare the locations of the radius valley, we fit the gap radius as a power-law function of orbital period (i.e. Equation \ref{eq:radius_gap}) using the \texttt{gapfit} package \citep{2020ApJ...890...23L} to each catalog and plot the median of each fit as the dashed lines. In the right panel, the bright red dashed line represents the fit to the simulated catalog shown, while the dull red dashed lines denote fits to other catalogs in the top 10\% by $\Delta_{\rm valley}$ from \HMC{}. We find that the slopes and offset locations of the fits to the simulated catalog are generally consistent with the fit to the \Kepler{} catalog.

The location of the radius valley is well predicted by both photoevaporation theory (where the amount of XUV radiation from the star received by the planet increases as separation decreases; e.g. \citealt{2018MNRAS.479.4786V}) and the core-powered mass-loss mechanism (where the planet's primordial heat of formation also increases with decreasing separation; \citealt{2018MNRAS.476..759G, 2019MNRAS.487...24G, 2020MNRAS.493..792G}). In fact, both theories predict a similar slope for the radius valley in radius--period space that is consistent with the \Kepler{} observations \citep{2020ApJ...890...23L, 2021MNRAS.508.5886R, 2023MNRAS.519.4056H, 2023arXiv230200009B}. While it is beyond the scope of this work to explore the latter or to attempt to differentiate between these formation mechanisms, our results further demonstrate that the photoevaporation model can closely explain the observed radius valley when implemented in a detailed multi-planet population model. Additionally, our \HMC{} model demonstrates a key finding that we emphasize here: when combining photoevaporation with a model in which the initial planet masses are clustered, we obtain a joint mass--radius--period distribution that can simultaneously explain both the observed radius valley and the intra-system size similarity patterns.

\subsection{The underlying distribution of planetary systems} \label{sec:results:model_underlying}


\begin{figure*}
\centering
\includegraphics[scale=0.45,trim={0.2cm 0.2cm 0.2cm 0.1cm},clip]{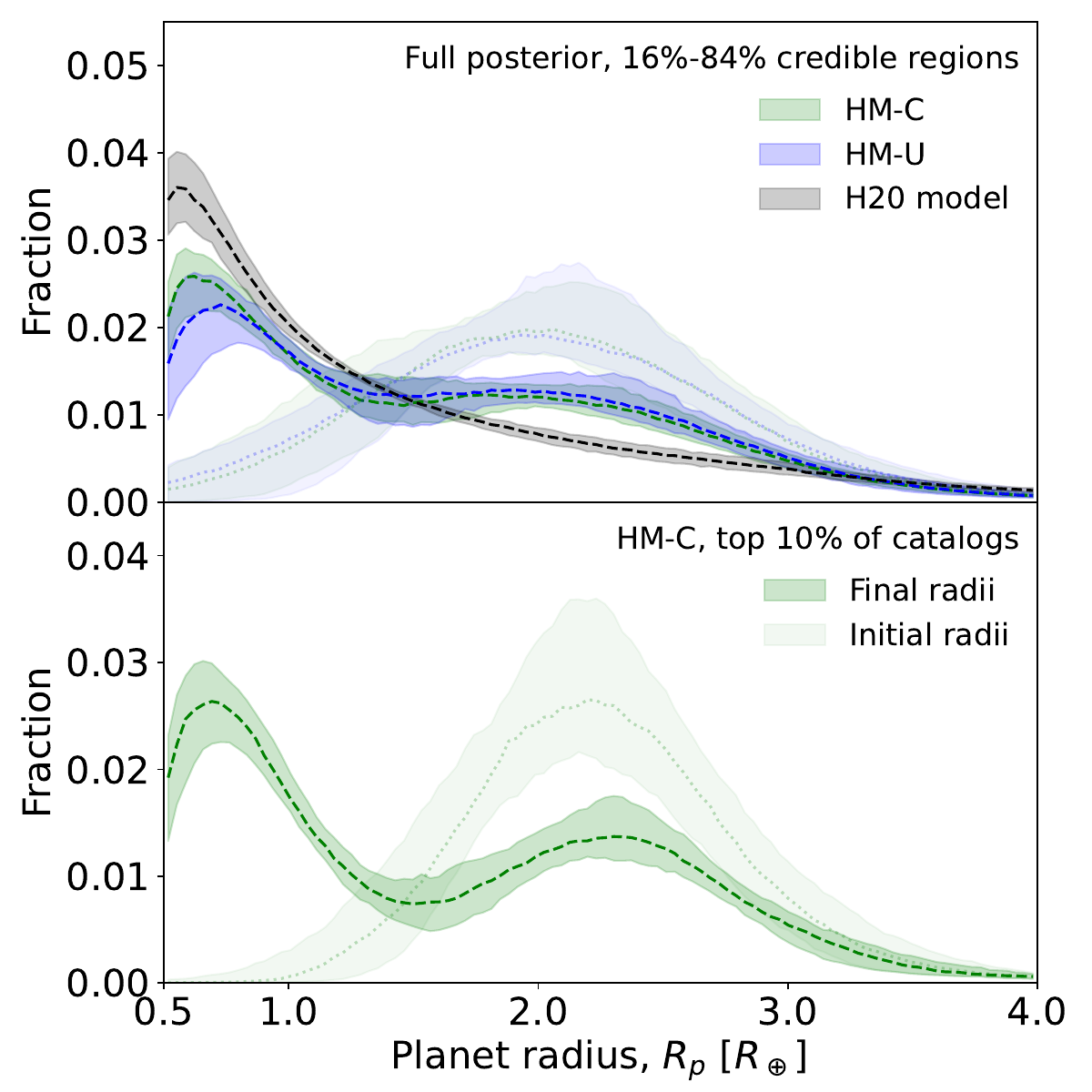} 
\includegraphics[scale=0.45,trim={0.2cm 0.2cm 0.2cm 0.1cm},clip]{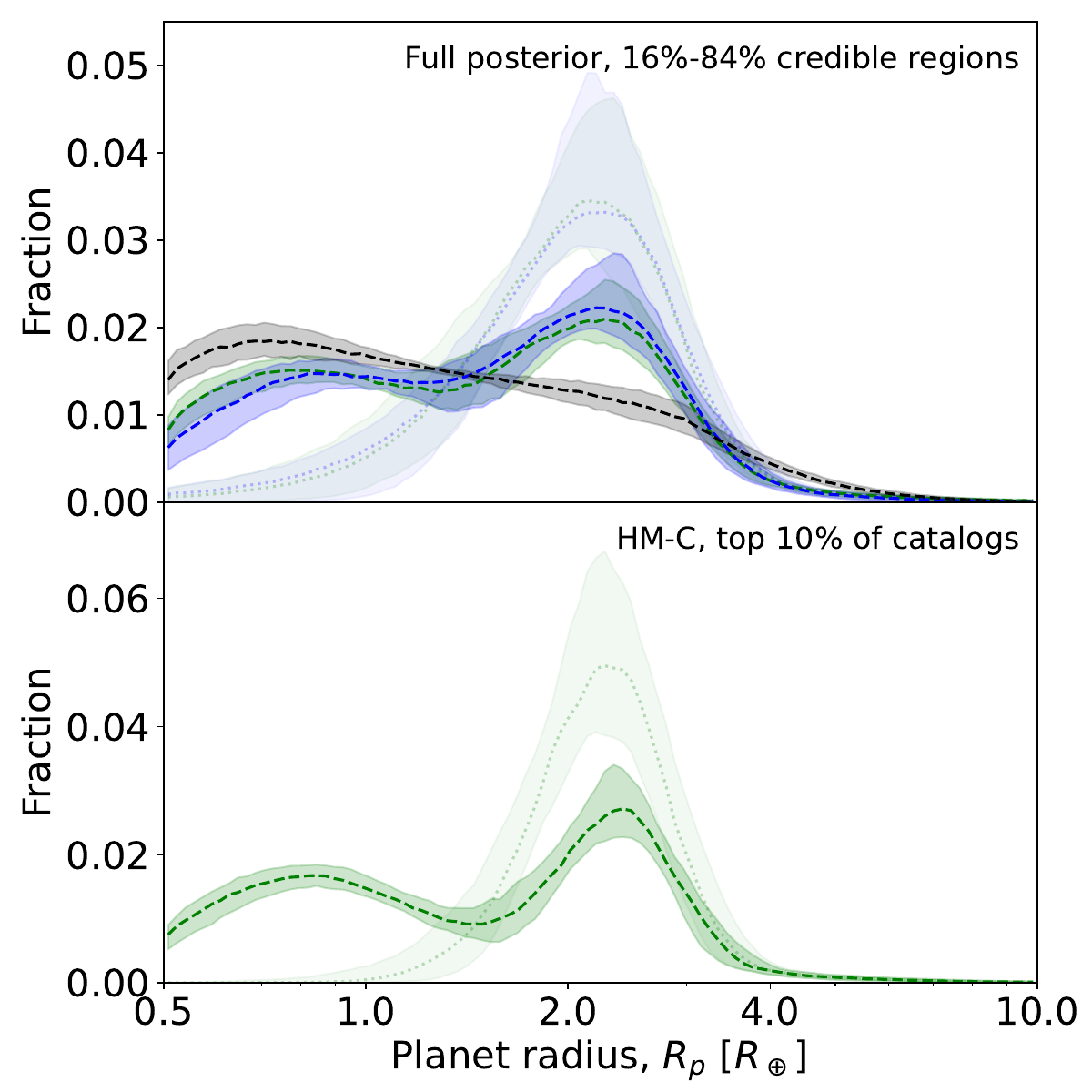} 
\caption{The underlying distributions of planet radii from each model. The left and right panels show the exact same simulated data; the left panel uses a linear x-axis (and is truncated to $R_p < 4 R_\oplus$ for clarity) while the right panel uses a log x-axis. In the top panels, the hybrid and H20 models are compared. The dashed lines and shaded regions denote the median and central 68.3\% credible regions, respectively, computed from 100 simulated catalogs drawn from the full posterior distribution of each model. For the hybrid models, the initial radius distributions (i.e. before envelope mass-loss) are also shown as the lighter shaded regions (the green and blue curves almost completely overlap). In the bottom panels, the initial and final radius distributions from the ``top 10\%" of catalogs from \HMC{}, in terms of generating the largest observed radius valley depths ($\Delta_{\rm valley}$; i.e. corresponding to the same catalogs shown in Figure \ref{fig:observed_radii_models_top_depths}), are shown.}
\label{fig:underlying_radii_models}
\end{figure*}

\begin{figure}
\centering
\includegraphics[scale=0.425,trim={0 0.2cm 0.2cm 0.1cm},clip]{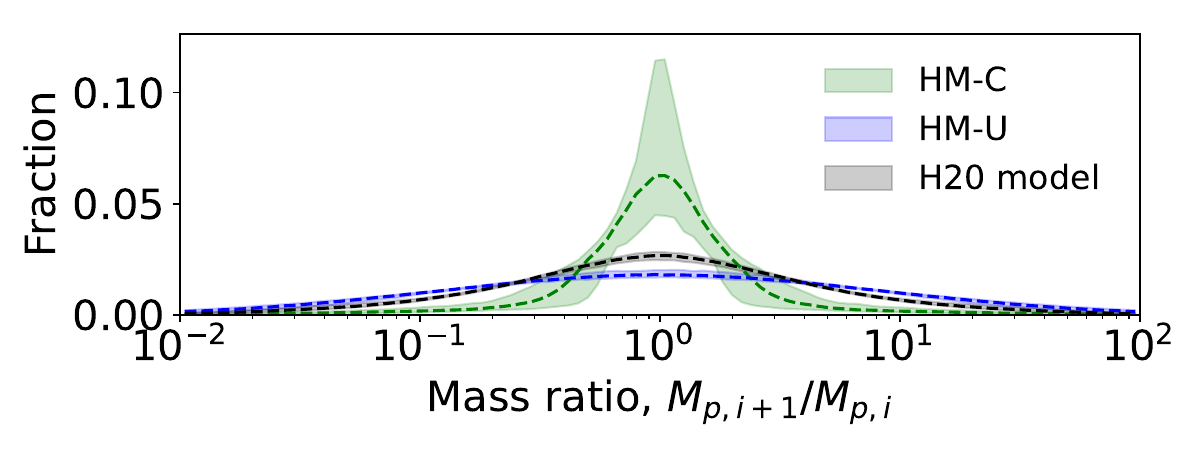} 
\includegraphics[scale=0.425,trim={0 0.2cm 0.2cm 0.1cm},clip]{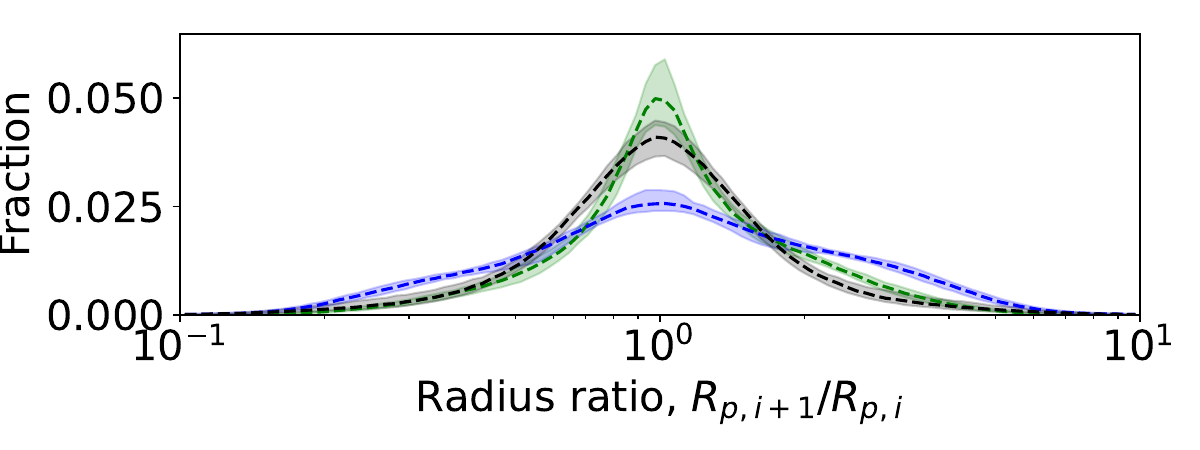} 
\caption{The underlying distributions of final planet mass ratios (top panel) and radius ratios (bottom panel), for adjacent planet pairs, from each model. \HMC{} (green) produces mass ratios that are significantly peaked around unity, due to the drawing of clustered initial masses which persists as a strong correlation in the final masses even after some planets undergo envelope mass-loss. This leads to an even stronger peak around unity in the final radius ratios compared to the H20 model (black), which draws clustered planet radii directly. On the other hand, \HMU{} (blue) produces the least correlated pairs of planet masses and radii. Both hybrid models also produce radius ratio distributions that are asymmetric, with more pairs of $R_{p,i+1}/R_{p,i} > 1$ than below, due to the effects of photoevaporation which is more likely to strip the atmosphere of the inner planet than the outer planet.}
\label{fig:underlying_mass_radii_ratios_models}
\end{figure}

\subsubsection{Uncovering the intrinsic radius distribution} \label{sec:results:model_underlying:radius}

How prominent is the radius valley in the underlying planetary population? The forward modeling of \SysSim{} allows us to directly address this question. In \S\ref{sec:results:radius_valley:radius_period}, we showed that our hybrid model can produce a bimodal distribution of observed planet sizes that is similar to what is seen in the \Kepler{} data, given an appropriate set of model parameters. Here, we use the accompanying \textit{physical catalogs} drawn from the hybrid model to examine the intrinsic distribution of planet radii.

In Figure \ref{fig:underlying_radii_models}, we plot the underlying, marginal distribution of planet radii from each model. The darker shaded curves are the final, ``present day" planet radius distributions (for the hybrid models, these are the distributions after sculpting from photoevaporation; for the H20 model, the radius distribution is directly modeled with a broken power-law). For each model, the dashed line and shaded regions denote the median and central 68.3\% credible regions, respectively, computed from the 100 simulated catalogs drawn from the posterior distribution. The left and right panels show the same simulated data, with a linear and a log x-axis, respectively. The top panels compare the distributions of catalogs drawn from the full posterior of each model. The hybrid models (blue and green) result in similar distributions, as expected. There is a hint of a bimodal distribution, though it is not very pronounced in the full posterior draws. However, the bottom panels show the top 10\% of catalogs from \HMC{}, in terms of generating the largest radius valley depths in the \textit{observed} distributions ($\Delta_{\rm valley}$; i.e. corresponding to the same catalogs shown in Figure \ref{fig:observed_radii_models_top_depths}). Thus, for catalogs that produce a strong observed radius valley, the valley is also seen in the underlying distribution, but appears at a smaller radius than in the observed distribution ($\sim 1.5 R_\oplus$ compared to $\sim 2 R_\oplus$). The peak above $\sim 2 R_\oplus$ is modest in linear $R_p$ (left panel), but is greatly emphasized when plotted on a log scale (right panel), along with the intrinsic ``radius cliff" marked by the rapid drop-off between $\sim 2.5 R_\oplus$ and $\gtrsim 4 R_\oplus$.
Compared to the H20 model, both hybrid models generate roughly twice as many planets with $R_p \sim 2.2 R_\oplus$. On the other hand, the H20 model includes more planets towards the smallest radii ($R_p < 1 R_\oplus$), and considerably more in the range $R_p \sim 4$-$5 R_\oplus$ due to the shallower drop-off.

We also plot the underlying distributions of initial planet radii (i.e. before photoevaporation) for the hybrid models, as the lighter shaded curves in Figure \ref{fig:underlying_radii_models}. The two hybrid models are very similar and exhibit a unimodal distribution that peaks at $\sim 2.1 R_\oplus$. Thus, the majority of the present-day small planets ($R_p \lesssim 1.5 R_\oplus$, especially the peak below $\sim 1 R_\oplus$) in the hybrid models are photo-evaporated cores.


\subsubsection{The intrinsic degree of planet mass similarity and radius similarity}
\label{sec:results:models_underlying:size_similarity}

In \S\ref{sec:results:new_model_fit:observed_size_clustering}, we showed that the hybrid model with clustered initial masses (\HMC{}) provides a significantly better fit to the \Kepler{} data than the model with un-clustered initial masses (\HMU{}), when considering the summary statistics that capture the radius similarity and ordering patterns in the observed multi-planet systems. Here, we examine the degree of similarity in the underlying distributions of planet masses and radii from the hybrid models. In other words, we can directly answer the question: to what extent are the intrinsic planet masses and radii clustered in \HMC{}, such that it reproduces the size similarity patterns of the \Kepler{} data which cannot be adequately reproduced by \HMU{}?

We consider two simple measures: the planet mass ratios and planet radius ratios, for physically adjacent planet pairs, and plot their distributions for each model in Figure \ref{fig:underlying_mass_radii_ratios_models} (top and bottom panels, respectively). Focusing first on the mass ratio distribution (top panel), the distribution from \HMC{} (green) is significantly more peaked around $M_{p,i+1}/M_{p,i} \simeq 1$ than either the \HMU{} (blue) or H20 model (black), indicating that adjacent planets often have significantly more similar masses in \HMC{}. Although this is to be expected from the drawing of clustered \textit{initial} masses (the sole difference between \HMC{} and \HMU{}; also see the typical cluster width in \HMC{} compared to the overall distribution in Figure \ref{fig:radius_mass_credible}), we note that the distributions plotted here are of the \textit{final} planet masses, i.e. after some planets have lost their envelopes due to photoevaporation. Nevertheless, the clustering in initial planet masses clearly leaves a strong imprint in the correlation of the final planet masses for planets in the same cluster. Interestingly, the mass ratio distribution for \HMU{} appears even slightly broader (suggesting that the planet masses are even less correlated) than for the H20 model. 

Finally, we focus on the underlying distribution of planet radius ratios (bottom panel of Figure \ref{fig:underlying_mass_radii_ratios_models}). This distribution relates more directly to the metrics that evidenced the size similarity and ordering patterns (i.e. Figure \ref{fig:observed_marginals_radii_metrics_compare}), since planet radius, not mass, is the observable property for all of the \Kepler{} planet candidates. We see that \HMC{} also produces the highest degree of similarity between adjacent planet radii; the peak around $R_{p,i+1}/R_{p,i} \simeq 1$ is the strongest from this model, even compared to the H20 model. We remind the reader that the H20 model directly draws clustered planet radii (see \S\ref{sec:methods:prev_models:mass_radius}, equations \ref{eq:radii_clustered}-\ref{eq:radii_scales}), whereas the intra-cluster similarity in final planet radii from \HMC{} results from the clustered planet masses described previously.
For the hybrid models, these distributions have been further sculpted by photoevaporation, producing an asymmetry with more planet pairs $R_{p,i+1}/R_{p,i} > 1$ than $R_{p,i+1}/R_{p,i} < 1$. This is easily understood because photoevaporation is a function of orbital period, and thus the inner planet is more likely to lose its atmosphere than the outer planet (resulting in $R_{p,i} < R_{p,i+1}$, provided they have similar masses). The asymmetry is subtle for \HMC{} (green), but is evident via comparison to the H20 model (black, which is perfectly symmetric in $\log{(R_{p,i+1}/R_{p,i})}$). Given that the \HMC{} model produces near perfect fits to the \Kepler{}-observed distributions as discussed in \S\ref{sec:results:new_model_fit:observed_size_clustering}, we consider the underlying distributions from \HMC{} to be the current best model for capturing the intrinsic intra-system planet-size similarity patterns.

\begin{figure*}
\centering
\begin{tabular}{cc}
 \includegraphics[scale=0.425,trim={0 0.5cm 0 0.2cm},clip]{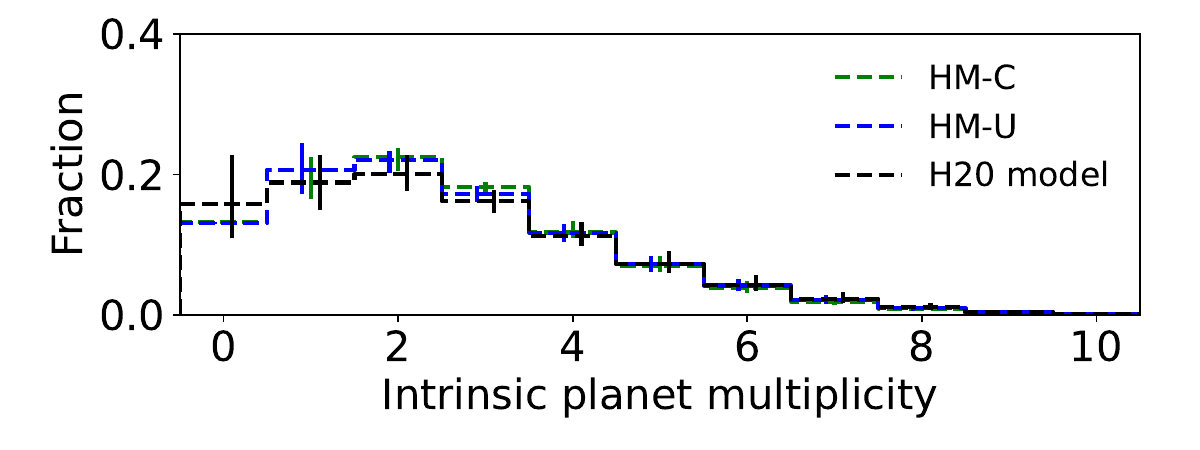} &
 \includegraphics[scale=0.425,trim={0 0.5cm 0 0.2cm},clip]{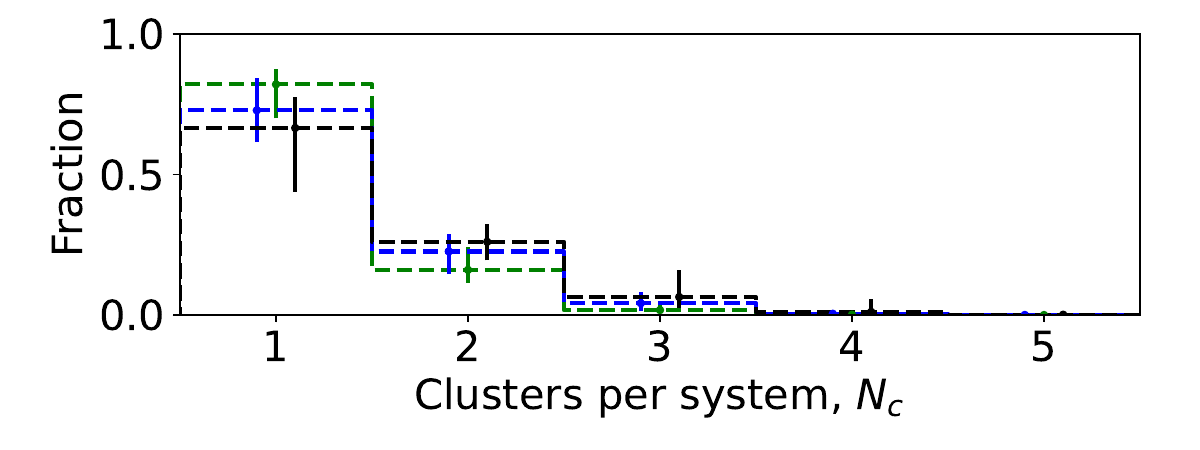} \\
 \includegraphics[scale=0.425,trim={0 0.5cm 0 0.2cm},clip]{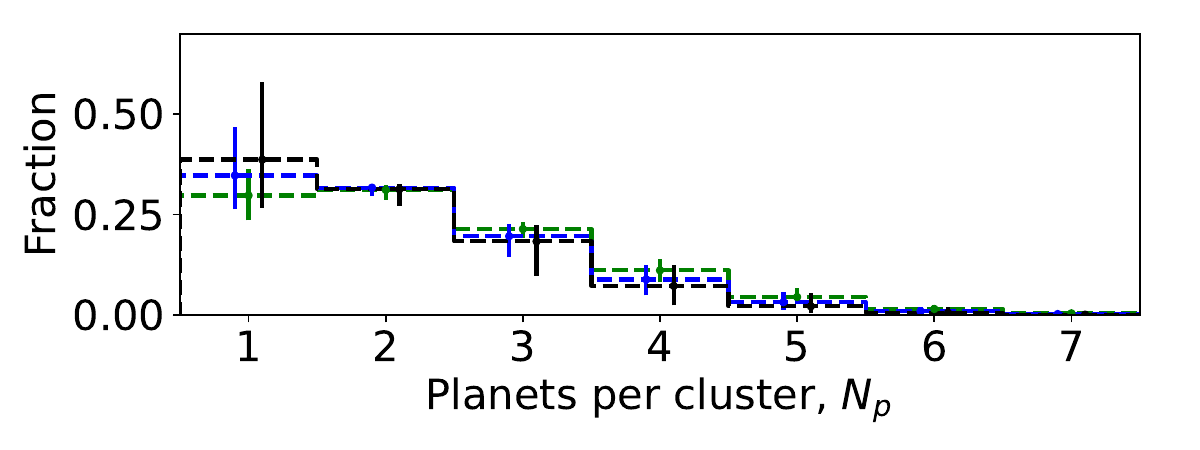} &
 \includegraphics[scale=0.425,trim={0 0.5cm 0 0.2cm},clip]{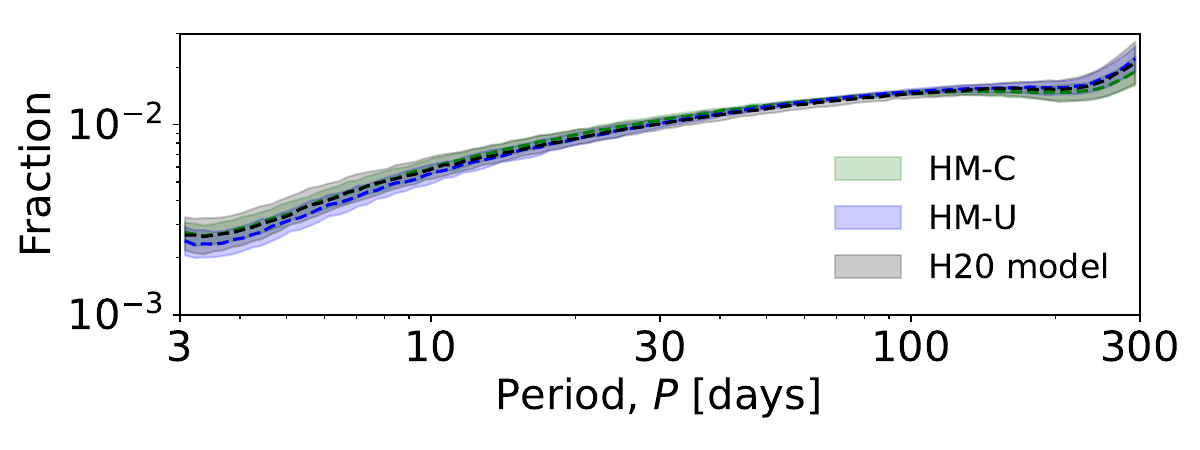} \\
 \includegraphics[scale=0.425,trim={0 0.5cm 0 0.2cm},clip]{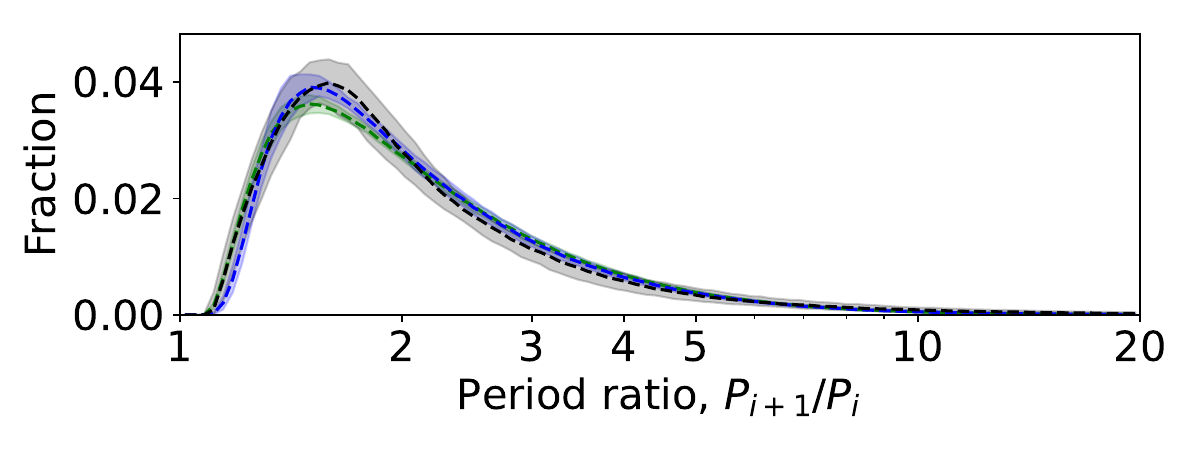} &
 \includegraphics[scale=0.425,trim={0 0.5cm 0 0.2cm},clip]{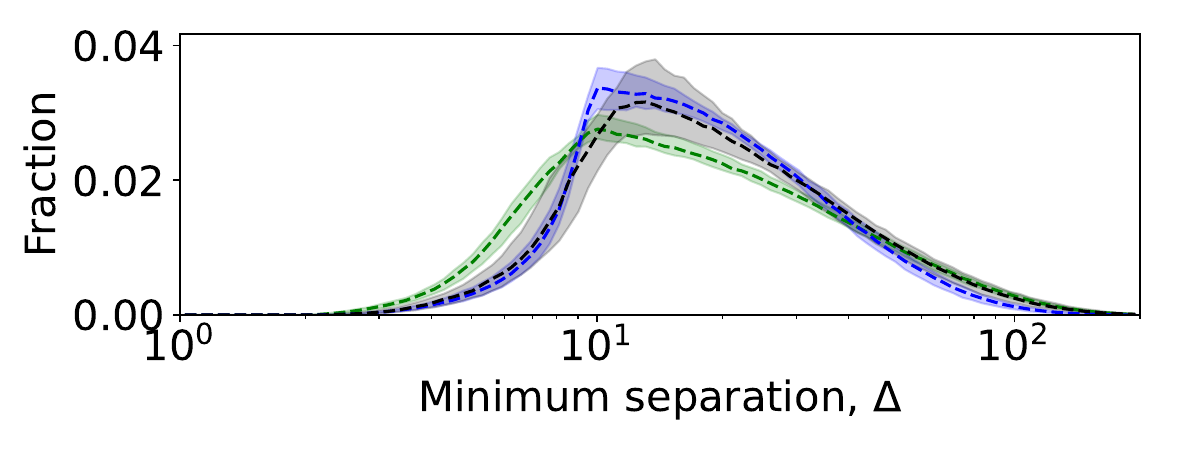} \\
 \includegraphics[scale=0.425,trim={0 0.5cm 0 0.2cm},clip]{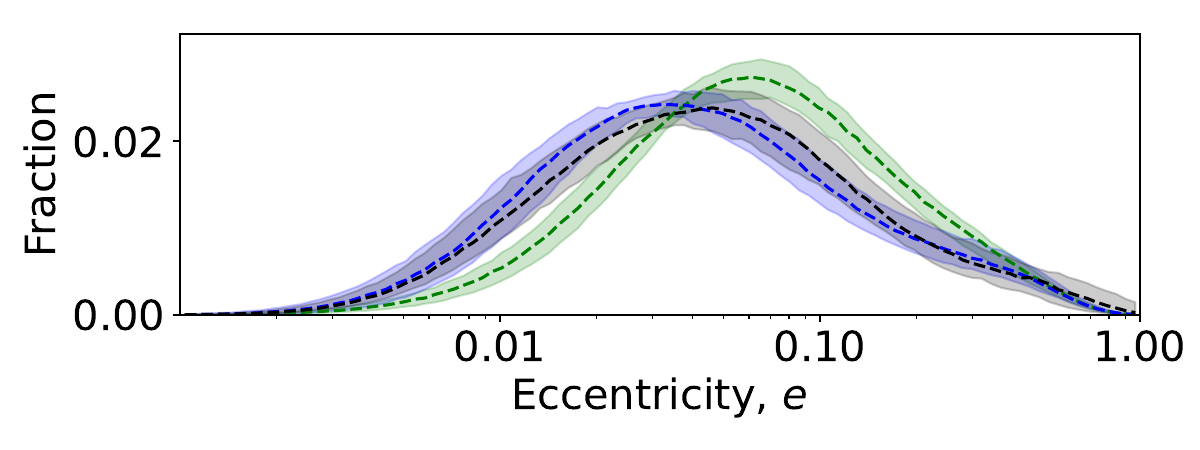} & 
 \includegraphics[scale=0.425,trim={0 0.5cm 0 0.2cm},clip]{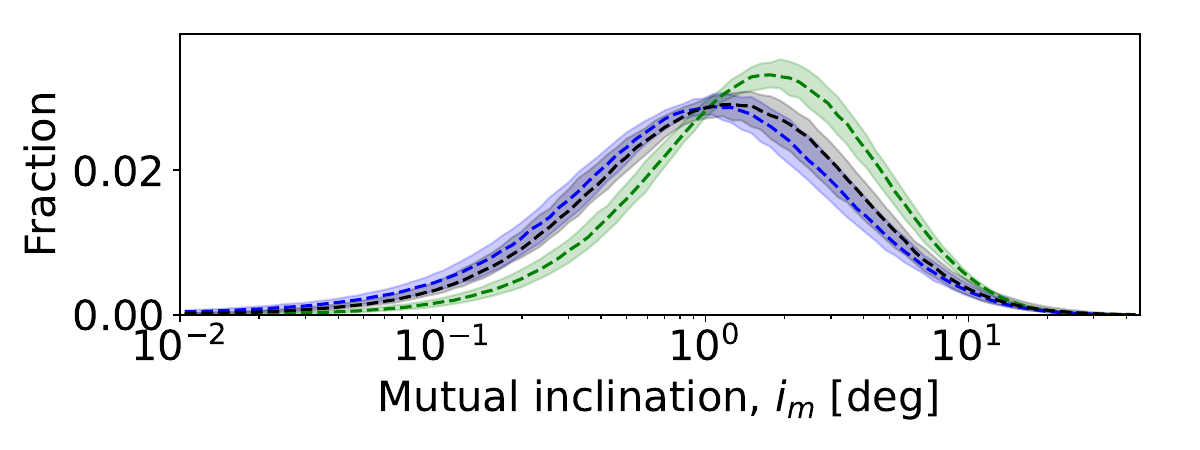} \\
\end{tabular}
\caption{The underlying distributions of other properties of the physical planetary systems from the hybrid models (blue and green) and the H20 model (black). As in Figures \ref{fig:underlying_radii_models} and \ref{fig:underlying_mass_radii_ratios_models}, the dashed lines and shaded regions in each panel denote the median values and the central 68.3\% credible regions, respectively, drawn from each of the model posterior distributions (for the cluster and planet multiplicity panels, these are the solid lines and error bars).
\textit{Planet and cluster multiplicities:} The distributions of the number of clusters per system ($N_c$) and planets per cluster ($N_p$), and thus the intrinsic planet multiplicity, are very similar between all three models.
\textit{Periods, period ratios, and minimum separations:} The distributions of periods and period ratios are also quite similar among the models, though the latter is partly by construction due to fixing $\sigma_P = 0.25$ in the hybrid models. There are more planet pairs with minimum separations $\Delta < 10$ (in units of mutual Hill radii) in \HMC{} (green) due to the moderately higher eccentricities.
\textit{Eccentricities and mutual inclinations:} While all three models involve distributing the critical AMD of each system to draw these orbital elements, and they lead to similar overall shapes of these distributions, they are shifted to higher $e$ and $i_m$ for \HMC{}, due to differences in the planet mass distributions within each system.}
\label{fig:underlying_properties_models}
\end{figure*}

\subsubsection{Comparison of other architectural properties between the hybrid and H20 models} \label{sec:results:model_underlying:model_comparison}

The primary differences between the hybrid and H20 models are in the planet radius and mass distributions, as discussed in the previous sections. The models share several other architectural properties. While the underlying distributions of the remaining properties are similar between all three models, there are some minor differences which we discuss here.
In Figure \ref{fig:underlying_properties_models}, we plot the underlying marginal distributions of these planetary system properties drawn from the hybrid and H20 models.

\textit{Planet and cluster multiplicities:} The three models share the same parameterization of the underlying planet multiplicity distribution (see \S\ref{sec:methods:prev_models:clusters}). We find that the distributions of the number of clusters per system and planets per cluster, and thus the intrinsic planet multiplicity, are also very similar between the models. While the underlying multiplicity distribution could have been different given the very different parameterizations of the joint planet radius-mass distribution, we find that the constraints on the multiplicity parameters $\lambda_c$ and $\lambda_p$ are generally in agreement, as also discussed in \S\ref{sec:results:new_model_fit:hm1_params}. The most common planetary system (within our simulation bounds) consists of two planets, and $0.61_{-0.06}^{+0.03}$ of all systems have between one to three planets in \HMC{} (the values are similar for the \HMU{} and H20 models). The majority of planet-hosting systems have just a single planet-cluster ($0.82_{-0.12}^{+0.06}$ in \HMC{}) while a sizable fraction have two clusters ($0.16_{-0.05}^{+0.08}$).

\textit{Periods, period ratios, and minimum separations:} The distributions of periods and period ratios are also very similar between all three models. We note that the hybrid model catalogs shown are from the runs in which the period cluster scale were fixed to $\sigma_P = 0.25$ (``Run 1" of each model), and thus have narrower credible regions around the median than the H20 model, for which the parameter was allowed to vary. For the separations between adjacent planets in mutual Hill radii ($\Delta$; equation \ref{eq:min_separation}), we also fixed the minimum allowed value to $\Delta_c = 10$ for the hybrid models, which is well within the range of best-fit values found in the H20 model in \PaperIII{} when $\Delta_c$ was allowed to vary. In both cases, the minimum separation is asserted before the eccentricities are drawn, which produces the tail towards smaller values. The tail is more pronounced in \HMC{} because the eccentricities are moderately higher in this model due to the differences in the planet masses, as further described below.

\textit{Eccentricities and mutual inclinations:} These distributions are effectively set by the dynamical limits imposed from distributing the critical AMD of each multi-planet system (see \S\ref{sec:methods:prev_models:amd_stability}, and \PaperIII{} for more details). For intrinsic single-planet systems, the eccentricity distribution is a Rayleigh distribution, where the Rayleigh scale parameter is fixed to $\sigma_{e,1} = 0.25$ for the hybrid models (guided by the best-fit values found in the H20 model). However, the AMD stability is also dependent on the planet masses (specifically, the distribution of planet masses \textit{within each system}), which are notably different between the three models. This explains the deviations of the \HMC{} model from the \HMU{} and H20 models: the \HMC{} model tends to produce systematically higher orbital eccentricities and mutual inclinations in multi-planet systems than the other two models, because the planet masses are most strongly clustered in this model (recall Figure \ref{fig:underlying_mass_radii_ratios_models}, top panel). In other words, since the critical AMD of a system is limited by the most compact dynamical pair of planets (a fundamental assumption of the AMD stability criteria is that the AMD can be freely exchanged between the planets), a system with planets of similar masses can have somewhat higher eccentricities and mutual inclinations relative to a comparable system in which the masses are more disparate (since the least massive planet would likely be smaller and thus more excitable). From Figure \ref{fig:underlying_properties_models} (bottom panels), we find that the distributions for \HMC{} peaks at approximately a factor of two higher than the \HMU{} or H20 models; the median $e \simeq 0.07$ for \HMC{} versus $e \simeq 0.03-0.04$ for \HMU{} (or H20); likewise, the median $i_m \simeq 2^\circ$ for \HMC{} compared to $i_m \simeq 1^\circ$ for \HMU{}/H20.

\subsubsection{The initial core mass distribution} \label{sec:results:model_underlying:core_masses}

We briefly remark on the initial core mass distribution. While the hybrid models are not specifically designed to infer this distribution, they do provide an implicit prediction for the underlying core mass distribution. As described in \S\ref{sec:methods:new_model}, the hybrid models are defined by a lognormal distribution of initial planet masses (recall the top panel of Figure \ref{fig:radius_mass_credible}) and the envelope mass of each planet is a fraction of the total mass. 
The remaining fraction of the planet mass is therefore assumed to be in the core. For planets with $M_{p,\rm init} \lesssim 10 M_\oplus$ (the bulk of the distribution), the envelope mass fraction is assumed to be 10\% (Equation \ref{eq:envelope_mass_low}), and thus 90\% of the planet mass is in the core, producing a broad distribution of core masses also centered close to the peak of the initial mass distribution ($\sim 1$-$2 M_\oplus$; note that this is comparable to the peak of the core mass distribution found by \citealt{2021MNRAS.503.1526R} of $\sim 4 M_\oplus$). However, the envelope mass fraction rises steeply for more massive planets (as expected for runaway gas accretion; \citealt{1996Icar..124...62P, 2016A&A...596A..90V}), thus stunting the growth of the core and producing a pileup of core masses between $\sim 7$-$12 M_\oplus$. We find that this leads to an additional $+9_{-3}^{+3}\%$ ($+8_{-3}^{+4}\%$) for the fraction of planets with core masses in this range, in \HMU{} (\HMC{}), corresponding to an enhancement by roughly a factor of two in this range.

\section{Discussion} \label{sec:discussion}


\begin{figure*}
\centering
\includegraphics[scale=0.46,trim={0.5cm 0.1cm 0.3cm 0.1cm},clip]{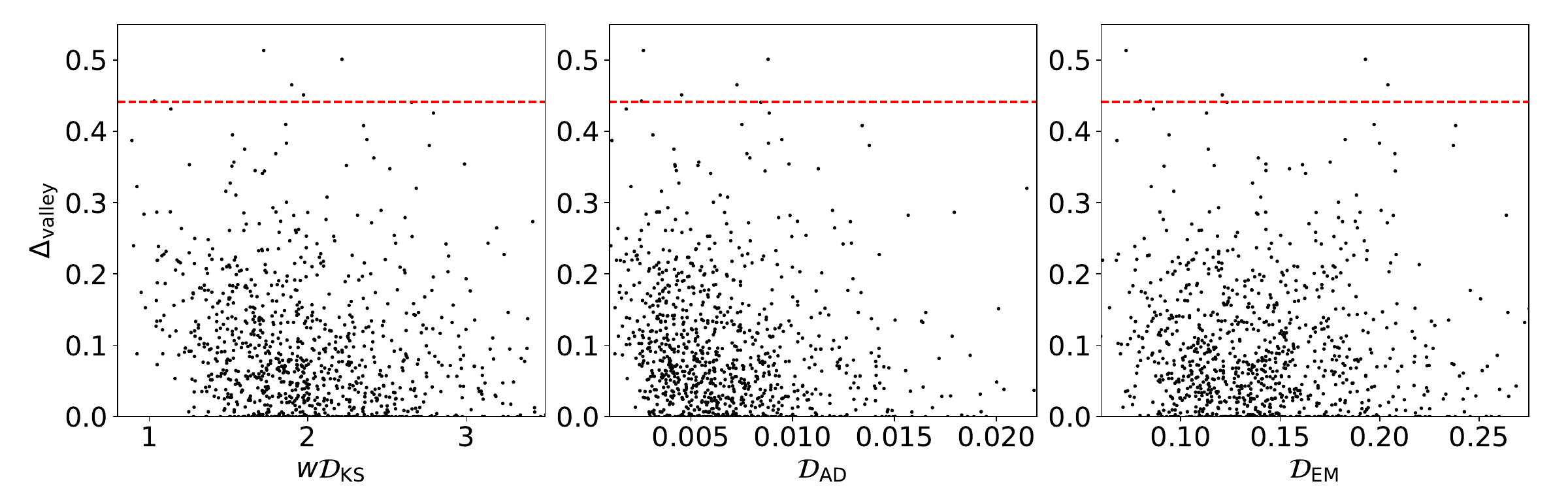} 
\caption{Measured radius valley depths ($\Delta_{\rm valley}$) vs. various distance metrics for the observed planet radius distribution, comparing simulated catalogs from \HMC{} to the \Kepler{} catalog. From left to right, the panels show the same values of $\Delta_{\rm valley}$ plotted against (1) the KS distance, (2) the AD distance, and (3) the EM distance (``Earth mover's" distance, also known as the Wasserstein distance or metric). The KS distances are multiplied by a constant weight $w$ (Equation \ref{eq:weights}), such that a perfect model subject to stochastic noise would give a distance of $w\mathcal{D}_{\rm KS} \simeq 1$; in any case, smaller values of the distance metrics indicate better fits to the \Kepler{}-observed distribution of planet radii. In each panel, each scatter point denotes a single simulated catalog drawn from the posterior distribution of \HMC{} (the results for \HMU{}, not shown, are similar). The horizontal red dashed line denotes $\Delta_{\rm valley}$ for the \Kepler{} catalog. Simulated catalogs with strong radius valleys (i.e. larger values of $\Delta_{\rm valley}$, closer to the \Kepler{} value) do not appear to give smaller distances for any of these metrics.}
\label{fig:radius_valley_depths_vs_distances}
\end{figure*}

\subsection{The choice of distance function}
\label{sec:discussion:distance_function}

In \S\ref{sec:results:radius_valley}, we showed that although the hybrid models are capable of generating a radius valley, this feature is not consistently reproduced via random draws from the posterior distributions of the models.
Here, we will show that this is not due to an inability of the model, but rather due to the reduced sensitivity to measuring the ``strength" of the radius valley given our distance function. The distance function we adopted (Equation \ref{eq:dist}) includes a term that is the KS distance for the marginal distribution of planet radii between the \Kepler{}- and simulated-observed catalogs. Unfortunately, we find that this metric does not adequately differentiate between models that exhibit a strong radius valley versus those that do not. In other words, both models with and without any observed valley in the marginal radius distribution can give similar KS distances when compared to the \Kepler{} distribution. In Figure \ref{fig:radius_valley_depths_vs_distances}, we plot the measured radius valley depths ($\Delta_{\rm valley}$) versus the KS distances for simulated catalogs from the posterior of \HMC{} (left panel; the results for \HMU{} are similar and are not shown). The KS distances are multiplied by a constant weight $w$ (Equation \ref{eq:weights}) such that a perfect model with stochastic noise would give $w\mathcal{D}_{\rm KS} \simeq 1$, as included in our weighted distance function. We find that catalogs with large radius valleys (i.e. high $\Delta_{\rm valley}$), close to the \Kepler{} value (denoted by the horizontal red dashed line), do not necessarily improve the distance.
We also experimented with other distance measures, including the Anderson-Darling distance (AD; \citealt{ad1952}) and the Wasserstein distance (also known as the ``Earth mover's distance" or EMD; \citealt{kantorovich1942translocation, rubner2000earth}), and find that they are also uncorrelated with the presence of a radius valley (middle and right panels of Figure \ref{fig:radius_valley_depths_vs_distances}, respectively). Thus, we conclude that these distance measures are insensitive to the observed radius valley.

In theory, we could have also included a distance metric that directly relates to our measure of $\Delta_{\rm valley}$ in our distance function. Future work is needed to explore an appropriate functional form, or how such a term should be weighted in combination with all of the other distance terms. The ranking of the simulated catalogs by $\Delta_{\rm valley}$ and selecting those above a threshold (e.g. for the top 10\%, as done in \S\ref{sec:results:radius_valley:depth}) can be thought of as imposing a sharp penalty to the distance function post-inference, and already provides some insight into which parameters produce the strongest radius valleys (\S\ref{sec:results:radius_valley:depth_large}).

\begin{figure}
\centering
\includegraphics[scale=0.44,trim={0.3cm 0.1cm 0.3cm 0.1cm},clip]{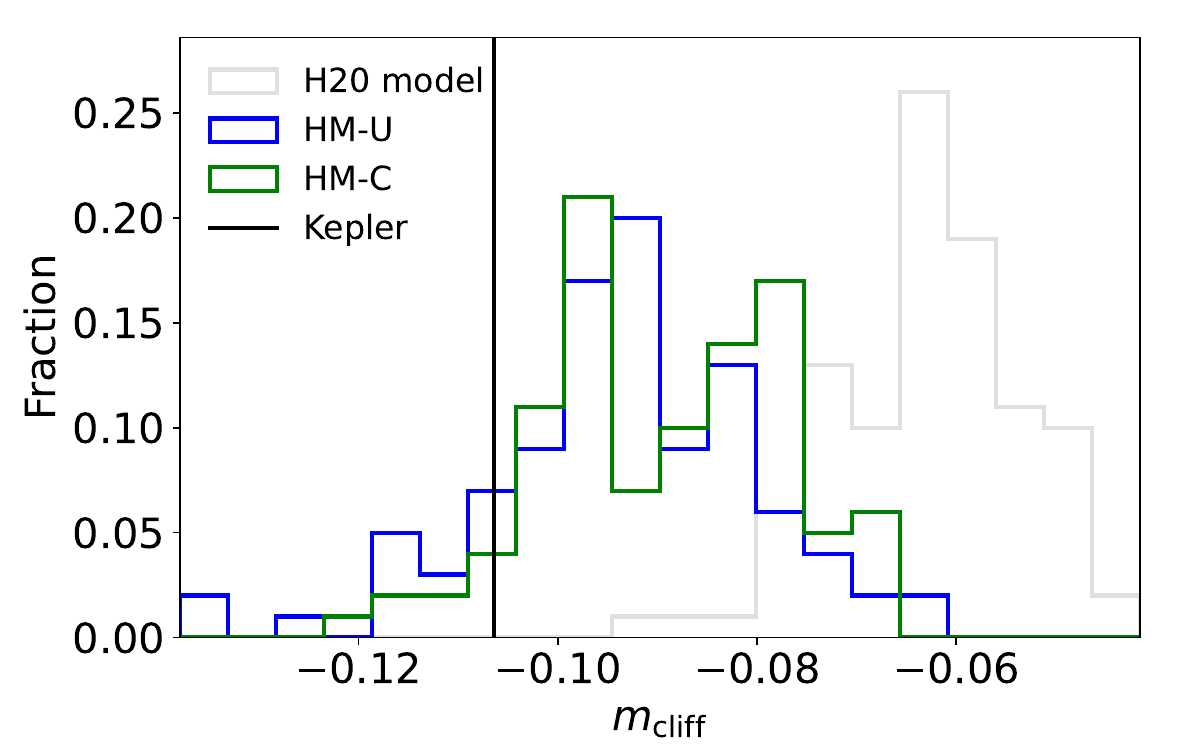} 
\caption{Distributions of observed radius cliff slopes, $m_{\rm cliff}$, as measured from fitting a line to the distribution of (log) planet radii between 2.5--$5.5 R_\oplus$ (Equation \ref{eq:radius_cliff}). The value for the \Kepler{} catalog is given by the vertical black line ($m_{\rm cliff} = -0.106$). The H20 model (gray histogram) produces slopes that are too shallow ($m_{\rm cliff} = -0.06 \pm 0.01$), indicating that the observed radius distribution does not drop off as fast as in the \Kepler{} data (as can be seen in the middle-left panel of Figure \ref{fig:observed_radii_models}). The hybrid models (blue and green histograms) produce very similar distributions of slopes, and are both consistent with the \Kepler{} value.}
\label{fig:observed_radius_cliff_slopes}
\end{figure}

\subsection{The planet radius ``cliff"}
\label{sec:discussion:radius_cliff}

We have seen from the observed marginal distribution of planet radii (Figures \ref{fig:observed_radii_models} and \ref{fig:observed_radii_models_top_depths}) that the hybrid models, unlike the H20 model, appear to produce a drop-off towards large planets that is strikingly similar to that of the \Kepler{} distribution. This steep drop above $\sim 2.5 R_\oplus$ has been referred to as the ``radius cliff" and has been explained by the dissolution of gas from the planet atmosphere into its molten interior, thereby stunting the radial growth of planets over a wide range of envelope masses (the ``fugacity crisis"; \citealt{2019ApJ...887L..33K}). Here, we briefly quantify the drop-off to compare the models and the \Kepler{} data. We follow the methodology of \citet{2023AJ....166..122D, 2024AJ....167..288D} and fit the slope of the radius cliff using a linear function of log-radius:
\begin{equation}
 f(R_p) = m_{\rm cliff} \log_{10}(R_p) + b, \label{eq:radius_cliff}
\end{equation}
where $f(R_p)$ is the ``occurrence" at a given radius, $m_{\rm cliff}$ is the slope of interest, and $b$ is the intercept point. Similar to the procedure in \S\ref{sec:results:radius_valley:depth}, we also use a KDE fit to the observed distribution of $\log_{10}(R_p)$ to estimate $f(R_p)$. We then fit Equation \ref{eq:radius_cliff} to the range 2.5--$5.5 R_\oplus$. This is repeated for the \Kepler{} catalog and each simulated observed catalog.

The resulting values of $m_{\rm cliff}$ are shown in Figure \ref{fig:observed_radius_cliff_slopes}. The \Kepler{} value is denoted by the vertical black line ($m_{\rm cliff} = -0.106$). The H20 model (gray histogram) produces radius cliff slopes that are significantly less steep than the \Kepler{} value ($m_{\rm cliff} = -0.062_{-0.011}^{+0.009}$). The hybrid models (blue and green histograms) generate very similar distributions of slopes, and are both consistent with the \Kepler{} value ($m_{\rm cliff} = -0.092_{-0.013}^{+0.012}$ and $-0.089_{-0.011}^{+0.012}$ for \HMU{} and \HMC{}, respectively). These results confirm that the hybrid models are significantly better than the H20 model for reproducing the observed radius cliff from \Kepler{}.

We note that we have fit the slope of the radius cliff to the \textit{observed} marginal distribution of planet radii, in contrast to \citet{2023AJ....166..122D, 2024AJ....167..288D} who fit Equation \ref{eq:radius_cliff} to the inferred occurrence of planets as a function of planet radii (i.e., to the \textit{de-biased} marginal distribution).\footnote{\citet{2023AJ....166..122D, 2024AJ....167..288D} further divided the \Kepler{} catalog into several period (and stellar type) bins and fit for $m_{\rm cliff}$ in each bin.} The magnitude of the slope also depends on the number/size of bins used for the radius distribution, and thus the normalizations also differ. Therefore, the values of $m_{\rm cliff}$ cannot be directly compared between these studies. The purpose of fitting to the observed radius distribution is to show that the hybrid models provide an excellent match to the \Kepler{}-observed radius cliff and are a substantial improvement over the H20 model.
However, we can also fit to the underlying radius distributions in Figure \ref{fig:underlying_radii_models}. We fit to the range 2.5-$4 R_\oplus$ (corresponding to the steep and nearly constant fall off seen in the right panel of Figure \ref{fig:underlying_radii_models}) using the same method above and find $m_{\rm cliff} = -0.090_{-0.022}^{+0.018}$ and $-0.085_{-0.022}^{+0.021}$ for \HMU{} and \HMC{}, respectively, compared to just $m_{\rm cliff} = -0.032_{-0.008}^{+0.006}$ for the H20 model.



\begin{deluxetable*}{lcccccc}
\centering
\tablecaption{Occurrence rates ($\eta$) of various planet types for each model, and comparisons to previous literature values.}
\tablehead{
 \colhead{Planet bounds} & \colhead{H20 model} & \colhead{\HMU{}} & \colhead{\HMC{}} & \colhead{\NR20{}*} & \multicolumn2c{Other literature value/reference} 
}
\startdata
 $R_p = 0.5$-10, $P = 3$-300 & $2.49_{-0.28}^{+0.37}$ & $2.53_{-0.20}^{+0.19}$ & $2.49_{-0.13}^{+0.23}$ & -- & -- & -- \\[5pt]
 $R_p = 2$-4, $P = 3$-100 & $0.34_{-0.03}^{+0.03}$ & $0.43_{-0.04}^{+0.06}$ & $0.42_{-0.03}^{+0.05}$ & $0.45_{-0.02}^{+0.02}$ & $0.37_{-0.02}^{+0.02}$ & \citet{2017AJ....154..109F}$^a$ \\[5pt]
 $R_p = 1.4$-2.8, $P = 3$-100 & $0.46_{-0.03}^{+0.04}$ & $0.60_{-0.05}^{+0.06}$ & $0.58_{-0.05}^{+0.03}$ & $0.47_{-0.03}^{+0.03}$ & $0.43_{-0.02}^{+0.02}$ & \citet{2017AJ....154..109F}$^a$ \\[5pt]
 $R_p = 0.75$-1.4, $P = 3$-100 & $0.54_{-0.08}^{+0.07}$ & $0.54_{-0.07}^{+0.07}$ & $0.55_{-0.04}^{+0.04}$ & -- & -- & -- \\[5pt]
 $M_p = 50$-1000, $P = 3$-11 & $0.002_{-0.001}^{+0.001}$ & $0.006_{-0.003}^{+0.006}$ & $0.008_{-0.004}^{+0.006}$ & $0.010_{-0.001}^{+0.002}$ & -- & -- \\[5pt]
 $M_p = 3$-10, $P = 3$-50 & $0.29_{-0.02}^{+0.02}$ & $0.23_{-0.02}^{+0.02}$ & $0.22_{-0.01}^{+0.02}$ & $0.22_{-0.01}^{+0.01}$ & $0.12_{-0.04}^{+0.04}$ & \citet{2012ApJS..201...15H} \\[5pt]
 $M_p = 1$-3, $P = 3$-50 & $0.23_{-0.03}^{+0.02}$ & $0.27_{-0.03}^{+0.05}$ & $0.27_{-0.02}^{+0.03}$ & -- & -- & -- \\[5pt]
 $R_p = 0.75$-2.5, $P = 50$-300 & $0.87_{-0.13}^{+0.16}$ & $0.98_{-0.11}^{+0.11}$ & $0.91_{-0.12}^{+0.13}$ & -- & $0.77_{-0.12}^{+0.12}$ & \citet{2015ApJ...809....8B} \\[5pt]
 $R_p = 0.8$-1.2, $P = 180$-270$^\dagger$ & $0.081_{-0.016}^{+0.030}$ & $0.038_{-0.009}^{+0.010}$ & $0.035_{-0.014}^{+0.018}$ & -- & $0.075_{-0.062}^{+0.225}$ & \citet{2015ApJ...809....8B} \\[5pt]
 $R_p = 0.8$-1.2, $P = 180$-270, $M_p = 0.65$-0.98$^\dagger$ & $0.022_{-0.005}^{+0.008}$ & $0.006_{-0.002}^{+0.004}$ & $0.006_{-0.002}^{+0.003}$ & -- & -- & -- \\[5pt]
 $M_p = 0.1$-4, $P = 180$-300, $R_p = (0.9$-$1.1)R_{\rm si}$$^\ddagger$ & $0.21_{-0.05}^{+0.09}$ & $0.075_{-0.027}^{+0.042}$ & $0.079_{-0.023}^{+0.038}$ & -- & -- & -- \\[5pt]
\enddata
\tablecomments{The units for the planet bounds are: $R_p$ [$R_\oplus$], $M_p$ [$M_\oplus$], and $P$ [days].}
\tablenotetext{*}{``Model 2" of \NR20{}; their values include periods down to 0.3 days.}
\tablenotetext{^a}{Includes all periods less than 100 days.}
\tablenotetext{^\dagger}{These bounds correspond to approximately within 20\% of Venus ($P_{\venus} = 224.7$ days and $M_{\venus} = 0.815 M_\oplus$).}
\tablenotetext{^\ddagger}{$R_{\rm si}$ is the radius, as a function of planet mass, given by the pure-silicate model (Equation \ref{eq:radius_mass_pure_silicate}; \citealt{2007ApJ...669.1279S}).}
\label{tab:occurrence_rates}
\end{deluxetable*}

\begin{deluxetable*}{lccccc}
\centering
\tablecaption{Fractions of stars with various planet types ($f_{\rm swp}$) for each model, and comparisons to previous literature values.}
\tablehead{
 \colhead{Planet bounds} & \colhead{H20 model} & \colhead{\HMU{}} & \colhead{\HMC{}} & \multicolumn2c{Other literature value/reference}
}
\startdata
 $R_p = 0.5$-10, $P = 3$-300 & $0.84_{-0.07}^{+0.05}$ & $0.87$* & $0.87$* & -- & -- \\[5pt]
 $R_p = 2$-4, $P = 3$-100 & $0.23_{-0.02}^{+0.03}$ & $0.34_{-0.02}^{+0.03}$ & $0.30_{-0.02}^{+0.02}$ & $0.24_{-0.02}^{+0.02}$ & \citet{2013PNAS..11019273P}$^a$ \\[5pt]
 $R_p = 1.4$-2.8, $P = 3$-100 & $0.30_{-0.02}^{+0.02}$ & $0.42_{-0.03}^{+0.03}$ & $0.37_{-0.02}^{+0.02}$ & $0.33_{-0.01}^{+0.01}$ & \citet{2013PNAS..11019273P}$^a$ \\[5pt]
 $R_p = 0.75$-1.4, $P = 3$-100 & $0.34_{-0.04}^{+0.03}$ & $0.36_{-0.03}^{+0.04}$ & $0.33_{-0.02}^{+0.02}$ & -- & -- \\[5pt]
 $M_p = 50$-1000, $P = 3$-11 & $0.002_{-0.001}^{+0.001}$ & $0.006_{-0.003}^{+0.006}$ & $0.007_{-0.004}^{+0.005}$ & $0.009_{-0.004}^{+0.004}$ & \citet{2011arXiv1109.2497M}$^b$ \\[5pt]
 $M_p = 3$-10, $P = 3$-50 & $0.22_{-0.01}^{+0.02}$ & $0.19_{-0.02}^{+0.02}$ & $0.15_{-0.01}^{+0.01}$ & -- & --\\[5pt]
 $M_p = 1$-3, $P = 3$-50 & $0.18_{-0.02}^{+0.02}$ & $0.22_{-0.03}^{+0.03}$ & $0.17_{-0.02}^{+0.03}$ & -- & -- \\[5pt]
 $R_p = 0.75$-2.5, $P = 50$-300 & $0.55_{-0.07}^{+0.07}$ & $0.63_{-0.05}^{+0.04}$ & $0.58_{-0.05}^{+0.05}$ & -- & -- \\[5pt]
 $R_p = 0.8$-1.2, $P = 180$-270$^\dagger$ & $0.079_{-0.016}^{+0.029}$ & $0.038_{-0.009}^{+0.010}$ & $0.035_{-0.014}^{+0.017}$ & -- & -- \\[5pt]
 $R_p = 0.8$-1.2, $P = 180$-270, $M_p = 0.65$-0.98$^\dagger$ & $0.022_{-0.005}^{+0.008}$ & $0.006_{-0.002}^{+0.004}$ & $0.006_{-0.002}^{+0.003}$ & -- & -- \\[5pt]
 $M_p = 0.1$-4, $P = 180$-300, $R_p = (0.9$-$1.1)R_{\rm si}$$^\ddagger$ & $0.19_{-0.05}^{+0.08}$ & $0.073_{-0.026}^{+0.039}$ & $0.073_{-0.020}^{+0.033}$ & -- & -- \\[5pt]
\enddata
\tablecomments{The units for the planet bounds are: $R_p$ [$R_\oplus$], $M_p$ [$M_\oplus$], and $P$ [days].}
\tablenotetext{*}{The uncertainties are not reported here because they are exceedingly low, due to fixing the parameters for the overall fraction of stars with planets in these models (see discussion at the end of \S\ref{sec:methods:model_inference:summary_stats}).}
\tablenotetext{^a}{Includes periods down to 5 days.}
\tablenotetext{^b}{Includes all periods less than 11 days, and the mass bounds are in $M_p\sin{i}$.}
\tablenotetext{^\dagger}{These bounds correspond to approximately within 20\% of Venus ($P_{\venus} = 224.7$ d and $M_{\venus} = 0.815 M_\oplus$).}
\tablenotetext{^\ddagger}{$R_{\rm si}$ is the radius, as a function of planet mass, given by the pure-silicate model (Equation \ref{eq:radius_mass_pure_silicate}; \citealt{2007ApJ...669.1279S}).}
\label{tab:fswp}
\end{deluxetable*}

\subsection{Occurrence and fractions of stars with planets}
\label{sec:discussion:occurrence_rates}

Here, we compute occurrence rates of various planet types from each of the hybrid models, as well as from the H20 model. A strength of our multi-planet population models is that we can measure and differentiate between the overall occurrence rate ($\eta$; defined as the mean number of planets per star) and the fraction of stars with planets ($f_{\rm swp}$). We compute both of these quantities for a variety of planet bounds of interest, for each model, by counting the number of planets in a given set of bounds (or the number of stars with at least one such planet, for $f_{\rm swp}$) and dividing by the total number of stars in a simulated catalog. The values and uncertainties are computed as the median and 16-84\% quantiles from 100 simulated catalogs drawn from the posteriors of each model, and thus includes both stochastic noise and the uncertainties in the model parameters. Tables \ref{tab:occurrence_rates} and \ref{tab:fswp} list the occurrence rates and fractions of stars with planets, respectively, along with comparisons to the relevant quantities found in the literature (\NR20{} and other references as listed). For the \NR20{} column, we again choose to focus our comparison to their ``Model 2" since our hybrid models are most analogous to that model (but see ``Table 2" of \NR20{} for comparisons to their other models).

First, we focus on the occurrence rates (Table \ref{tab:occurrence_rates}). The first row includes the full simulation bounds of the \SysSim{} models; we find that the overall mean number of planets per star is very similar between all three models ($\eta \simeq 2.5$), despite the different underlying joint mass-radius-period distributions of the hybrid models compared to the H20 model.
The hybrid models predict an occurrence of planets with $R_p = 2$-$4 R_\oplus$ and $P = 3$-100 days that is similar to the \NR20{} model, but higher than in the H20 model. For $R_p = 1.4$-$2.8 R_\oplus$ (i.e. super-Earths to sub-Neptune sized planets), the occurrences from the hybrid models are higher than both the H20 and \NR20{} models.
\NR20{} remarked that their values reflect a similar rate of planets in $R_p = 1.4$-$2.8 R_\oplus$ and $R_p = 2$-$4 R_\oplus$, in contrast to \citet{2017AJ....154..109F} who found that the former are somewhat more common. Interestingly, both of our hybrid models predict an even higher occurrence of the former bin compared to the latter bin (e.g., $\eta = 0.58_{-0.05}^{+0.03}$ vs. $0.42_{-0.03}^{+0.05}$ for \HMC{}) despite being similar in construction to the \NR20{} model.
We also compute the occurrence of smaller planets with $R_p = 0.75$-$1.4 R_\oplus$ and find that it remains the same across the hybrid and H20 models.

The occurrence of planets more massive than $50 M_\oplus$ on short orbits (i.e. hot sub-Saturns and Jupiters) is lowest in the H20 model, and comparable between the hybrid and \NR20{} models. For less massive close-in planets ($M_p = 3$-$10 M_\oplus$, $P = 3$-50 days), their occurrence is highest in the H20 model and somewhat lower in the hybrid models, which are very similar to the \NR20{} model, and all of which are notably higher than \citet{2012ApJS..201...15H} who reported RV survey results.
We also compute the occurrence of even less massive, Earth-massed close-in planets ($M_p = 1$-$3 M_\oplus$, $P = 3$-50 days) and find slightly higher values in the hybrid models compared to the H20 model. Comparing the $M_p = 1$-$3 M_\oplus$ and $M_p = 3$-$10 M_\oplus$ bins, the occurrence seems to flip for the hybrid models (which has slightly higher $\eta$ for the lower masses) compared to the H20 model (slightly higher $\eta$ for the higher masses).

Next, \citet{2015ApJ...809....8B} computed the occurrence of small planets $R_p = 0.75$-$2.5 R_\oplus$ with $P = 50$-300 days and found $\eta = 0.77_{-0.12}^{+0.12}$. We also directly compute the occurrence of these planets as the ranges are fully contained within our simulation bounds, and find somewhat higher values ($\eta = 0.98_{-0.11}^{+0.11}$ and $0.91_{-0.12}^{+0.13}$ for \HMU{} and \HMC{}, respectively). \citet{2015ApJ...809....8B} also computed the occurrence of Venus-type planets, defined as those with radii and periods within roughly 20\% of Venus' values, and reported $\eta_{\venus} = 0.075$ with a large acceptable range of 0.013-0.30. While the H20 model predicts a similar median value (0.081), the hybrid models predict about a factor of two fewer ($0.038_{-0.009}^{+0.010}$ and $0.035_{-0.014}^{+0.018}$ for \HMU{} and \HMC{}, respectively). Moreover, we echo a sentiment expressed in \NR20{}: a strength of modeling the joint mass-radius-period distribution is that we can further restrict the criteria to also require planet masses similar to that of Venus (e.g. also within 20\% of $M_{\venus} = 0.815 M_\oplus$), thus enabling us to compute the occurrence of Venus analogs; we find even lower values ($\eta_{\venus} = 0.006_{-0.002}^{+0.003}$ for the hybrid models).
Lastly, we compute the occurrence of small, long-period planets ($M_p < 4 M_\oplus$, $P = 180$-300 days) with masses and radii that would be consistent with an ``Earth-like" composition (i.e. within 10\% of the radii given by the pure-silicate model of \citealt{2007ApJ...669.1279S}, $R_{\rm si}$ as a function of mass; this is given by Equation \ref{eq:radius_mass_pure_silicate} and is shown as the brown curve in Figures \ref{fig:radius_mass} and \ref{fig:radius_mass_credible}). We find $\eta = 0.075_{-0.027}^{+0.042}$ ($0.079_{-0.023}^{+0.038}$) for \HMU{} (\HMC{}), which is roughly a factor of $\sim 3$ times lower than in the H20 model.

Finally, we discuss the model predictions for the fractions of stars with planets (Table \ref{tab:fswp}). By definition, $f_{\rm swp}$ must be between zero and one, and these values differ most substantially from $\eta$ when multiple planets in a given range can exist in the same system. The fraction of stars with planets in the full simulation bounds was effectively fixed in the hybrid models in order to reduce the number of free parameters (see the discussion at the end of \S\ref{sec:methods:model_inference:summary_stats}), and thus they have arbitrarily small uncertainties (arising only from stochastic noise), and are similar to the H20 model. For super-Earth to sub-Neptune sized planets in broad period bounds, $f_{\rm swp}$ is generally lower than $\eta$ in the corresponding bin by up to $\sim 50\%$ (i.e. comparing Table \ref{tab:fswp} to Table \ref{tab:occurrence_rates}). This implies that a substantial fraction of systems with these types of planets have two or more such planets. While the H20 model appears to agree well with \citet{2013PNAS..11019273P} for $R_p = 1.4$-$2.8 R_\oplus$ and $R_p = 2$-$4 R_\oplus$, the hybrid models predict notably higher values (although \HMC{} roughly splits the differences). As expected, there is very little difference between $f_{\rm swp}$ and $\eta$ for hot sub-Saturns to Jupiters, and the hybrid models are consistent with the results from \citet{2011arXiv1109.2497M} (unlike the H20 model, which produces too few such planets). Likewise, the fraction of stars with Venus-sized planets/analogs are virtually unchanged from their occurrence rates in Table \ref{tab:occurrence_rates}.
Similarly, there is little difference when distinguishing between $f_{\rm swp}$ and $\eta$ for ``Earth-like"/pure-silicate composition planets for the period range we have chosen (limited by the upper simulation bound of 300 days), and again the hybrid models predict significantly lower values than the H20 model.

In choosing the specific occurrence rates and fractions of stars with planets to compute, we have avoided extrapolating our models to longer periods beyond our simulation bounds ($P > 300$ days), such as would be required to estimate the occurrence of Earth-sized planets in the habitable zones of their host stars (i.e. ``$\eta_\oplus$"). This is because of (1) the difficulty of extrapolating our population models (due to the nature of drawing a number of planet clusters within the pre-defined period bounds, and the AMD stability criteria which assumes that the system's total AMD budget can be freely exchanged between all the planets), and (2) the risks of extrapolation in general, which assumes that the model definition (e.g., the power-law in orbital period) holds in a region of parameter space in which there is little to no available data to constrain the model.
However, our upper period bound of 300 days overlaps with the inner region of the optimistic habitable zone for Solar type stars (inner boundary of 237 days; \citealt{2013ApJ...765..131K}) which is widely used for estimates of $\eta_\oplus$ (e.g., \citealt{2019AJ....158..109H, 2021AJ....161...36B, 2020AJ....159..248K}). Thus, our computed occurrence rates for the planets of ``Earth-like" composition (the bottom row of Tables \ref{tab:occurrence_rates} and \ref{tab:fswp}) can be considered as a proxy for $\eta_\oplus$, and our model comparisons provide a similar finding to \NR20{}.
In particular, \NR20{} showed that their estimates of $\eta_\oplus$ drops by nearly an order of magnitude when including envelope mass-loss in their models compared to a baseline model with no envelope mass-loss (their ``Model 1"). A similar conclusion was also drawn by \citet{2019ApJ...883L..15P}, who fit broken power-laws to the planet radius distribution and found a factor of $\sim 4$-8 drop in the inferred $\eta_\oplus$ when excluding planets below $P = 25$ days (which contains much of the observed population of rocky cores below the radius valley) compared to including them. Likewise, we find roughly a factor of $\sim 2$-4 drop in $\eta_{\venus}$ (and a factor of $\sim 3$ fewer for planets of ``Earth-like" composition in $P = 180$-300 days) when comparing our hybrid models to the H20 model (in which there is also no envelope mass-loss).
Thus, we reiterate a key point previously stated by \NR20{}: occurrence rates are highly model dependent and care must be taken to avoid over-interpretation. Models that are missing key physical processes/correlations (such as ``Model 1" of \NR20{}, or the H20 model, both of which assume period distributions that are independent of the planet radius and mass distributions) can potentially lead to significantly over-estimated occurrence rates for small planets at longer periods where there is less data (e.g. for $\eta_{\venus}$, and even more so, for $\eta_\oplus$). These results underscore the importance of including a physically motivated model for envelope mass-loss in a population model.

In addition to the occurrence rates and fractions of stars with planets presented in Tables \ref{tab:occurrence_rates} and \ref{tab:fswp}, we provide a functionality for computing $\eta$ and $f_{\rm swp}$ for any given type of planet (given user-defined bounds in planet radius, mass, and/or period, within the simulation bounds) as part of the publicly available \texttt{SysSimPyPlots}\footnote{\url{https://syssimpyplots.readthedocs.io/en/latest/}} package in Python.

\subsection{Limitations and avenues for future improvements}
\label{sec:discussion:limitations}



Perhaps the biggest weakness of the previous H20 model was in how it modeled the planet radius distribution, with an underlying simple broken power-law that could not produce any form resembling a radius valley. \NR20{} provided a framework for modeling the underlying joint mass-radius-period distribution subject to photoevaporation for sculpting the radius valley, though they treated planets independently and thus any correlations between planets in multi-planet systems were not considered. In this paper, we have significantly improved the leading \SysSim{} model for describing the planet radius distribution, by constructing a hybrid model that combines the ``best of both worlds" in order to reproduce both the observed radius valley and the size similarity patterns in multi-planet systems, as well as the broad orbital architectures of these multi-planet systems. However, there are also several limitations of the new model, motivating potential avenues for future improvements, which we describe here.

As discussed in \S\ref{sec:discussion:distance_function}, our distance function is not very sensitive to the observed radius valley, despite including terms for the marginal radius distribution and gap-subtracted radius distribution (KS distances between the \Kepler{} and simulated catalogs). While the models could be further tuned by including a term involving $\Delta_{\rm valley}$ (in order to further extract valley-like features in the radius distribution), another approach could be to also include fits to the 2-d period-radius distribution. In Figure \ref{fig:observed_period_radius}, we showed that the hybrid models as currently constrained can produce observed period-radius distributions that qualitatively resemble that of the \Kepler{} catalog, though there are differences and not all simulated catalogs are this similar or even exhibit a clear radius valley. Directly including a distance term for the 2-d period-radius distribution (such as the KS distance generalized to two dimensions) may provide an even better match and thus stronger constraints on the models.

Another limiting factor in our model inference is the absence of planet masses from the data used to constrain the models. Although the vast majority of the \Kepler{} planet candidates do not have mass measurements, a small fraction of them (a few hundred planets) have masses measured via either radial velocity (RV) observations or transit timing variations (TTVs). However, these masses are biased in ways that are difficult if not impossible to quantify (e.g., they are highly incomplete for the smallest planets, the RV masses are strongly biased to bright and nearby stars for which precise RV measurements are possible, and there are also systematic differences in the densities of the planets with RV versus TTV masses; \citealt{2014ApJ...785...15J, 2016MNRAS.457.4384S, 2017ApJ...839L...8M, 2024A&A...687L...1L}). 
Despite these factors, \NR20{} included mass constraints from a small subset of planets in the California-\Kepler{} Survey (CKS; \citealt{2017AJ....154..107P, 2017AJ....154..109F, 2018AJ....156..264F})-\Gaia{} sample with RV masses (53 planets) in their Bayesian inference to help constrain the planet radius-mass relation, along with reasonably informative priors for the relevant model parameters. By excluding mass measurements altogether, we also had to choose rather restrictive optimizer bounds/priors (including fixing the break mass and the radius-mass relation above it, $M_{p,\rm break}$, $\gamma_1$, and $\sigma_1$, which were poorly constrained when allowed to vary in the initial run) and there is some reliance on these bounds to avoid unphysical models. Future work is needed to explore how constraints from the sample of planet masses can be incorporated in a way that adequately accounts for the biases of the mass measurements.

A simplification of our model is that we have assumed the same stellar age (a nominal value of 5 Gyr) for all stars when estimating the probability for envelope mass-loss, as was also done in \NR20{}. However, stellar ages for the majority of the \Kepler{} FGK dwarfs are available and updated in the GKSPC (\citealt{2020AJ....159..280B}; though the uncertainties can be large, with a median fractional uncertainty of 56\%). A more detailed implementation would be to use the individual stellar ages for calculating the envelope mass-loss probability of the simulated planets, which are assigned to specific \Kepler{} target stars. While the other specific stellar properties (namely the stellar mass, radius, and effective temperature) affecting the bolometric fluxes received by the planets, as well as their detectability by a \Kepler{}-like mission, are accounted for in our forward model, there may be additional correlations with stellar age that could provide a more accurate simulation. In particular, there is evidence of a shift in the radius distribution across the radius valley, with a larger fraction of super-Earths relative to sub-Neptunes for planets around older stars ($\tau \gtrsim 1$ Gyr) than younger stars (\citealt{2020AJ....160..108B}; see also \citealt{2021AJ....161..265D, 2021ApJ...911..117S, 2023AJ....166..248C, 2024AJ....167..210V}).

As in \NR20{}, we have also simulated conditions for envelope mass-loss due solely to photoevaporation. Other physical processes have been shown to be also capable of producing a planet radius valley. Two competitive theories are the mechanisms of core-powered mass-loss (CPML; \citealt{2018MNRAS.476..759G, 2019MNRAS.487...24G, 2020MNRAS.493..792G}) and envelope stripping due to giant impacts \citep{2022ApJ...939L..19I}. Recent studies have shown that CPML may explain the radius valley just as well as photoevaporation \citep{2020ApJ...890...23L, 2021MNRAS.508.5886R, 2023MNRAS.519.4056H, 2023arXiv230200009B}. Though outside the scope of this paper, future work could attempt to simulate these processes in a detailed multi-planet population model.

Finally, we have attempted to keep the hybrid models simple by incorporating only a ``single" primordial population subject to photoevaporation (i.e., defined by a single mass-radius-period distribution from which all planets are drawn) for explaining the bulk of the \Kepler{} planet candidates. However, this does not preclude other populations of planets from existing, or even from describing the same set of \Kepler{}-observed planets. Indeed, \NR20{} also explored the need for mixture models involving multiple underlying populations of planets, with models of increasing complexity including an intrinsically rocky population (their ``model 3") and additionally even two initially gaseous populations with separate lognormal mass distributions (their ``model 4"). In fact, \NR20{} found that these more complex models are equally preferred over the simpler model with a single population subject to photoevaporation driven envelope mass-loss (their ``model 2", which our hybrid models encapsulate). Including several model components (and therefore having even more model parameters) would be challenging to constrain with our ABC methodology due to the increased computational cost of adding additional model parameters. More modern approaches to simulation-based inference (SBI) involving neural posteriors or normalizing flows may offer a practical approach to exploring such models \citep[e.g.,][]{2017arXiv170507057P, 2019arXiv190507488G, 2019MNRAS.488.4440A, 2020PNAS..11730055C}. Thus, future work could further explore the need for additional underlying populations, perhaps including water worlds or planets of varying compositions.

\section{Summary and Conclusions} \label{sec:summary}

With the goal of finding a model that can simultaneously reproduce the \Kepler{}-observed planet radius valley and intra-system size similarity patterns, we have constructed a ``hybrid" model between the H20 and \NR20{} models by combining features of a clustered multi-planet model in which systems are at the AMD-stability limit (\PaperIII{}) with a joint mass-radius-period distribution sculpted by photoevaporation (\NR20{}). We compared two versions of the hybrid model, one with unclustered initial masses (\HMU{}) and one with clustered initial masses (\HMC{}). Using \SysSim{}, we constrained the parameters of the hybrid models by forward modeling the \Kepler{} catalog via Approximate Bayesian Computation in order to fit to multiple observed marginal distributions. Our main findings are summarized below:
\begin{itemize}
    \item The hybrid models are capable of generating an observed planet radius valley similar to that seen in the \Kepler{} data (i.e. a local minimum at $\sim 2 R_\oplus$), given appropriate choices of the model parameters. However, these parameters are difficult to constrain given our choice of distance function; the KS distance for the marginal radius distribution (or gap-subtracted radius distribution) is not sensitive enough to adequately distinguish between catalogs that exhibit a radius valley versus equally close distributions that do not.
    \item We devise a simple measure of the ``depth" of the radius valley, $\Delta_{\rm valley}$, which ranges from 0 (no valley) to 1 (the valley is completely cleared out), to rank simulated catalogs by the strength of the radius valley. The \Kepler{} catalog exhibits a depth of $\Delta_{\rm valley} = 0.44$. The top 10\% of simulated catalogs with the strongest radius valleys, drawn from the hybrid model posteriors, have $\Delta_{\rm valley} \gtrsim 0.31$ (0.26) for \HMU{} (\HMC{}).
    \item The hybrid model with clustered initial planet masses (\HMC{}) provides a substantially better fit to the \Kepler{} catalog than the hybrid model with unclustered masses (\HMU{}). Compared to \HMU{}, \HMC{} significantly improves the fits to the marginal distributions of the metrics designed to capture the observed patterns of planet size similarity and ordering in multi-planet systems. 
    Thus, we find strong evidence for intra-system mass similarity. These results also demonstrate that the observed preference for similar planet radii within a system can be fully explained by a primordial clustering in the initial planet masses (before the onset of envelope mass-loss), signatures of which remain even after sculpting by photoevaporation.
    \item The hybrid models also naturally reproduce the observed ``radius cliff" (the steep drop-off beyond $\sim 2.5 R_\oplus$). The slope of the observed radius cliff (measured by a linear fit to the distribution of planet radii between $2.5$-$5.5 R_\oplus$; Equation \ref{eq:radius_cliff}) is $m_{\rm cliff} = -0.092_{-0.013}^{+0.012}$ ($-0.089_{-0.011}^{+0.012}$) for \HMU{} (\HMC{}), compared to the \Kepler{} value of $-0.106$. This is a significant improvement over the H20 model, which produces a drop-off that is too shallow ($m_{\rm cliff} = -0.062_{-0.011}^{+0.009}$).
\end{itemize}
Altogether, HM-C is the first model that can simultaneously reproduce three defining features of the distribution of planet sizes from \Kepler{}'s transiting exoplanets: (1) the radius valley, (2) the radius cliff, and (3) the degree of radius similarity in multi-planet systems.

Lastly, we used the hybrid models to compute occurrence rates and fractions of stars with planets for a variety of planet bounds of interest. 
For super-Earth to sub-Neptune sized planets, we find a modestly higher occurrence of planets in $R_p = 1.4$-$2.8 R_\oplus$ than in 2-$4 R_\oplus$, within $P = 3$-100 days ($\eta = 0.58_{-0.05}^{+0.03}$ versus $0.42_{-0.03}^{+0.05}$ for \HMC{}), in contrast to \NR20{} who find a similar occurrence in both ranges, and which are systematically higher than in the H20 model and also in \citet{2017AJ....154..109F}.
The occurrence of hot sub-Saturn to Jupiter mass planets ($M_p = 50$-$1000 M_\oplus$, $P = 3$-11 days) is consistent between the hybrid models and \NR20{} (and the fraction of stars with these planets is also consistent between the hybrid models and \citealt{2011arXiv1109.2497M}), but much lower in the H20 model.
Finally, we compute the occurrence of Venus-type planets ($\eta_{\venus}$) and ``Earth-like" composition planets within $P = 180$-300 days (a shorter-period proxy for $\eta_\oplus$ that does not require extrapolation). We find that these occurrence rates drop by a factor of $\sim 2$-4 when comparing the hybrid models to the H20 model, further underscoring the importance of accounting for the role of envelope mass-loss in shaping the radius distribution of short period planets compared to those further out. This implies that a significantly larger number of stars would potentially need to be surveyed by the Habitable Worlds Observatory (HWO; \citealt{2024JATIS..10c4006S}) in order to reach its desired sample size of 25 exo-Earth candidates. We also find that $\eta_{\venus}$ drops by a factor of $\sim 6$ when further requiring the planet masses (in addition to the radii and periods) to be within 20\% of Venus' values, highlighting the importance of jointly modeling the underlying mass-radius-period distribution.

We provide a new functionality in the publicly available \texttt{SysSimPyPlots} package that allows users to compute occurrence rates and fractions of stars with planets from the hybrid models, for any given bounds in planet radius, mass, and/or period within the \SysSim{} simulation bounds.

\begin{acknowledgments}

We thank the entire \Kepler{} team for years of work leading to a successful mission and data products critical to this study.  
We acknowledge many valuable contributions with members of the \Kepler{} Science Team's working groups on multiple body systems, transit timing variations, and completeness working groups.  
We thank Darin Ragozzine, Danley Hsu, Robert Morehead, and Keir Ashby for contributions to the broader \SysSim{} project.
We thank Lauren Weiss, Daniel Fabrycky, Leslie Rogers, Angie Wolfgang, Steve Bryson, Aritra Chakrabarty, Sarah Millholland, James Rogers, Ilaria Pascucci, Songhu Wang, Jack Lissauer, and Daniel Jontof-Hutter for useful discussions.
%
%
M.Y.H. acknowledges support from the NASA Exoplanets Research Program NNH22ZDA001N-XRP (grant \#80NSSC23K0269).
This work was enabled in part by the Center for Research Computing (CRC) at the University of Notre Dame and by the NASA Advanced Supercomputing (NAS) resources at the NASA Ames Research Center.
This manuscript incorporates ideas and code developed with support of NASA XRP award NNX15AE21G.
This research was also sponsored by NASA through a contract with Oak Ridge Associated Universities (ORAU). The views and conclusions contained in this document are those of the authors and should not be interpreted as representing the official policies, either expressed or implied, of NASA or the U.S. Government. The U.S. Government is authorized to reproduce and distribute reprints for Government purposes notwithstanding any copyright notation herein.
The Center for Exoplanets and Habitable Worlds is supported by the Pennsylvania State University and the Eberly College of Science.
The citations in this paper have made use of NASA's Astrophysics Data System Bibliographic Services.
This work has made use of the NASA Exoplanet Archive, which is operated by the California Institute of Technology, under contract with NASA under the Exoplanet Exploration Program.
%
%

\software{NumPy \citep{2020Natur.585..357H},
          Matplotlib \citep{2007CSE.....9...90H},
          ExoplanetsSysSim \citep{eric_ford_2022_5915004},
          SysSimData \citep{eric_ford_2019_3255313},
          SysSimExClusters \citep{matthias_yang_he_2022_5963884},
          SysSimPyPlots \citep{matthias_yang_he_2022_7098044},
          gapfit \citep{2020ApJ...890...23L}
          }

\end{acknowledgments}


\bibliographystyle{aasjournal}
\bibliography{main}



\appendix

\renewcommand{\thefigure}{A\arabic{figure}}
\renewcommand{\thetable}{A\arabic{table}}
\setcounter{figure}{0}
\setcounter{table}{0}

\begin{figure*}
\centering
\includegraphics[scale=0.35,trim={0.1cm 0.1cm 0.2cm 0.1cm},clip]{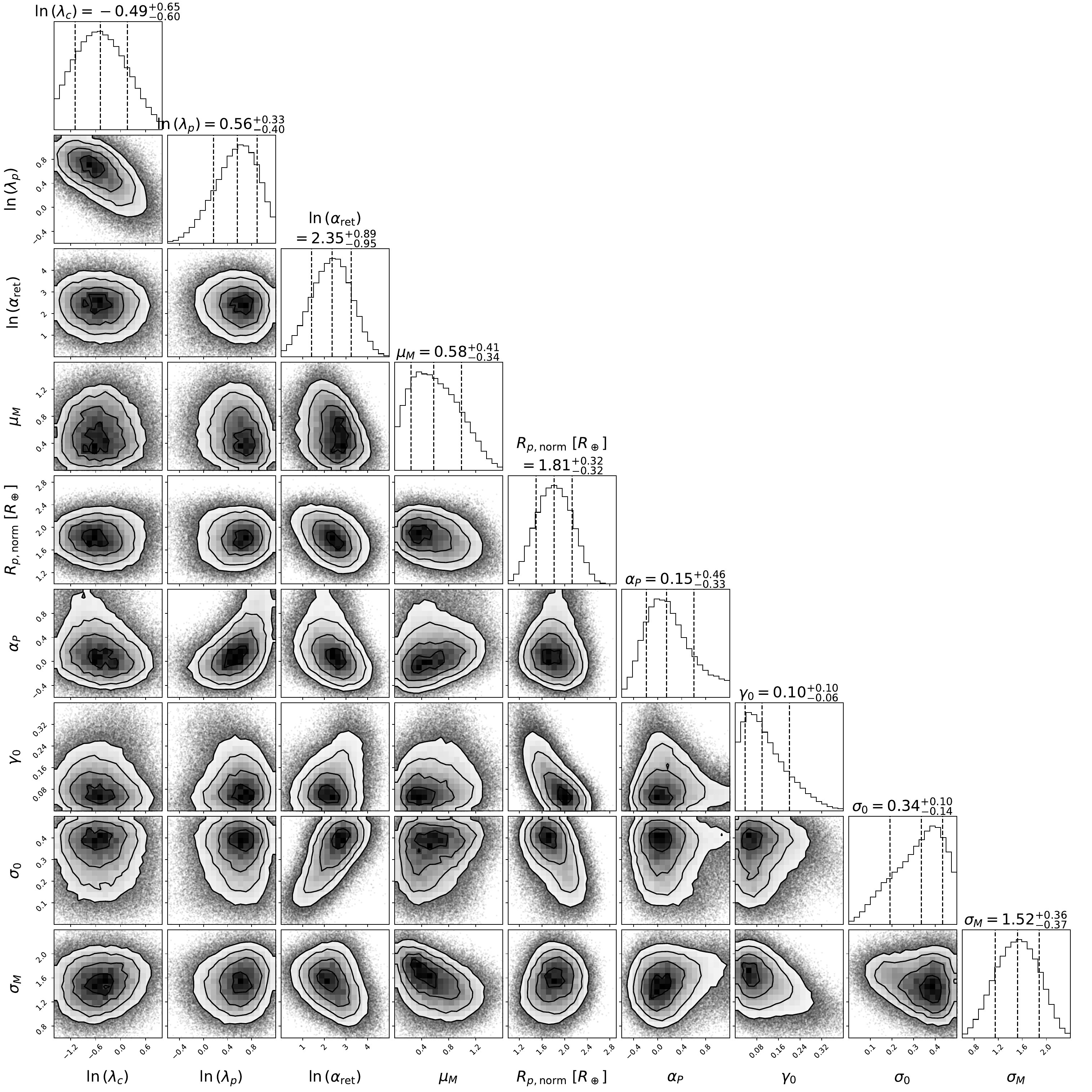} 
\caption{ABC posterior distribution of the free parameters from Run 1 of \HMU{}. A distance threshold of $\mathcal{D}_W \leq 20$ is used and $10^5$ points passing this threshold as evaluated using the GP emulator are plotted. The median and central 68.3\% credible intervals for each parameter are labeled above each histogram, and are also listed in Table \ref{tab:param_fits}.}
\label{fig:hm1_posterior_params9}
\end{figure*}

\begin{figure*}
\centering
\includegraphics[scale=0.45,trim={0.5cm 0.3cm 0.3cm 0.1cm},clip]{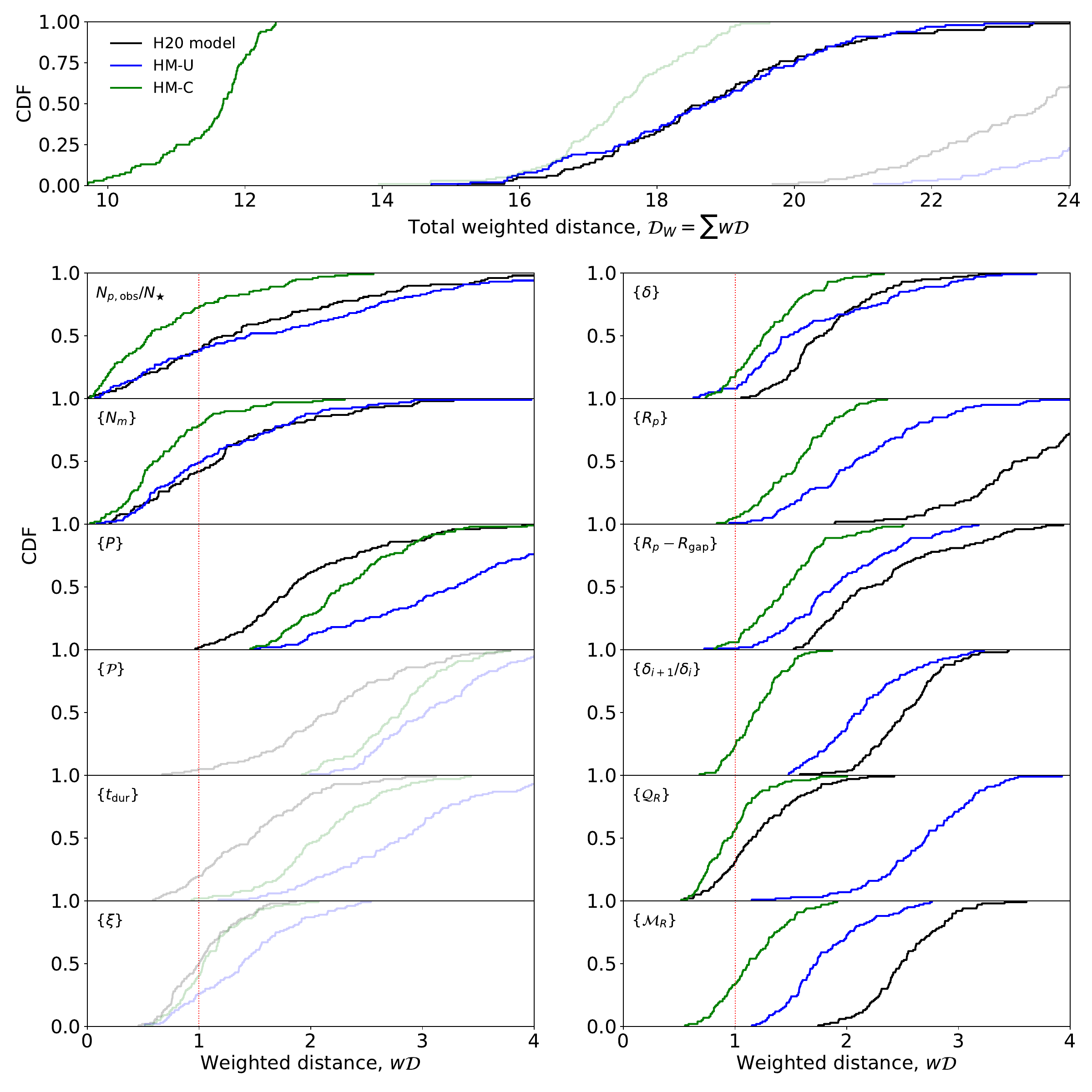} 
\caption{Distributions of weighted distances for each summary statistic, of each model compared to the \Kepler{} catalog. The total weighted distance (as computed using Equation \ref{eq:dist}) is shown in the top panel, while the individual (weighted) distance terms are shown in the lower sub-panels (with each summary statistic as labeled). Three distance terms ($\{\mathcal{P}\}$, $\{t_{\rm dur}\}$, and $\{\xi\}$) were not included in the total distance function during the final optimization runs (``Run 1" of each model) described in \S\ref{sec:methods:model_inference:optimization}, and are thus plotted as faded lines. Likewise, the faded lines in the top panel indicate the distributions of total weighted distances with these three distance terms also included. \HMC{} (green) is a significantly better fit to the \Kepler{} catalog than the other two models, with improvements to the fits of most of the summary statistics included (except the period distribution, $\{P\}$). Compared to the H20 model (black), the most dramatic improvements are the fits to the observed distributions of planet radii ($\{R_p\}$), depth ratios ($\{\delta_{i+1}/\delta_i\}$), and radii monotonicity ($\{\mathcal{M}_R\}$). \HMU{} (blue) also improves the fits to some of these summary statistics but worsens others, leading to a similar distribution of the total weighted distance compared to the H20 model.}
\label{fig:dists}
\end{figure*}

\end{document}